\documentclass[prd,aps,twocolumn,a4paper,floatfix,nofootinbib]{revtex4-1}

\usepackage[utf8]{inputenc}
\usepackage{graphicx,psfrag}
\usepackage{mathrsfs}
\usepackage{amsmath,amsfonts,amssymb}
\usepackage{multirow}
\usepackage{diagbox}
\usepackage{comment}
\usepackage{xcolor}
\usepackage{enumerate}
\usepackage{booktabs}
\usepackage[normalem]{ulem}
\usepackage{url}

\DeclareMathOperator*{\argmax}{arg\,max}

\definecolor{BerlinU9}{HTML}{F18800}

\usepackage[colorlinks=true]{hyperref}
\hypersetup{
    colorlinks = true,
    linkcolor = BerlinU9,
    urlcolor = BerlinU9,
    linkbordercolor = {white},
    citecolor = BerlinU9,
    }
\usepackage{orcidlink}

\usepackage{color}
\definecolor{cyan}{rgb}{0,0.9,0.9}
\definecolor{orange}{rgb}{0.9,0.5,0}
\definecolor{magenta}{rgb}{1,0,1}
\definecolor{purple}{rgb}{0.8,0.4,0.8}
\definecolor{gray}{rgb}{0.8242,0.8242,0.8242}
\definecolor{green}{rgb}{0.,0.8,0.}

\def\bam{{\texttt{BAM}}}
\def\possis{{\texttt{POSSIS}}}
\def\fuka{{\texttt{FUKA}}}
\def\lorene{{\texttt{LORENE}}}
\def\imrphenomnsbh{{\texttt{IMRPhenomNSBH}}}
\def\seobnsbh{{\texttt{SEOBNRv4T\_ROM\_NRTidalv2\_NSBH}}}
\def\imrphenomdnrtidal{{\texttt{IMRPhenomD\_NRTidalv2}}}
\def\seobnr{{\texttt{SEOBNRv4T}}}

\def\lalsuite{{\texttt{LALSuite}}}
\def\surfinBH{{\texttt{surfinBH}}}
\def\nrsurremnant{{\texttt{NRSur7dq4Remnant}}}

\newcommand{\NSbh}[1]{$\mathrm{NSbh}_{#1}$}
\newcommand{\Tms}[1]{$t=#1\ \mathrm{ms}$}

\begin{document}

\title{General-relativistic hydrodynamics simulation of a neutron star -- sub-solar-mass black hole merger}

\author{Ivan \surname{Markin}\,\orcidlink{0000-0001-5731-1633}$^{1,2}$}
\author{Anna \surname{Neuweiler}\,\orcidlink{0000-0003-3205-8373}$^{1}$}
\author{Adrian \surname{Abac}\,\orcidlink{0000-0003-4786-2698}$^{1,3}$}
\author{Swami Vivekanandji \surname{Chaurasia}$^{4}$}
\author{Maximiliano \surname{Ujevic}\,\orcidlink{0000-0003-2869-4449}$^{5}$}
\author{Mattia \surname{Bulla}\,\orcidlink{0000-0002-8255-5127}$^{6,7,8}$}
\author{Tim \surname{Dietrich}\,\orcidlink{0000-0003-2374-307X}$^{1,3}$}
\affiliation{${}^1$Institut f\"{u}r Physik und Astronomie, Universit\"{a}t Potsdam, Haus 28, Karl-Liebknecht-Straße 24/25, 14476, Potsdam, Germany}
\affiliation{${}^2$Leibniz-Institute for Astrophysics Potsdam (AIP), An der Sternwarte 16, 14482 Potsdam, Germany}
\affiliation{${}^3$Max Planck Institute for Gravitational Physics (Albert Einstein Institute), Am M\"uhlenberg 1, Potsdam 14476, Germany}
\affiliation{${}^4$The Oskar Klein Centre, Department of Astronomy, Stockholm University, AlbaNova, SE-10691 Stockholm,
Sweden}
\affiliation{${}^5$Centro de Ciências Naturais e Humanas, Universidade Federal do ABC, Santo André 09210-170, SP, Brazil}
\affiliation{${}^6$Department of Physics and Earth Science, University of Ferrara, via Saragat 1, I-44122 Ferrara, Italy}
\affiliation{${}^7$INFN, Sezione di Ferrara, via Saragat 1, I-44122 Ferrara, Italy}
\affiliation{${}^8$INAF, Osservatorio Astronomico d’Abruzzo, via Mentore Maggini snc, 64100 Teramo, Italy}

\date{\today}

\begin{abstract}
Over the last few years, there has been an increasing interest in sub-solar mass black holes due to their potential to provide valuable information about cosmology or the black hole population. Motivated by this, we study observable phenomena connected to the merger of a sub-solar mass black hole with a neutron star. 
For this purpose, we perform new numerical-relativity simulations of a binary system composed of a black hole with mass $0.5M_\odot$ and a neutron star with mass $1.4 M_\odot$. 
We investigate the merger dynamics of this exotic system and provide information about the connected gravitational-wave and kilonova signals. Our study indicates that current gravitational-waveform models cannot adequately describe such systems and that phenomenological relations connecting the binary parameters with the ejecta and remnant properties do not apply to our system. Furthermore, we find a dependence of the kilonova signal on the azimuthal viewing angle due to the asymmetric mass ejection. This first-of-its-kind simulation opens the door for studying sub-solar mass black hole - neutron star mergers and could serve as a testing ground for future model development.
\end{abstract}

\maketitle

\section{Introduction}
\label{sec:intro}
The first detection of gravitational waves (GWs) from a binary black hole (BBH) merger in 2015~\cite{LIGOScientific:2016aoc} inaugurated a new era in astronomy. Since then, almost one hundred compact binary mergers have been detected, including the observation of a binary neutron star (BNS) system (GW170817) accompanied by electromagnetic (EM) counterparts~\cite{LIGOScientific:2017vwq,LIGOScientific:2017ync} and the black hole - neutron star (BHNS) detection GW200115\footnote{In the absence of an EM counterpart, the classification as BHNS merger is based on the fact that the mass of the secondary component is consistent with expectations for an NS. However, the possibility that the secondary component was a light BH instead remains.}~\cite{LIGOScientific:2021qlt}. Due to the increasing sensitivity of the existing GW observatories~\cite{LIGOScientific:2014pky,VIRGO:2014yos} and the planned next generation of GW detectors~\cite{Punturo:2010zz,Reitze:2019iox,LIGO:2020xsf,Ackley:2020atn}, we expect to detect many more compact binary mergers in the near future. 

Until now, the BHs that have been detected via GWs have masses that are typically larger than those discovered in X-ray binaries~\cite{Ozel:2010su,Farr:2010tu,Jonker:2021rkb,LIGOScientific:2018mvr,LIGOScientific:2020ibl,LIGOScientific:2021djp}. 
The lightest compact object that has been observed via GWs and was very likely a BH, was the secondary component in GW190814 with a mass of about $2.6 M_\odot$~\cite{LIGOScientific:2020zkf}. While such light BHs might form through previous compact binary mergers, cf.\ GW170817, hardly any astrophysical evolutionary processes predict the formation of BHs with even smaller masses, particularly with sub-solar mass (SSM). Nevertheless, SSM BHs are particularly interesting as they might indicate new formation mechanisms and potentially new physics.

One possible scenario for the formation of SSM BHs is the gravitational collapse of overdensities in the early Universe that could result in primordial BHs (PBHs), e.g.,  ~\cite{Garcia-Bellido:1996mdl,Dolgov2000,Khlopov:2008qy,Green:2020jor}. Specifically, these PBHs can form during quantum chromodynamic (QCD) phase transition, producing a population with a peak around $1~M_{\odot}$~\cite{Jedamzik:1996mr,Byrnes:2018clq,Carr:2019kxo,Carr:2020xqk}. This formation channel might be supported by recent GW detections~\cite{Clesse:2020ghq} of compact binary coalescences that challenge the astrophysical formation scenarios of BHs.
 SSM BHs can also form as a result of dynamical capture of small PBHs ($10^{-16} < M/M_{\odot} < 10^{-7}$) by white dwarfs (WDs) or NSs, where a significant part of the compact star matter falls onto the BH~\cite{Capela:2013yf,Sasaki:2018dmp,Bramante:2017ulk,Takhistov:2017nmt}. This can result in a remnant BH mass of $0.5 M_{\odot} - 1.4 M_{\odot}$ after interaction with a WD, which can lead to the formation of a BHNS binary with a SSM BH~\cite{Takhistov:2017bpt,Sasaki:2021iuc}. Another possible mechanism for SSM BH formation might be the gravitational collapse of dark matter halos~\cite{Shandera:2018xkn}.

So far, searches for compact binaries with at least one
SSM component have been unsuccessful in finding any evidence for this class of objects, e.g., \cite{LIGOScientific:2018glc,LIGOScientific:2019kan,Nitz:2021vqh,LIGOScientific:2021job,Nitz:2022ltl}, but similar searches are planned for the next observing runs, and, with increasing sensitivity and redshift reach of the GW detectors, the chances of success are continuously rising. 

In light of the large interest in SSM BHs and possible multimessenger sources, we are focusing in this paper on new numerical-relativity (NR) simulations of BHNS systems for which the BH has a sub-solar mass, which we are referring to as NSbh hereafter. NSbh simulations could also be of particular interest for the future development of GW models for compact binary systems containing at least one NS since they provide a testing ground for calibration and validation of existing models, e.g.,~\cite{Hinderer:2016eia,Nagar:2018zoe,Thompson:2020nei,Matas:2020wab}, outside their original calibration region. \\

The article is structured as follows. In Sec.~\ref{sec:methods}, we discuss the employed numerical methods and highlight changes between our previous BHNS studies~\cite{Chaurasia:2021zgt,Rashti:2021ihv} and this study.
In Sec.~\ref{sec:simulation}, we present the physical configuration that we performed and our simulation results. In Sec.~\ref{sec:observables}, we discuss observable GW and kilonova signatures connected to our simulations. We conclude in Sec.~\ref{sec:conclusion}. Furthermore, in the appendix, we provide information about simulations with a different grid setup and the input variables for computing the kilonova light curves. Unless otherwise stated, this article uses geometric units, with $c=G=1$ and $M_\odot=1$.

\section{Methods and Setups}
\label{sec:methods}

In this paper, we employ the NR code \bam\ for our dynamical simulations. Throughout this work, we follow mainly Ref.~\cite{Chaurasia:2021zgt} unless stated otherwise. 

\subsection{Upgrades to use \fuka\ Initial data}

In contrast to Refs.~\cite{Chaurasia:2021zgt,Rashti:2021ihv}, simulations shown in this paper use initial data computed with the publicly available \fuka\ code~\cite{Papenfort:2021hod}. \fuka\ is a spectral solver to construct consistent and constraint-solved initial data using the eXtended Conformal Thin-Sandwich formulation of Einstein's field equations~\cite{Wilson:1995uh,Wilson:1996ty,York:1998hy}. The advantage of \fuka~compared to \lorene~\cite{LoreneCode} is that the code allows us to compute initial configurations for a large variety of configurations, including extremely compact, asymmetric, and spinning binaries. Similarly, the code is noticeably more tested than \texttt{Elliptica}~\cite{Rashti:2021ihv} and has been employed in numerous other NR studies, e.g., \cite{Papenfort:2021hod,Demircik:2022uol,Papenfort:2022ywx,Tootle:2021umi}. Finally, using \fuka\ also leads to more accurate initial data, wrt.\ constraint violations. Due to this improvement, we find that we can perform more reliable simulations with smaller constraint damping parameters ($\kappa_1=0.02, \kappa_2=0.0$) compared to Ref.~\cite{Chaurasia:2021zgt}.

Thanks to the modular architecture of \bam\ and \fuka, both codes have been easily extended to load the initial data. We computed \fuka~initial data at resolution 15, i.e., 15 collocation points in $r$ and $\theta$ directions, and 14 in $\phi$ direction for every computational grid domain. We used two additional shells around the BH, which increases the resolution in this region without changing the global one and thus further reduces the constraint violations. Then, the spectral data are imported from \fuka\ onto \bam's Cartesian grid by first constructing the simulation grid. Then we evaluate the geometric and hydrodynamic fields at these Cartesian coordinates utilizing \texttt{exporter} provided in \fuka. \texttt{exporter} handles the excised interior of the BH by filling it with constraint-violating initial data using polynomial extrapolation of the fields outside the horizon. 

\subsection{Tracking the Black Hole}

We also modify the tracking method to follow the BH motion, compared to~\cite{Chaurasia:2021zgt}. In Ref.~\cite{Chaurasia:2021zgt}, we used the shift $\beta^i$ to track the position $x^i _{\rm punc}$ of the puncture by integrating:
\begin{align}
\label{eqn:track_puncture}
\partial _t {x ^i _{\rm punc}} = -\beta ^i (x ^j _{\rm punc});
\end{align}
cf.~\cite{Campanelli:2005dd}. Here, we track both compact objects (the NS and the SSM BH) by locating the minimum of the lapse within the finest refinement levels that cover each of the compact objects. This approach allows us to set a lower limit on the change in the trajectories that can be tracked, which was not possible with the shift-integrating method used in Ref.~\cite{Chaurasia:2021zgt}, as the tracking accuracy is set by the time-step that affects the iterative Crank-Nicholson (ICN) method. Using the minimum of the lapse allows us to avoid failures in tracking the compact BH close to the merger, where using the ICN method, the puncture was not always located in the center of the finest refinement box due to a too-large integration time step.

\section{NSbh Simulation}
\label{sec:simulation}

\subsection{Configuration}

\begin{table}[tp]
\centering
{\renewcommand{\arraystretch}{1.5}
\setlength{\tabcolsep}{3pt}
\begin{tabular}{ccccccccccc}
\toprule
EoS & $M^{\rm NS}_b$  & $M^{\rm NS}_g$ & $M/R$ & $M^{\rm BH}$ & $\chi^{\rm BH}$ & $d_0$ & $e$ & $M  \Omega$  \\ 
\hline 
 SLy & 1.55748 & 1.4 & 0.180 & 0.5 & 0 & 25.1 & 0.019 & 0.0188 \\
 \hline 
\hline
\end{tabular}
}
\caption{NSbh physical configuration: the EoS of the NS, its baryonic and gravitational masses, $M^{\rm NS}_b$ and $M^{\rm NS}_g$, its compactness $M/R$, the BH gravitational mass $M^{\rm BH}$, its dimensionless spin $\chi^{\rm BH}$, initial coordinate separation $d_0$, residual eccentricity $e$ of the system in the initial data, and the initial orbital frequency $M \Omega$. The values are given in geometric units with $c=G=1$ and $M_\odot=1$.}
\label{tab:config}
\end{table}

\begin{figure}[tb]
    \centering    \includegraphics[width=\linewidth]{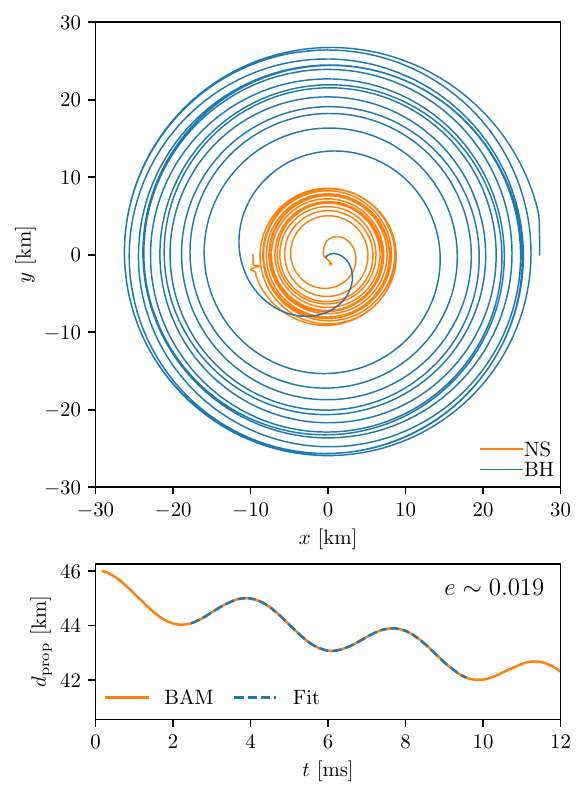}
    \caption{\textit{Top}: Orbits of the compact objects for NSbh$_{\rm R3}$ using the simulation coordinates ($x,y$). \textit{Bottom}: Initial eccentricity estimate, $e$, using the proper distance separation, $d_{\rm prop}$, within the first milliseconds of the simulation.} 
    \label{fig:orbits}
\end{figure}

In this work, we study a single physical configuration, in which the SSM BH has a mass of $0.5M_\odot$ and the NS has a gravitational mass in isolation of $1.4M_\odot$; cf.~Tab.~\ref{tab:config}. We employ a piecewise-polytropic representation of the SLy equation of state (EoS) ~\cite{Read:2008iy}. 
The initial GW frequency for our setup is $M \omega_{22}^0=0.0377$, which results in approximately $25$ GW cycles from the beginning of the simulation up to the merger\footnote{We define merger time as the time when the amplitude of strain $h$ reaches its maximum: $$ t_{\rm{merger}} = \argmax{\lvert h(t) \rvert} $$.}; cf.\ top panel of Fig.~\ref{fig:orbits}. The residual eccentricity is $e \approx 0.019$, as visible from the bottom panel of Fig.~\ref{fig:orbits}. The estimated eccentricity was calculated following \cite{Dietrich:2015pxa}, in particular Sec. IV C.\par 

To quantify uncertainties and validate our results, we have performed simulations with six different grid setups, in which we varied the grid spacing and the number of refinement levels surrounding the BH. We use $8$ refinement levels to cover the NS for all our simulations, while we use $11$ or $12$ refinement levels to resolve the BH. This way, the BH is resolved with an $8$ or $16$ times higher resolution than the NS, respectively. The larger number of refinement levels leads to a drastic increase in the computational costs, cf.~\cite{Bruegmann:2006ulg} and see also the last column in Tab.~\ref{tab:grid}. To reduce the computational costs, we use bitant symmetry, i.e., we assume reflection symmetry relative to the equatorial plane, $z=0$. 

The additional refinement levels are necessary due to the steeper gradients of the metric components around the puncture compared to the NS. We find more reliable results for the setups in which we use a total of $12$ refinement levels to cover the BH, cf.~Sec.~\ref{sec:monitoring} and Appendix~\ref{app:convergence}. 

In the following text, we denote the maximum number of refinement levels as $L$ and refer to a specific refinement level by its order number $l$ starting from the coarsest level with $l=0$. We summarize all grid configurations in Tab.~\ref{tab:grid}. 

\begin{table}[tp] 
\centering
{\renewcommand{\arraystretch}{1.5}
\setlength{\tabcolsep}{2.2pt}
\begin{tabular}{c|cccccccc|cc}
\toprule
Name & $L$ & $L_{\rm mv}$ & $n$ & $n_{\rm mv}$ & $h_{7}$ & $h_{10,11}$ & $h_0$ & $R_0$ & $T_{\rm{c}}$ & $N_{\rm{c}}$\\
\hline
NSbh$_{\rm R1}^{\rm L11}$ & 11 & 4 & 192 & 96 & 0.188 & 0.023 & 24 & 2316 & 0.19 & 768\\
NSbh$_{\rm R2}^{\rm L11}$ & 11 & 4 & 288 & 144 & 0.125 & 0.016 & 16 & 2312 & 0.62 & 768\\
NSbh$_{\rm R3}^{\rm L11}$ & 11 & 4 & 384 & 192 & 0.094 & 0.012 & 12 & 2310 & 1.58 & 1056\\
NSbh$_{\rm R1}$ & 12 & 5 & 192 & 96 & 0.188 & 0.012 & 24 & 2316 & 0.41 & 864\\
NSbh$_{\rm R2}$ & 12 & 5 & 288 & 144 & 0.125 & 0.008 & 16 & 2312 & 1.34 & 1056\\
NSbh$_{\rm R3}$ & 12 & 5 & 384 & 192 & 0.094 & 0.006 & 12 & 2310 & 4.11 & 1920\\
\hline 
\hline 
\end{tabular}
}

\caption{Grid configurations. The first column gives the configuration name. The next eight columns give the number of levels $L$, the number of moving box levels $L_{\rm mv}$ (for the BH, only one for NS),
the number of points in the non-moving boxes $n$, the number of points 
in the moving boxes $n_{\rm mv}$, the grid spacing $h_{7}$ ($l=7$) in the finest level
covering the NS, the grid spacing $h_{10}$ ($l=10$), $h_{11}$ ($l=11$) in the finest level
covering the BH, the grid spacing $h_0$ ($l=0$) in the coarsest level,
and the outer boundary position $R_0$. The grid spacing and the
outer boundary position are given in units of $M_{\odot}$. 
The last two columns show the amount of spent computational time $T_{\rm{c}}$ in millions of CPU-hours, and the total number of computational Intel Cascade Lake Platinum 9242 cores at HLRN Lise $N_{\rm{c}}$ for each run.}
\label{tab:grid}
\end{table}

\begin{figure*}[htpb]
\centering
\includegraphics[width=\linewidth]{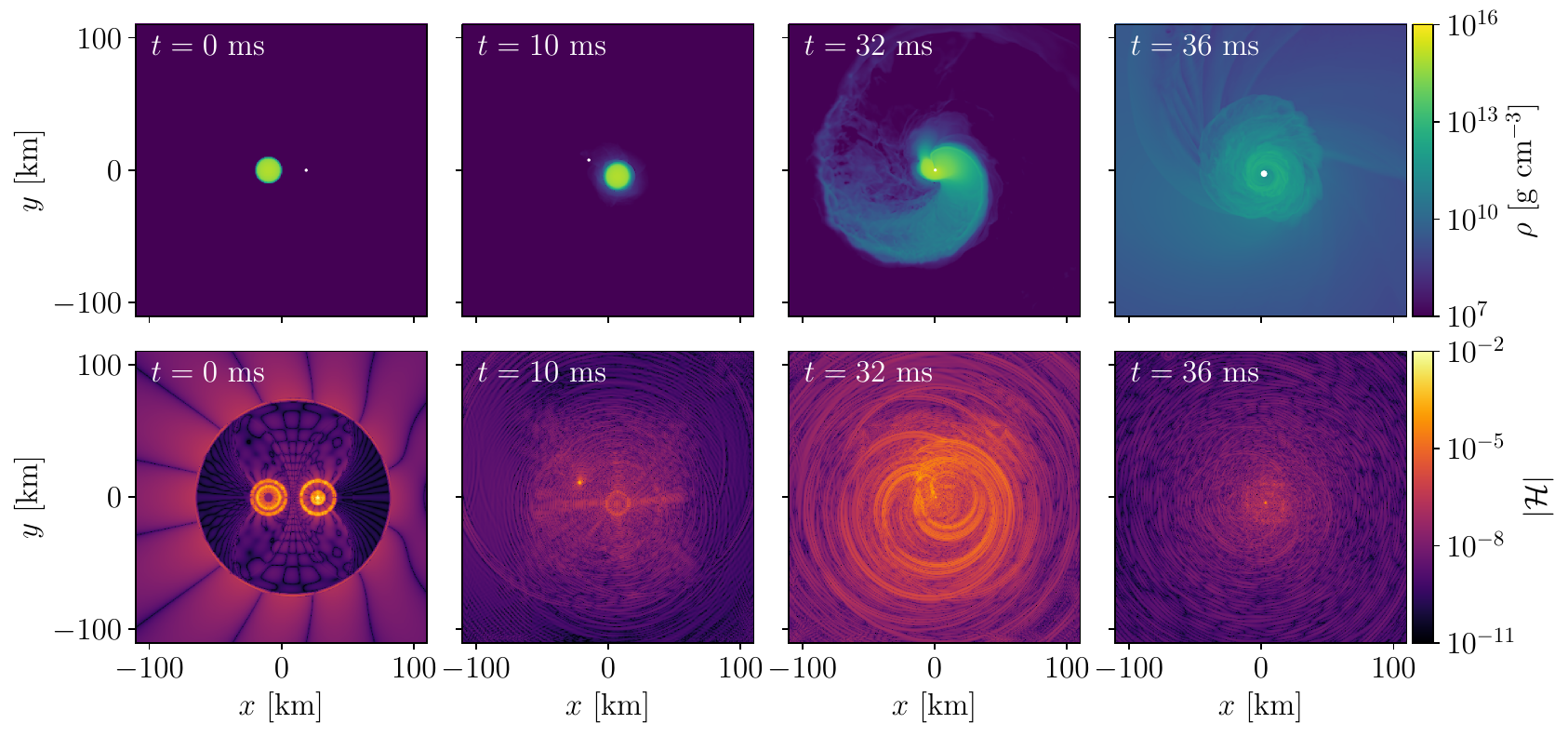}
\caption{Evolution of the matter density $\rho$ (top row) and the Hamiltonian constraint $|\mathcal{H}|$ (bottom row) at the beginning of the simulation, inspiral, merger, and post-merger, respectively for NSbh$_{\rm R3}$. The white circle on the density plots represents the apparent horizon of the BH. The values are computed for $l = 5$, where both compact objects are covered by the same refinement level.}
\label{fig:2d_ham_rho}
\end{figure*}

\subsection{Qualitative Discussion}
\label{sec:QualitativeDiscussion}

In Fig.~\ref{fig:2d_ham_rho}, we present 2D snapshots of the density and Hamiltonian constraint for the dynamical evolution of the configuration \NSbh{\rm R3} in the equatorial plane.  

At the beginning of the simulation ($t=0\ \mathrm{ms}$), the objects are separated by a coordinate distance of $37.0\ \mathrm{km}$. The radius of the NS is $9.1\ \mathrm{km}$\footnote{The evolution coordinates are not Schwarzschild coordinates, but rather close to isotropic coordinates due to our usage of the moving puncture gauge.}, and is around $15$ times larger than the radius of the apparent horizon with $0.6\ \mathrm{km}$.
For the Hamiltonian constraint (bottom panel of Fig.~\ref{fig:2d_ham_rho}), we find a specific pattern of the spectral grid used in \fuka, cf.~\cite{Papenfort:2021hod}. There, the largest constraint violations reside in the grid cells surrounding the NS and the BH. \par

The second column of Fig.~\ref{fig:2d_ham_rho} shows the system at \Tms{10}, i.e., $\sim 3.5$ orbits after the beginning of the simulation. The Hamiltonian constraint violation has significantly decreased compared to the one at the first time step (\Tms{0}). This fact is also visible at Fig.~\ref{fig:convergence_ham_mom_Mb_Mbh} (see for details Sec.~\ref{sec:monitoring}). We attribute the decrease in constraint violation to the constraint-damping properties of the Z4c evolution scheme~\citep{Hilditch:2012fp, Bernuzzi:2009ex} that we used for the simulations. Nevertheless, the largest constraint violations happen for the BH, not the NS. For the regions outside the BH or NS, there are circular structures propagating outwards. Around the NS, there is a cross-shaped grid artifact, which appears due to our choice of Cartesian coordinates.

At the time \Tms{32}, the NS undergoes tidal disruption, and accordingly, the Hamiltonian constraint violation increases during that time. 

In the last column, we show the system at \Tms{36}, when a torus has formed and the Hamiltonian constraint violation is again reduced. As in all the cases before, the highest Hamiltonian constraint violation occurs for the BH.

\begin{figure*}[htpb]
\centering
\includegraphics[width=\linewidth]{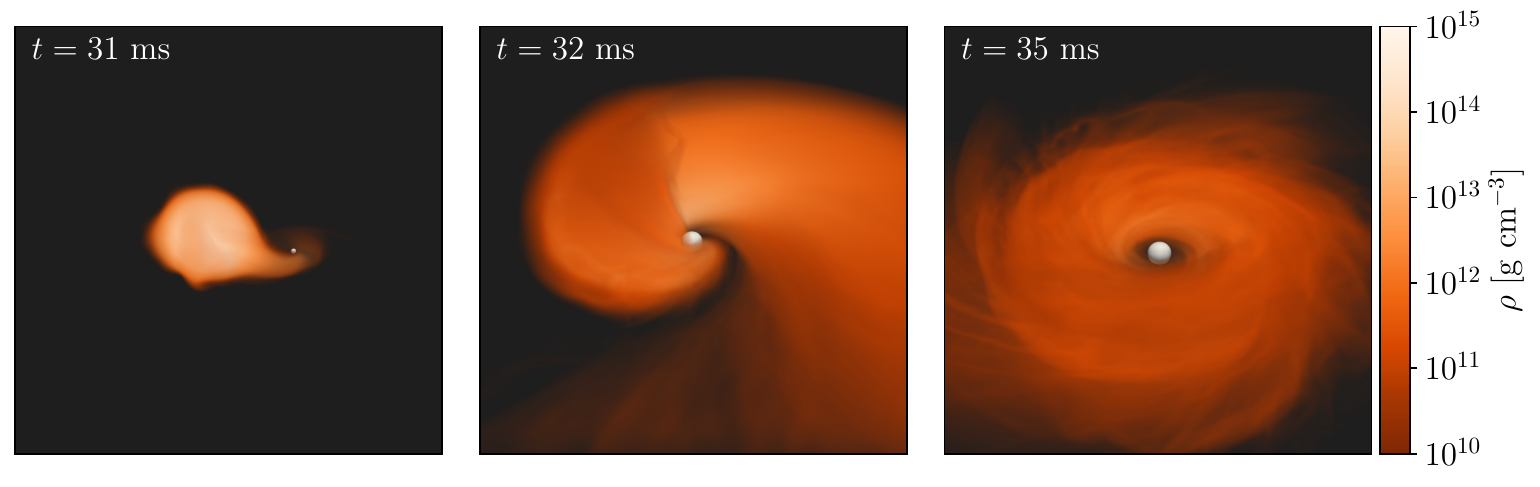}
\caption{Matter density distribution of NSbh$_{\rm R2}$ in 3D for three stages: mass transfer, merger, and disk formation. The grey sphere represents the apparent horizon of the BH. The full-length animation for the simulation is available at~\cite{NSbhDensityVideo}.}
\label{fig:3d_rho}
\end{figure*}

To further illustrate the dynamics of the system, we produce a three-dimensional (3D) visualization of the matter density evolution for the entire simulation NSbh$_{\rm R2}$ for which we output 3D data~\cite{NSbhDensityVideo}. The notable frames are outlined in Fig.~\ref{fig:3d_rho}: mass transfer (\Tms{31}), merger (\Tms{32}), and torus formation (\Tms{35}).

In the early inspiral, there is effectively no mass transfer onto the BH, and the NS is being tidally deformed. The mass transfer commences at around \Tms{29} and is depicted in progress at \Tms{31} on the first panel.

Soon after the mass transfer, at \Tms{32} (second panel), the NS is tidally disrupted and twisted around the BH, while most of the matter rapidly falls into the BH. Once the remaining material makes its first orbit around the BH, it forms a torus. At that time, the torus still has a density of $\sim 3\times 10^{13}\ \mathrm{g\ cm^{-3}}$. Then, it begins to expand (see the video, ~\cite{NSbhDensityVideo}) at roughly \Tms{34} and its density drops to $\sim 1.4 \times 10^{12}\ \mathrm{g\ cm^{-3}}$ by \Tms{38}.

The stable yet expanding torus is shown on the third panel at \Tms{35}. From then onward, a steady accretion onto the BH occurs at a rate of around $10^{-3}\ \mathrm{M_{\odot}\ ms^{-1}}$ (see also Sec.~\ref{sec:postmerger}). As can be seen in the video, the BH experiences a noticeable kick after the merger. We measure its recoil velocity in the simulation coordinates to be about $1140\ \mathrm{km\ s^{-1}}$ by performing a linear fit for the radial coordinate of the BH puncture after \Tms{34}.

\subsection{Monitoring the quality of the simulation}
\label{sec:monitoring}

\begin{figure}[htbp]
    \centering
    \includegraphics[width=\linewidth]{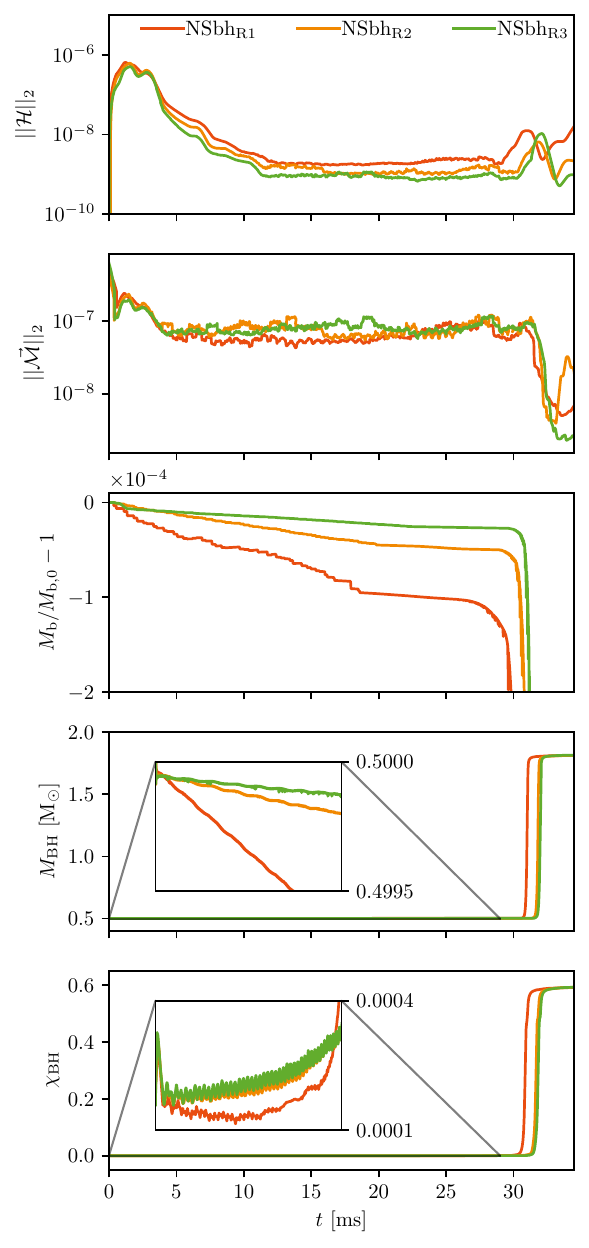}
    \caption{Evolution of selected metrics for NSbh$_{\rm R1}$, NSbh$_{\rm R2}$, and NSbh$_{\rm R3}$ at $l=2$.
    Here, $||\mathcal{H}||_2$ is the $L^{2}$ volume norm of the Hamiltonian constraint,
    $||\vec{\mathcal{M}}||_2$ is the Euclidean norm of the $L^{2}$ volume norms of the Cartesian components of the momentum constraint, $||\vec{\mathcal{M}}||_2 = [\sum_{i \in \{x, y, z\}} ||\mathcal{M}^i||^2]^{\frac{1}{2}}$,
    $M_b/M_{b,0}-1$ is the relative difference of the total baryonic mass $M_b$ from its initial value $M_{b,0}$, $M_{\mathrm{BH}}$ is the mass of the BH, and $\chi_{\rm{BH}}$ is the dimensionless spin parameter for the BH.
    The values for $||\mathcal{H}||_2$ and $||\vec{\mathcal{M}}||_2$ were smoothed out with 1st-order Savitzky-Golay filter with the window length of 151 sample, and $\chi_{\rm{BH}}$ with 61 sample.}
    \label{fig:convergence_ham_mom_Mb_Mbh}
\end{figure}

To test the validity of the results, we monitor the vital metrics of the system during the simulation.

In the first two panels of Fig.~\ref{fig:convergence_ham_mom_Mb_Mbh}, we plot $L^{2}$ norms of the Hamiltonian and momentum constraints. Both norms have their highest values at the beginning of the simulation and decrease soon after 4~ms until 10~ms; after that, they remain roughly constant throughout the inspiral.
For the Hamiltonian constraint, as discussed above in relation to Fig.~\ref{fig:2d_ham_rho}, it increases at the time of the merger. We attribute this behavior to tidal disruption and mass transfer. In contrast, the momentum constraint drops at the time of the merger.

Throughout the inspiral, the Hamiltonian constraint decreases with higher resolutions, converging to a certain value. This is an indicator of the robustness of the evolution scheme, as well as its constraint-damping properties. Meanwhile, the momentum constraint remains roughly constant and sufficiently small across all resolutions.

In the third panel, we show the relative difference between the total baryonic
mass $M_b$ from its initial value, demonstrating its conservation throughout the simulation before the merger. From \Tms{0} until $t \approx 20\ \mathrm{ms}$, the baryonic mass experiences a linear loss due to numerical errors. The mass loss error remains below 0.015\% until the beginning of the mass transfer (\Tms{27}), after which most of the baryonic mass falls into the BH. As we employ shift gauge speed of $\mu_{S} = 1$ for $\Gamma$-driver shift condition, the matter leaves the computational grid close to the puncture due to large stretching of the spatial coordinates, cf. Section IIIB of ~\cite{Thierfelder:2010dv}.

\subsection{Postmerger}
\label{sec:postmerger}

We continue the simulations up to $\sim 7$~ms after the merger. The post-merger properties are summarized in Tab.~\ref{tab:remnant}. To characterize the remnant, we compute the BH mass based on its apparent horizon to be $M_{\rm BH}~\approx~1.8 M_\odot$ and its dimensionless spin $\chi_{\rm BH}~\approx~0.60$. 

These values are in good agreement with previous studies and fitting formulas for BHNS and BBH remnant mass and its dimensionless spin parameter. In particular, applying the formula from \cite{Pannarale:2013jua} for BHNS systems results in $M_{\rm BH} = 1.71 M_\odot$ and $\chi_{\rm BH} \approx 0.55$, while applying the \nrsurremnant\ \cite{Taylor:2020bmj} model using \surfinBH\ \cite{Varma:2019csw} for BBH systems results in $M_{\rm BH} = 1.84 M_\odot$ and $\chi_{\rm BH} \approx 0.56$. Both models predict a slower-spinning remnant than in this study.

We also used the latter model for BBH systems to compute the expected recoil velocity resulting in $\sim 170\ \mathrm{km\ s^{-1}}$. Since the model assumes two BHs and not a NS as a secondary companion, we can consider this value as an upper limit of the GW kick contribution, as our NSbh simulation merges earlier than a comparable BBH system with two BHs of the same mass. Consequently, less momentum is emitted by GWs. However, the recoil velocity determined from our simulation is around $1140\ \mathrm{km\ s^{-1}}$, i.e., more than six times larger, see Sec.~\ref{sec:QualitativeDiscussion}. As in other NSBH systems, e.g., \cite{Foucart:2014nda}, the kick is mainly caused by the ejected matter rather than by the radiation of linear momentum via GWs. To verify this, we computed the net linear momentum of the ejected material. More explicitly, we use the 3D ejecta data from NSbh$_{\rm R2}$ on $l=1$ at the last time step, i.e., $39.16$\,ms. At this time, the mass-weighted average velocity of the ejecta is $0.16c$, due to the asymmetric mass ejection. Combining the ejecta contribution and the previously computed contribution of the asymmetric GW emission is sufficient to explain a recoil velocity for the BH of about $1100\ \mathrm{km\ s^{-1}}$, which is in almost perfect agreement with the remnant velocity after the merger.

\par

\begin{figure}[tp]
    \centering
    \includegraphics[width=\linewidth]{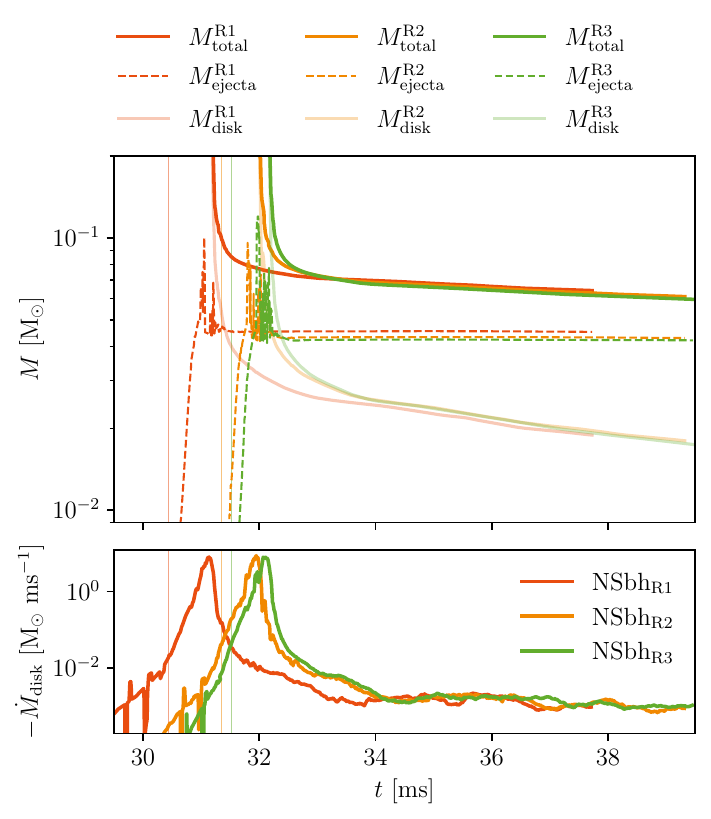}
    \caption{\textit{Top}: Evolution of the rest-mass of each simulation. The results are extracted at $l=1$. Solid lines show the total rest-mass $M_{\rm total}$, dashed lines the ejected mass $M_{\rm ejecta}$, and faint lines the disk mass $M_{\rm disk}$. \textit{Bottom}: The accretion rate, i.e., reduction of the disk mass extracted at $l=1$. We show the corresponding merger times as thin vertical lines in the background of both panels. The values for the accretion rate were smoothed out with a 1st-order Savitzky-Golay filter with a window length of 10 samples.} 
    \label{fig:masses}
\end{figure}

We evaluated the properties of the ejecta and the mass of the disk on $l=1$. To select the matter that is gravitationally unbound from the system, we use the geodesic criterion of $u_t < -1$, where $u_t$ is the time component of the four-velocity, and demand a positive radial velocity. We denote hereafter the unbound matter with sub-index $u$, e.g., the unbound rest-mass density $D_u$\footnote{We use here the conserved rest-mass density $D$, which is related to the proper rest-mass density $\rho$ in the fluid rest-frame by $D= W \rho$ with $W=(1- v_i v^i)^{-1/2}$ as Lorentz factor. }. That way, the mass of the ejecta and the disk are defined as:
\begin{equation}
    M_{\rm ejecta} = \int D_u d^3x, \quad 
    M_{\rm disk}   = \int D d^3x - M_{\rm ejecta}.
\end{equation}
We show the time evolution of the total baryonic, the ejecta, and the disk masses in Fig.~\ref{fig:masses}. The values converge at roughly the same values for all resolutions. Specifically, we find the total rest-mass of $\sim 0.06 M_\odot$, the ejecta mass of $\sim 0.04 M_\odot$, and the disk mass of $\sim 0.02 M_\odot$ at about $7$~ms after the merger. In contrast to the BHNS studies to date, cf.~\cite{Foucart:2014nda,Kyutoku:2015gda,Most:2020exl,Chaurasia:2021zgt,Hayashi:2022cdq}, the disk mass in NSbh system is lower than the mass of the dynamical ejecta. Thereby, the disk mass decreases with time due to accretion onto the BH and mass ejection. We show the evolution of the derivative of the disk mass in the lower panel of Fig.~\ref{fig:masses}, which we consider as an upper bound of the accretion rate. In the range between $34$\,ms and $39$\,ms, i.e., about $2.5$\,ms and $7.5$\,ms after the merger, we get values in the order of $10^{-3}$~M$_\odot\ \mathrm{ms^{-1}}$.

To compare our results with other studies, we used our initial parameters for the BH and the NS in several fitting formulas for the ejecta and disk masses of BHNS mergers, e.g., \cite{Kawaguchi:2016ana,Foucart:2018rjc,Kruger:2020gig}, although the formulas do not cover our parameter range. Therefore, it is not surprising that the results of the models do not agree with our simulations since they are employed well outside their calibration region. 

For instance, the models of \cite{Kawaguchi:2016ana} and \cite{Kruger:2020gig} both predict $M_{\rm ej} = 0 M_\odot$, while applying the model of \cite{Foucart:2018rjc} yields a remnant baryonic mass comprising ejecta and disk masses of $M_{\rm rem} = 0.29 M_\odot$, which is about five times larger than in our simulations.

Hence, further numerical-relativity simulations in larger regions of the parameter space are needed for calibration and extension of the models to enable their usage to interpret EM signatures connected to NSbh mergers.

\begin{table}[tp]
{\renewcommand{\arraystretch}{1.5}
\setlength{\tabcolsep}{4pt}
 
\centering

\begin{tabular}{ccccc}
\toprule
Name                & $M_{\rm BH}$ [M$_\odot$] & $\chi_{\rm BH}$  & $M_{\rm ejecta}$ [M$_\odot$]  & $M_{\rm disk}$ [M$_\odot$] \\
\hline
NSbh$_{\rm R1}$   & 1.814         & 0.59529         & 0.045         & 0.019 \\
NSbh$_{\rm R2}$   & 1.816         & 0.59716         & 0.043         & 0.019 \\
NSbh$_{\rm R3}$   & 1.819         & 0.59763         & 0.042         & 0.018 \\
\hline \hline
\end{tabular}
}
\caption{Properties of the remnant: configuration name, the gravitational mass of the BH $M_{\rm BH}$, its dimensionless spin parameter $\chi_{\rm BH}$, mass of the ejected material $M_{\rm ejecta}$, and the mass of the disk (torus) surrounding the BH $M_{\rm disk}$. The quantities are extracted $\sim 7$\,ms after the merger from $l=1$.}
\label{tab:remnant}
\end{table}

\section{Observable Signatures}
\label{sec:observables}

\subsection{Gravitational-wave emission}

\begin{figure}
\centering
\includegraphics[width=\linewidth]{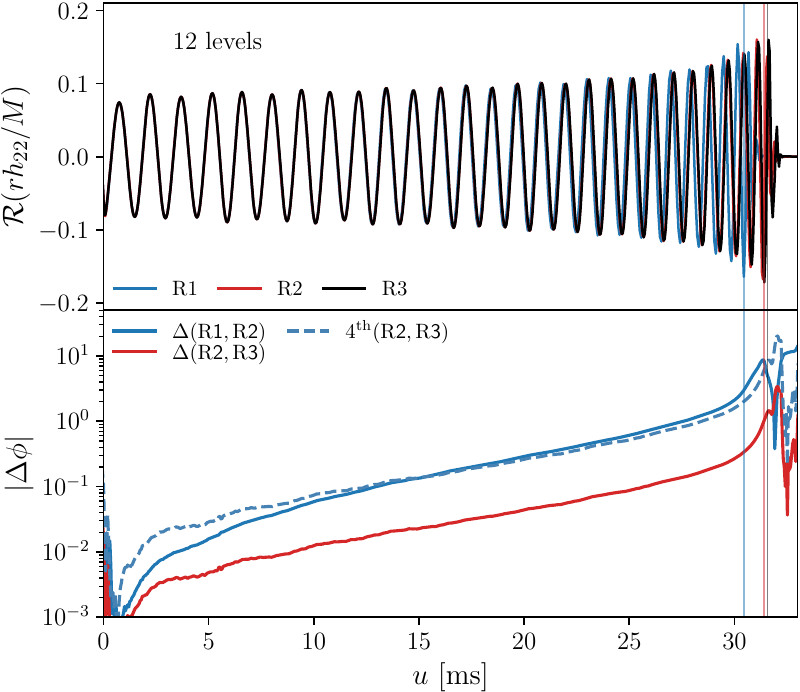}
\caption{{\it Top}: Real part of the (2,2)-mode of the GW strain, $rh _{22}$, against the retarded coordinate time $u$ for all resolutions of the NSbh system with 12 levels. {\it Bottom}: Convergence for the GW (2,2)-mode. We show the phase differences between different resolutions (solid lines) and the
rescaled phase difference, assuming fourth-order convergence (dashed line). The vertical lines in the plot indicate the merger time.} 
\label{fig:phaseconvergence}
\end{figure}

\begin{figure}
    \centering
    \includegraphics[width = \columnwidth]{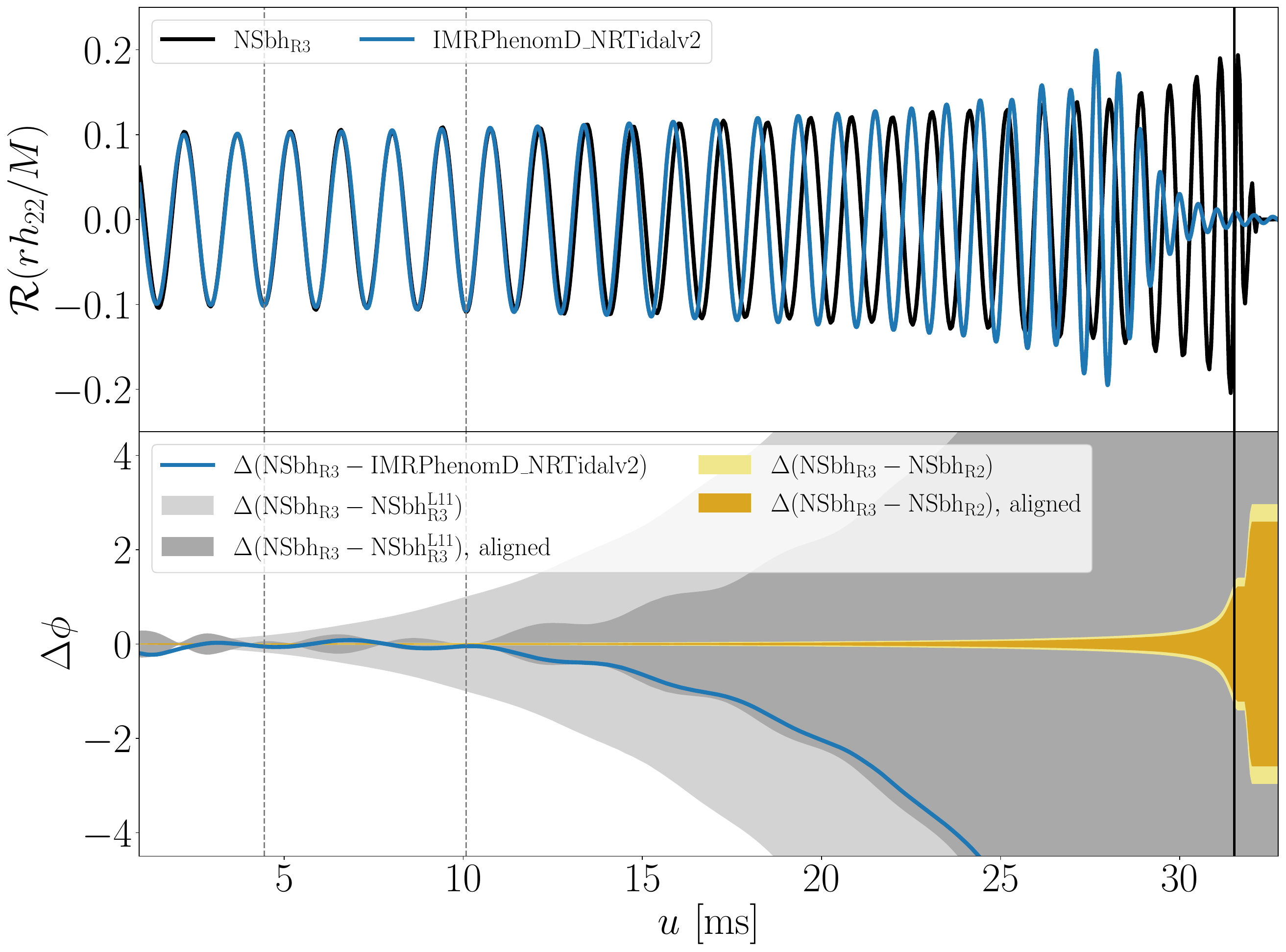}
    \caption{{\it Top}: The NR-simulated NSbh waveform compared with the \imrphenomdnrtidal\ model, for the same configuration.The two dashed vertical lines in the early inspiral constitute the alignment window, while a solid vertical black line indicates the merger. Note that \imrphenomdnrtidal\ yields a shorter waveform than the NSbh simulation. {\it Bottom}: The phase differences between the NR and model waveforms. The gray bands correspond to the error or phase difference between the highest and second-highest resolutions of the $L=12$ simulation (both taken raw and aligned with each other), while the yellow bands correspond to the phase difference between the highest resolutions of the $L$=12 and $L$=11 simulations. There is a noticeable deviation beyond the maximum tolerance before the alignment window; significant dephasing also occurs beyond the alignment window, up to $\mathcal{O}[10] \, \rm rad$ near the merger. }
    \label{fig:waveformcomparison}
\end{figure}

In the following discussion, we extract GW signal at the radius $r_{\rm{extr}} = 1200$ and in retarded time coordinate $u$, defined as
\begin{equation}
		u = t - r_{\rm{extr}} - 2 M \ln \left[\frac{r_{\rm{extr}}}{2M} - 1\right],
\end{equation}
where $t$ is the simulation time and $M$ is the total mass of the system.

In the top panel of Fig.~\ref{fig:phaseconvergence}, the real part of the (2,2)-mode of the GW strain $rh_{22}$ is plotted against the retarded coordinate time $u$ for $L = 12$ and all resolutions. The configurations with lower resolutions result in earlier merger times. This is caused by the additional numerical dissipation for low-resolution setups, e.g., \cite{Bernuzzi:2012ci}. To check the convergence of the GW signals, we plot the phase differences (bottom panel of Fig. \ref{fig:phaseconvergence}) between the different resolutions as well as the rescaled phase difference between the two highest resolutions assuming fourth-order convergence (dashed line). Notice that this rescaled phase difference $|\Delta\phi|$ $\rm{4th(R2, R3)}$ behaves similarly to that of $\Delta$(R1, R2), though there are noticeable differences near the beginning (and at the very end) of the simulation. Nevertheless, this contrasts with the convergence failure of the BHNS simulations in Ref. \cite{Chaurasia:2021zgt} and demonstrates that the \fuka\ initial data solves this convergence issue. In contrast to BNS simulations done with \bam\ for which we find only second-order convergence when we employ the WENOZ limiter~\cite{Bernuzzi:2016pie}, which is also employed here, we find a higher convergence order. This indicates that the leading order error is not connected to the hydrodynamical evolution of the matter but to the simulation of the puncture. 
This assumption is further supported by the investigations performed in Appendix~\ref{app:l11runs}, where we investigate the performance of simulations with 11 levels covering the BH ($L$=11). 
However, further tests for a larger set of binary parameters are necessary to confirm this hypothesis.

After assessing the accuracy of the GW signal, we compare the resulting waveform to existing GW waveform models. Most models assume that the BH mass is the primary mass and must be greater than the NS mass. Some existing BHNS models, such as \imrphenomnsbh\ \cite{Thompson:2020nei} and \seobnsbh\ \cite{Dietrich:2018uni}, explicitly prohibit BH masses smaller than the NS mass. Other models, such as \imrphenomdnrtidal\ \cite{Dietrich:2017aum, Dietrich:2019kaq}, may not yield a reliable waveform even if they do not explicitly prohibit it. 

To demonstrate this, we provide a comparison between the highest resolution in the $L = 12$ simulations ($\mathrm{NSbh_{R3}}$) with \imrphenomdnrtidal~\cite{Dietrich:2019kaq}, which is invoked using \lalsuite~\cite{lalsuite}. \imrphenomdnrtidal\ uses a closed-form, analytical expression of the tidal contributions of a binary system that is calibrated to NR data \cite{Dietrich:2017aum, Dietrich:2019kaq}, which is then added to the phase of a BBH baseline model \cite{Khan:2015jqa}. For our comparison, we align the NR waveform with the model waveform by finding the appropriate time and phase shifts $\delta t, \,\, \delta \phi$ that minimize the integral 
\begin{equation}
		\mathcal{I}(\delta t, \delta \phi) = \int_{t_1}^{t_2} dt |\phi_{\text{NR}}(t) - \phi_{\text{Model}}(t + \delta t) + \delta \phi|, 
\end{equation}
over some chosen frequency interval or alignment window, typically near the beginning of the NR waveform~\cite{Hotokezaka:2016bzh}. 

The result is shown in the top panel of Fig.~\ref{fig:waveformcomparison}, with the dashed vertical lines corresponding to the alignment window, and the solid black vertical line corresponding to the merger in the NR simulation. The top panel contains the comparisons of the waveforms for the duration of the simulation; the bottom panel indicates the corresponding phase differences between the waveform model and $\mathrm{NSbh_{R3}}$. We show different types of dephasing errors: the error between the highest and second-highest resolutions of the $L=12$ configuration, $\Delta(\mathrm{NSbh_{R3}} - \mathrm{NSbh_{R2}})$, and the error between the highest resolutions between the configurations with $L=11$ and $L=12$, $\Delta(\mathrm{NSbh_{R3}} - \mathrm{NSbh_{R3}^{L11}})$ designating them using the corresponding colored bands. For both cases, we also compute errors where we first align the lower resolution ($\mathrm{NSbh_{R2}}$) or $L=11$ configuration ($\mathrm{NSbh_{R3}^{L11}}$) with respect to $\mathrm{NSbh_{R3}}$ in the same manner as we aligned the \imrphenomdnrtidal~model.

Compared with \imrphenomdnrtidal, the tidal effects in the model seem greater than in the NR simulation, as seen by the shorter time to the merger. Furthermore, the value obtained from the merger frequency function $f_{\text{merger}}$, which scales as $f_{\text{merger}} \propto 1/\sqrt{q}$, is attained earlier in the evolution so that the \imrphenomdnrtidal\ waveform gets tapered before the merger of the NR waveform. Overall, similar observations remain true also for other waveform models such as \seobnr\ \cite{Lackey:2018zvw} or \texttt{TEOBResumS}~\cite{Riemenschneider:2021ppj}, which we have tested as well. 

There is a noticeable amount of dephasing before and after the alignment window, beyond the limits of the estimated numerical uncertainty. Within the alignment window, the dephasing oscillates in and out of the $\Delta(\mathrm{NSbh_{R3}} - \mathrm{NSbh_{R3}^{L11}})$ bands, which can be attributed to the eccentricity of our system. There is a considerable amount of dephasing $|\Delta \phi|$ starting at $u \sim 11 \,\, \text{ms}$ (which can also be seen in the top panel as \imrphenomdnrtidal\ dephase from NSbh) in the phases from the resolution error $\Delta(\mathrm{NSbh_{R3}} - \mathrm{NSbh_{R2}})$ which increases as we approach the merger time up to $|\Delta\phi| \sim \mathcal{O}(10 \mathrm{rad})$. This relatively large dephasing before and after the alignment window still persists despite numerous attempts to change the size of the window or its location entirely. For example, using a narrower hybridization window covering one GW cycle and starting at $u\approx 0$, we observe the dephasing even inside this window, and it still exceeds the tolerance regions computed between the two levels in the simulation. This implies that more accurate models are needed to describe a system consisting of an NS and an SSM BH by properly taking into account the contribution of the tidal effects into the GW phase and the merger frequency. 

For completion, we also do the same analysis for the simulation with level $L=11$ ($\mathrm{NSbh_{R3}^{L11}}$), and the results are shown in Appendix~\ref{app:l11runs}. Interestingly, the $L=11$ NR waveform is shorter than the $L=12$ waveform, making the merger time similar to the one for \imrphenomdnrtidal. We also note that in this case, all the phase differences fall within the dark gray resolution error $\Delta(\mathrm{NSbh_{R3}} - \mathrm{NSbh_{R2}})$. However, the error between the resolutions is larger than that of $L = 12$, and is around twice in value near the merger. 

\subsection{Kilonova Light-Curves}

\begin{figure*}[htbp]
    \centering
    \includegraphics[width=\linewidth]{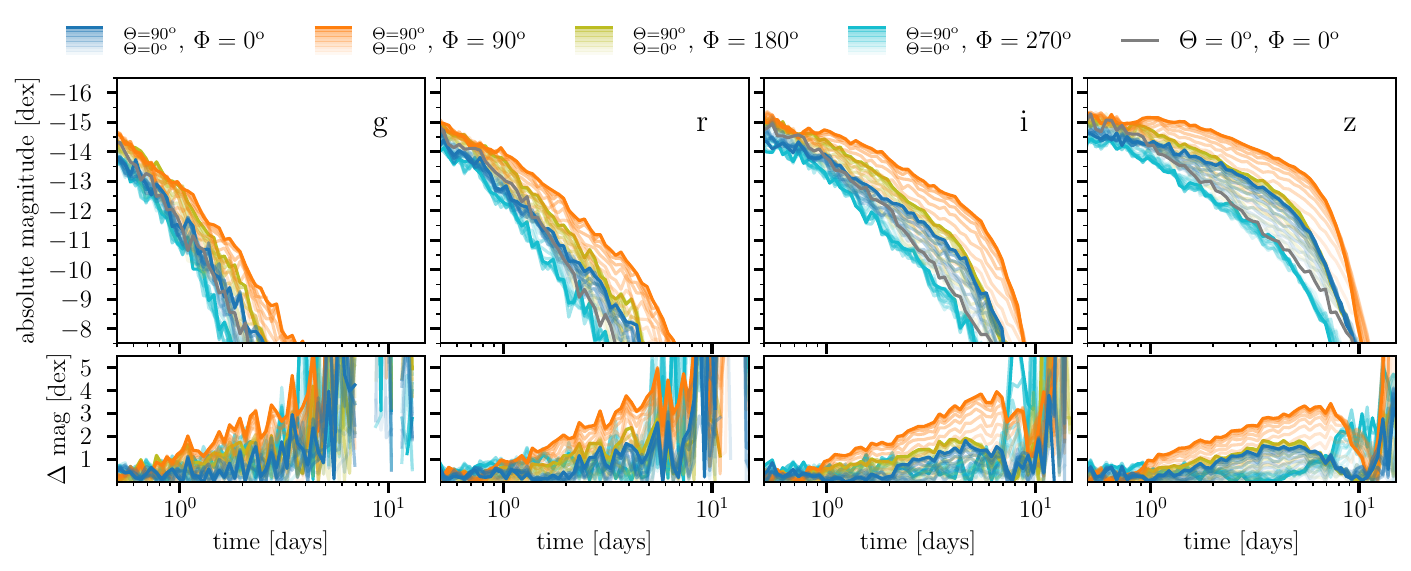}
    \includegraphics[width=\linewidth]{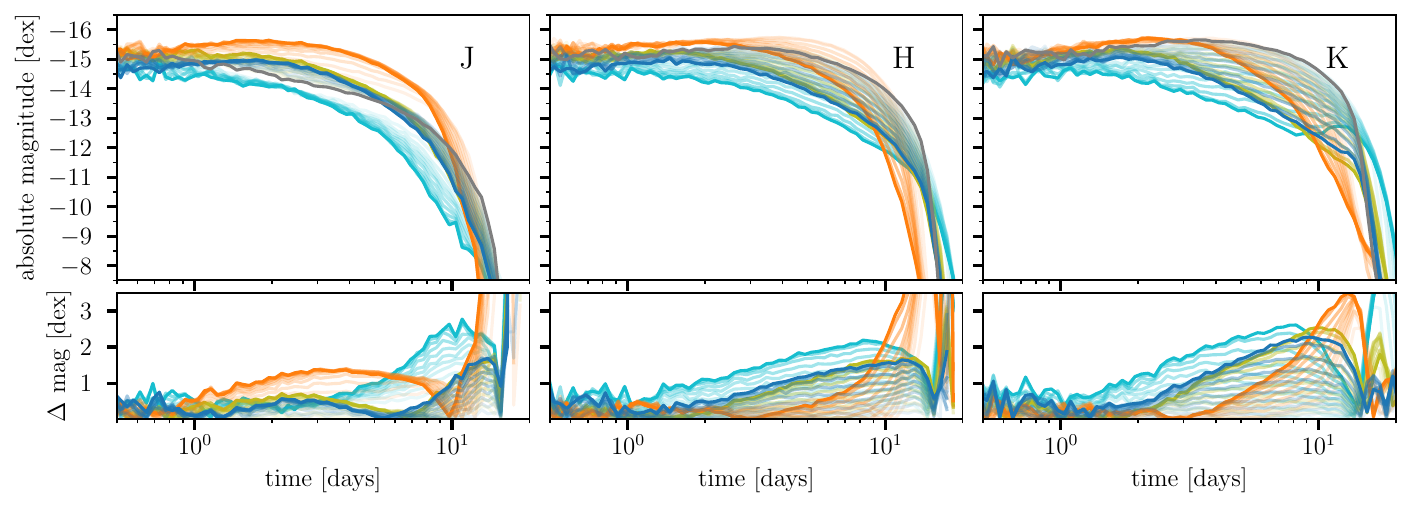}
    \includegraphics[width=\linewidth]{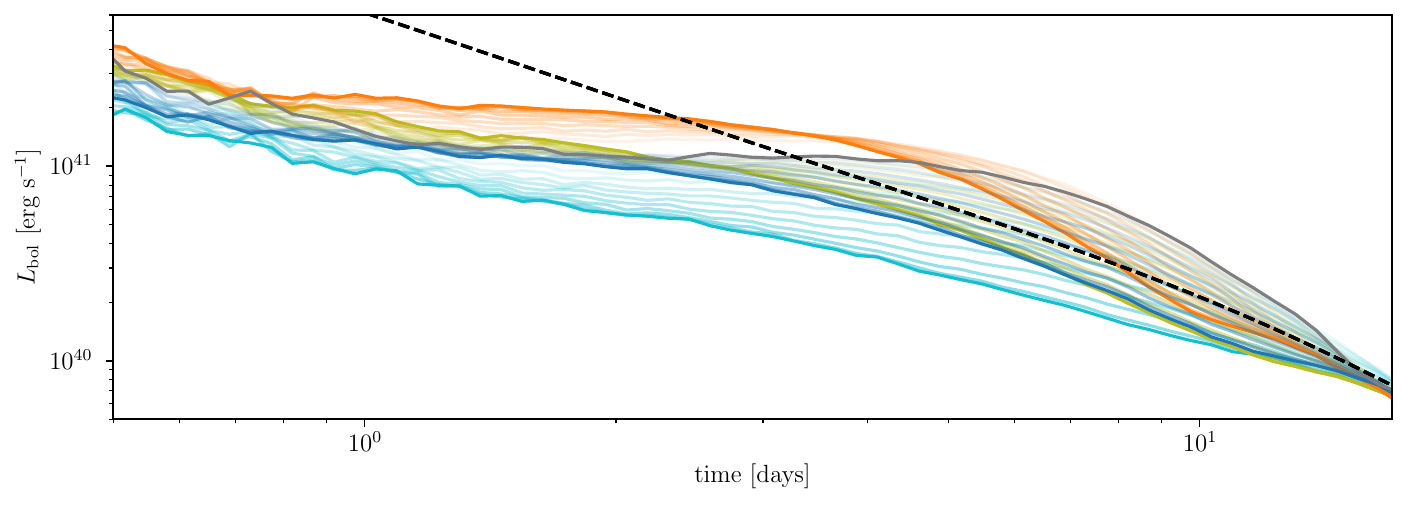}
    \caption{{\it Top panels}: Light curves for the NSbh$_{\rm R2}$\footnote{We use NSbh$_{\rm R2}$ instead of the higher resolution NSbh$_{\rm R3}$ because we did not store all relevant 3D data for this simulation.} simulation for the g, r, i, z, J, H, and K frequency bands at different observation angles $\Phi$ and $\Theta$. We show in faint lines the light curves for the variation of $\Theta$ from $\cos{\Theta} = 0$ to $\cos{\Theta} = 1$ in $0.1$ steps and additionally, in grey lines, the light curves for $\Theta = 0^\circ$ and $\Phi = 0^\circ$. Below each light curve plot, we show the differences between the light curves for different observation angles compared to the curve observed at $\Phi = 0^\circ$ and $\Theta = 0^\circ$. {\it Bottom panel}: Bolometric light curves for the NSbh$_{\rm R2}$ simulation at different observation angles $\Phi$ and $\Theta$. The deposition curve, derived from the amount of energy available, is shown in the dashed line.} 
    \label{fig:lc}
\end{figure*}

To analyze the kilonova signal associated with our NSbh simulation, we perform radiative transfer simulations using the 3D Monte Carlo code \possis~\cite{Bulla:2019muo}. Specifically, we use the latest version of the code \citep{Bulla:2022mwo} that employs heating-rate libraries \citep{Rosswog2022}, thermalization efficiencies \citep{Barnes:2016umi}, and wavelength- and time-dependent opacities \citep{Tanaka:2019iqp} that depend on local properties of the ejecta as density, temperature, $Y_{\rm e}$, and velocity. We extract the ejecta at $\sim 8$~ms after the merger from the NHbh$_{\rm R2}$ simulation and use it as input to \possis\ (see Appendix~\ref{app:possis}). The light curves are presented in Fig.~\ref{fig:lc}.  We show the light curves in optical and infrared bands as well as the bolometric luminosity $L_{\rm bol}$. While the light curves in the infrared peak (about $-15$~mag) after about $2$ to $3$~days and decrease sharply after about $10$~days, the light curves in the optical bands are much fainter (about $-12$~mag to $-13$~mag at $1$~day after the merger) and decline quite early. Since the numerical noise for the optical light curves is relatively high, we will focus our discussion mostly on the J, H, and K bands. These light curves are the more prominent ones because we consider only the dynamical ejecta, which is lanthanide-rich (see Appendix~\ref{app:possis}) and thus causes the emission to peak in infrared bands. Other ejecta components that form at later time scales and may emit more in optical filters are neglected at this point. Also, due to the low mass of the disk compared to the mass of the tidal ejecta, the disk
outflow ejecta will be a subdominant component of the kilonova emission. We note that \possis\ assumes a homologous expansion to model the outflowing material, which, however, may not be present already at $8$~ms after the merger \cite{Kawaguchi:2020vbf,Kawaguchi:2022bub,Neuweiler:2022eum}. \par 

Nevertheless, we use our simulations to investigate the angular dependence of the light curves. In our simulations, the ejecta is fairly asymmetric, as shown in Fig.~\ref{fig:2d_ham_rho} (and in Appendix~\ref{app:possis}). While only a few studies carried out full 3D radiative transfer calculations~\cite{Darbha:2021rqj, Collins:2022ocl, Neuweiler:2022eum}, most kilonova light curve models are restricted to spherical or axial symmetry, e.g., \cite{Kasen:2014toa, Kawaguchi:2020vbf, Curtis:2021guz, Just:2021vzy, Kawaguchi:2022bub, Klion:2021jzr, Wu:2021ibi,Kedia:2022onl,Bulla:2022mwo}. Utilizing 3D simulations of \possis, we investigate differences in the light curves originating from the deviations of axisymmetry. To do that, we consider different azimuthal observation angle $\Phi$, going from $0^\circ$ to $360^\circ$ starting from the positive $x$ axis, and polar observational angle $\Theta$, going from $0^\circ$ to $90^\circ$ starting from the positive $z$-axis. Fig.~\ref{fig:lc} shows deviations from about $1$~mag up to $2$~mag for the different observation angles. Compared to light curves for the other $\Phi$ angles the light curves for $\Phi = 90^\circ$ are brighter, have a later peak, and yet decrease faster. In contrast, the light curves for $\Phi = 270^\circ$ are compared to the others less bright, peak earlier, yet decrease more slowly. Both correspond to the $y-z$ plane of our simulation, but along different directions, specifically, $\Phi = 90^\circ$ corresponds to the positive $y$ axis and $\Phi = 270^\circ$ to the negative $y$ axis. We explain the bright light curves for $\Phi = 90^\circ$ by the denser and faster material emitted in this direction (see Appendix~\ref{app:possis}). In the opposite direction associated with $\Phi = 270^\circ$, we have less and slower ejecta than the average, which explains the fainter light curves here. \par

\begin{figure}[t]
    \centering
    \includegraphics[width=\linewidth]{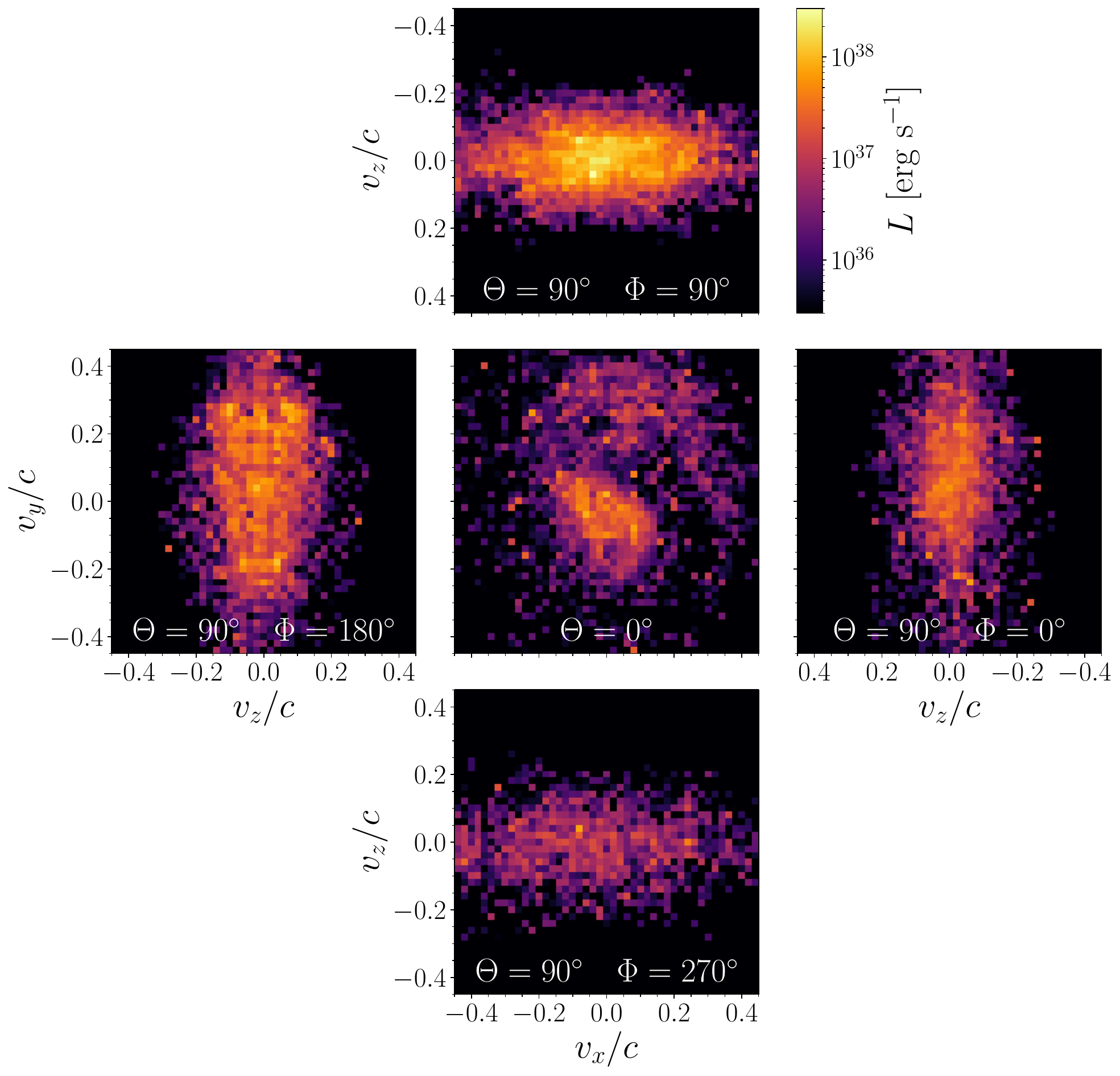}
    \caption{Luminosity maps as seen by observers from the pole with $\Theta = 0^\circ$ and from the four angles $\Phi = 0^\circ$, $\Phi = 90^\circ$, $\Phi = 180^\circ$ and $\Phi = 270^\circ$ in the equatorial plane with $\Theta = 90^\circ$. The maps show the luminosity from each region of the ejecta integrated along the line of sight and are calculated in $\left(6000-8000\right)$ \r{A} band at $1$~day after the merger. An animation of this figure for the first 10 days after the merger is available at~\cite{NSbhLuminosityVideo}.} 
    \label{fig:maps}
\end{figure}

Regarding the $\Theta$-dependence, we observe the tendency that light curves for small polar angles are brighter than in the equatorial plane, except for the $\Phi = 90^\circ$ light curves (compare in Fig.~\ref{fig:lc} the curve for the bolometric luminosity in the pole with $\Theta = 0^\circ$, i.e., the gray curve, with the ones in the equatorial plane with $\Theta = 90^\circ$, i.e., the blue, orange, green, and cyan curves). Since most of the mass is concentrated around the merger plane and photons can travel more freely to higher latitudes, we also expect brighter light curves at the pole. However, the $\Phi = 90^\circ$ light curves deviate from this behavior. In the first $4$~days after the merger, the bolometric luminosity for $\Theta = 90^\circ$ is higher than for $\Theta = 0^\circ$. 
For a correct interpretation of this behavior, we show intensity maps in Fig.~\ref{fig:maps} from the perspective of an observer at the pole and at the four different angles in the equatorial plane for $1$ day after the merger. The intensities are computed in the $\left(6000-8000\right)$ \r{A} band and are integrated along the line of sight from each region of the ejecta to generate these maps representing what an observer would see if it were possible to resolve the ejecta. 
The maps demonstrate that the radiation observed in the equatorial plane for $\Phi =90^\circ$ is much stronger than for the pole or the other observation angles. This contrasts with the intensities for $\Phi =270^\circ$ in the equatorial plane, which are much lower.
We explain this by the high-velocity material along the line of sight at this observation angle. At early times, the emission originates mainly from the outermost regions of the ejecta. Thus, this emission comes from the front of the fast-moving part of the ejecta. On later time scales, radiation from deeper regions also becomes visible. Because this emission can escape more easily toward the pole than along the equatorial plane, the bolometric luminosity for $\Phi = 90^\circ$ becomes then higher for smaller polar angles, see Fig.~\ref{fig:lc}.

We emphasize here that the discussed findings are not limited to our presented \NSbh{} system. In fact, Ref.~\cite{Darbha:2021rqj} has obtained similar results for kilonova in BHNS systems: the brightest light curves are observed along the vector of the total momentum of the ejecta.

The results show that in our case, an axisymmetric assumption would be rather broad and lead to systematic uncertainties of $\sim 1$~mag. The secondary ejecta is expected to spread more isotropically and therefore contribute to a more axisymmetric distribution of the total ejecta. We, therefore, expect that by including this component, the differences in the light curves for different $\Phi$ angles will generally decrease.

\section{Conclusion}
\label{sec:conclusion}
Despite the interest in the search for sub-solar-mass black holes, there has not been any numerical-relativity simulation of such a black hole merging with a neutron star. This lack of simulations was mainly due to issues with constructing initial data for such a system and the high computational costs of performing its dynamical evolution.  
In this work, we overcame these issues and performed, up to our knowledge, the first numerical-relativity simulations of a system composed of a neutron star and a sub-solar-mass black hole.

In contrast to our previous studies, where we used \lorene\ or \texttt{Elliptica}, we used the initial data code \fuka, which allowed us to construct such an exotic configuration. We have simulated this system with a total of six different grid setups and have studied the accuracy of our simulation, finding a mass conservation of $\sim2 \times 10^{-5}$ for the baryonic mass and the horizon mass of the black hole and Hamiltonian constraint violations of the order of $\sim 10^{-8}$. 
Due to the large mass ratio of the system, we find the ejecta mass of the order of $4\times 10^{-2}M_\odot$ and the disk mass of $2\times 10^{-2}M_\odot$. These findings are consistent across resolutions but noticeably outside of predictions using phenomenological relations for the ejecta and disk mass that are derived from typical black hole-neutron star simulations with stellar-mass black holes. 

We also computed the gravitational-wave signal connected to the simulation of our system and, as for the disk and ejecta properties, found the existing gravitational-wave models performing poorly in this region of the parameter space. Hence, for a search for similar systems using gravitational-wave observations, further waveform development work is needed. 

Finally, investigating the light curves of kilonova from the dynamically ejected matter, we found that the ejected material is not axisymmetric, which leads to differences of up to 2 magnitudes in the infrared bands depending on the polar and azimuthal viewing angle.

Overall, we hope that our simulations can serve as the first testing ground for future gravitational-wave and electromagnetic modeling of sub-solar-mass black hole-neutron star mergers. However, more simulations for various mass ratios, equations of state, and spins are needed to draw a more complete picture. 

\section{Acknowledgements}
\label{sec:acknowledgements}

We thank Alessandra Buonanno, Vsevolod Nedora, Peter James Nee, Vijay Varma, 
and Sebastian V\"olkel for helpful discussions, and Juan García-Bellido and William E.~East for their sagacious comments. S.V.C. was supported
by the research environment grant “Gravitational Radiation and Electromagnetic Astrophysical Transients
(GREAT)” funded by the Swedish Research Council (VR)
under Grant No.~Dnr.~2016-06012. TD acknowledges support from the Daimler Benz Foundation for the project `NUMANJI'.
The simulations were performed on the national supercomputer HPE Apollo Hawk at the High Performance
Computing (HPC) Center Stuttgart (HLRS) under the
grant number GWanalysis/44189, on the GCS Supercomputer SuperMUC NG at the Leibniz Supercomputing Centre (LRZ) [project pn29ba], and on the HPC systems Lise/Emmy of the North German Supercomputing
Alliance (HLRN) [project bbp00049] for the final production runs.

\section{Data availablility}
\label{sec:data_availablity}
We make publicly available the gravitational waveform ~\cite{NSbh_waveform} and the 3D ejecta data~\cite{NSbh_3D_data}.
The other simulation data and configuration parameters can be provided upon a reasonable request.

\appendix

\section{Mass convergence across different resolutions}
\label{app:convergence}
In this section, we compare the evolution of the masses of the compact objects across different setups that share the same highest grid resolution covering the interior of each object.

\begin{figure}[htpb]
\centering
\includegraphics[width=\linewidth]{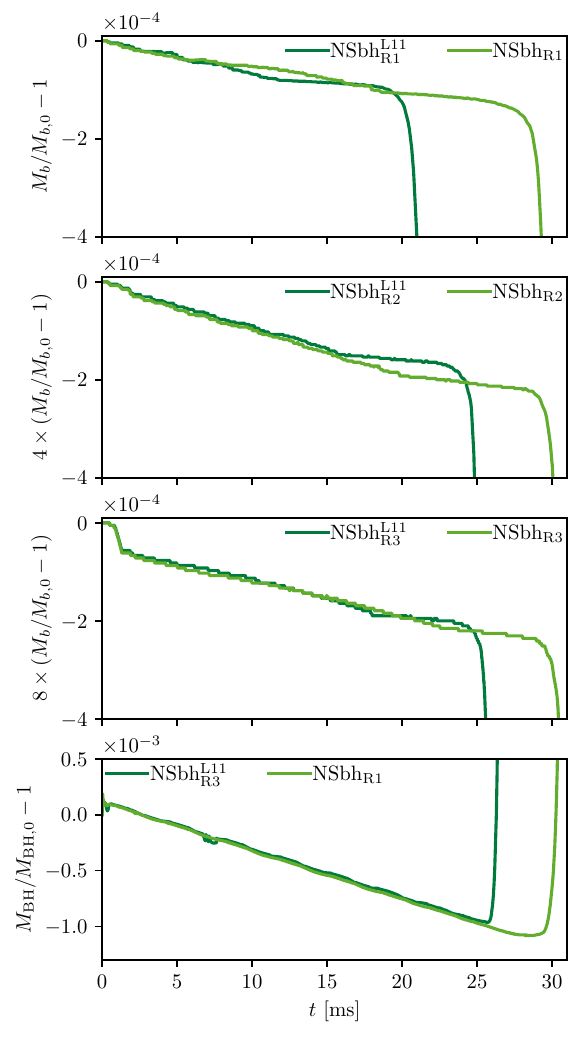}
\caption{\textit{Top three panels}: Evolution of the baryonic mass $M_b$ inside a sphere of coordinate radius of $14.0$~km around the NS relative to its initial value $M_{b,0}$. Each panel compares the same resolutions of the grid covering the NS across different resolutions of the grid around the BH.  The plotted values are rescaled with corresponding error scaling factors to make them comparable. The values are evaluated at $l=7$. \textit{Bottom panel}: Evolution of the BH mass $M_{\mathrm{BH}}$ relative to its initial mass $M_{\mathrm{BH,0}}$ for the same resolutions of the grid covering the BH and different resolutions of the grid covering the NS.}
\label{fig:convergence_masses}
\end{figure}

In the top three panels of Fig.~\ref{fig:convergence_masses}, we compare the baryonic mass evolution inside a sphere of coordinate radius $R_{\mathrm{sphere}} = 14.0$ km around the NS for the configurations with matching resolutions of the grid covering the NS. There are three such pairs, see Tab.~\ref{tab:grid}. For most of the inspiral, up until around \Tms{15}, baryonic mass undergoes consistent evolution regardless of the presence of an additional refinement level around the BH, confirming the correctness of the employed evolution scheme. After the beginning of the mass transfer (which is different across the setups, see Sec.~\ref{sec:monitoring}), the baryonic mass starts to rapidly leave the aforementioned sphere at a considerably higher rate than the mass loss due to numerical error.

We perform a similar analysis for the mass of the BH. In the bottom panel of Fig.~\ref{fig:convergence_masses}, we compare the configurations with matching resolutions of the grid covering the BH. There is only one such combination of \NSbh{} configurations, i.e.,  $\rm{NSbh_{R3}^{L11}}$ and $\rm{NSbh_{R1}}$, see Tab.~\ref{tab:grid}. As in the case of the NS mass, the BH mass evolves consistently across different setups for the duration of the inspiral and until the mass transfer into the BH sets in. This indicates that the evolution scheme performs consistently across different grid setups.

\section{Convergence metrics for $L=11$ runs}
\label{app:metrics_l11}

\begin{figure}[htbp]
\centering
\includegraphics[width=\linewidth]{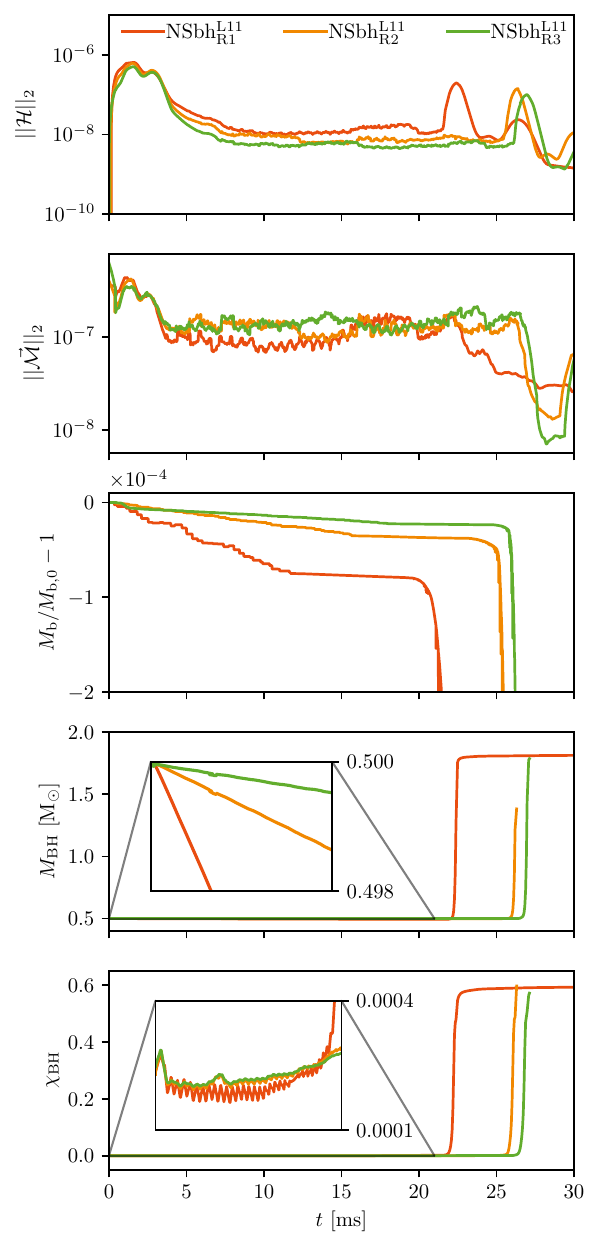}
\caption{Evolution of selected metrics for the configurations $\mathrm{NSbh_{R1}^{L11}}$, $\mathrm{NSbh_{R2}^{L11}}$, and $\mathrm{NSbh_{R3}^{L11}}$ at $l=2$.
    Here, $||\mathcal{H}||_2$ is the $L^{2}$ volume norm of the Hamiltonian constraint,
    $||\vec{\mathcal{M}}||_2$ is the Euclidean norm of the $L^{2}$ volume norms of the Cartesian components of the momentum constraint,
    $M_b/M_{b,0}-1$ is the relative difference of the total baryonic mass $M_b$ from its initial value $M_{b,0}$, $M_{\mathrm{BH}}$ is the mass of the black hole, and $\chi_{\rm{BH}}$ is the dimensionless spin parameter for the BH.
    The values for $||\mathcal{H}||_2$ and $||\vec{\mathcal{M}}||_2$ were smoothed out with a 1st-order Savitzky-Golay filter with the window length of 151 sample points, and $\chi_{\rm{BH}}$ with 21 sample points.}
\label{fig:convergence_ham_mom_Mb_Mbh_L11}
\end{figure}

We plot in Fig.~\ref{fig:convergence_ham_mom_Mb_Mbh_L11} metrics similar to Fig.~\ref{fig:convergence_ham_mom_Mb_Mbh} for the $L=11$ runs. Both $L=11$ and  $L=12$ runs have identical Hamiltonian constraint evolution up until $\sim\ 4$ ms. After that, the Hamiltonian constraints for $L=11$ decrease to about $10^{-8}$, an order of magnitude higher than  $L=12$ runs, while this value is higher for progressively lower resolutions. At the time of the merger, $L=11$ runs experience significantly higher Hamiltonian constraint violations when compared to $L=12$ runs.

The momentum constraint remains on the same order of magnitude as for $L=12$ runs, though it is generally slightly higher at the merger than for $L=12$.

Finally, the BH experiences linear mass loss at the early inspiral (before \Tms{20}) at higher rates for progressively lower resolutions. For a more detailed comparison of the baryonic mass behavior, see Appendix~\ref{app:convergence}.

\section{Gravitational Wave Signals and Comparisons for {$L = 11$}}
\label{app:l11runs}

\begin{figure}[t]
\centering
\includegraphics[width=\linewidth]{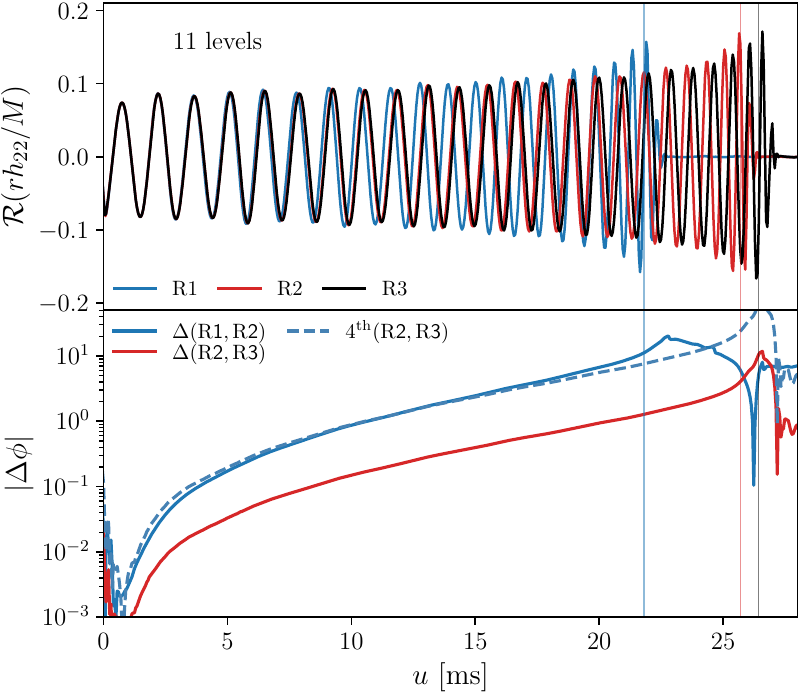}
\caption{{\it Top}: Real part of the (2,2)-mode of the gravitational wave strain, $rh_{22}$, as a function of the retarded time $u$ for all resolutions of the NSbh system with $11$ levels. {\it Bottom}: Convergence for (2,2)-mode of the GW strain. We show the phase differences between different resolutions (solid lines) and the rescaled phase difference, assuming fourth-order convergence (dashed line). The vertical lines in the plot indicate the merger time. The phase difference between different resolutions is at least one order of magnitude higher than the ones found in the 12-level simulations.} 
\label{fig:phaseconvergence_l11}
\end{figure}

In this section, we perform the convergence analysis of the simulated NSbh waveforms at different resolutions for $L = 11$. The main results are shown in Fig.~\ref{fig:phaseconvergence_l11}. Note the failure of the waveforms to converge properly in the last quarter of the simulation. Nevertheless, the overall behavior is still significantly better than the one found in Ref.~\cite{Chaurasia:2021zgt}. 

\begin{figure}[t!]
    \centering
    \includegraphics[width=\columnwidth]{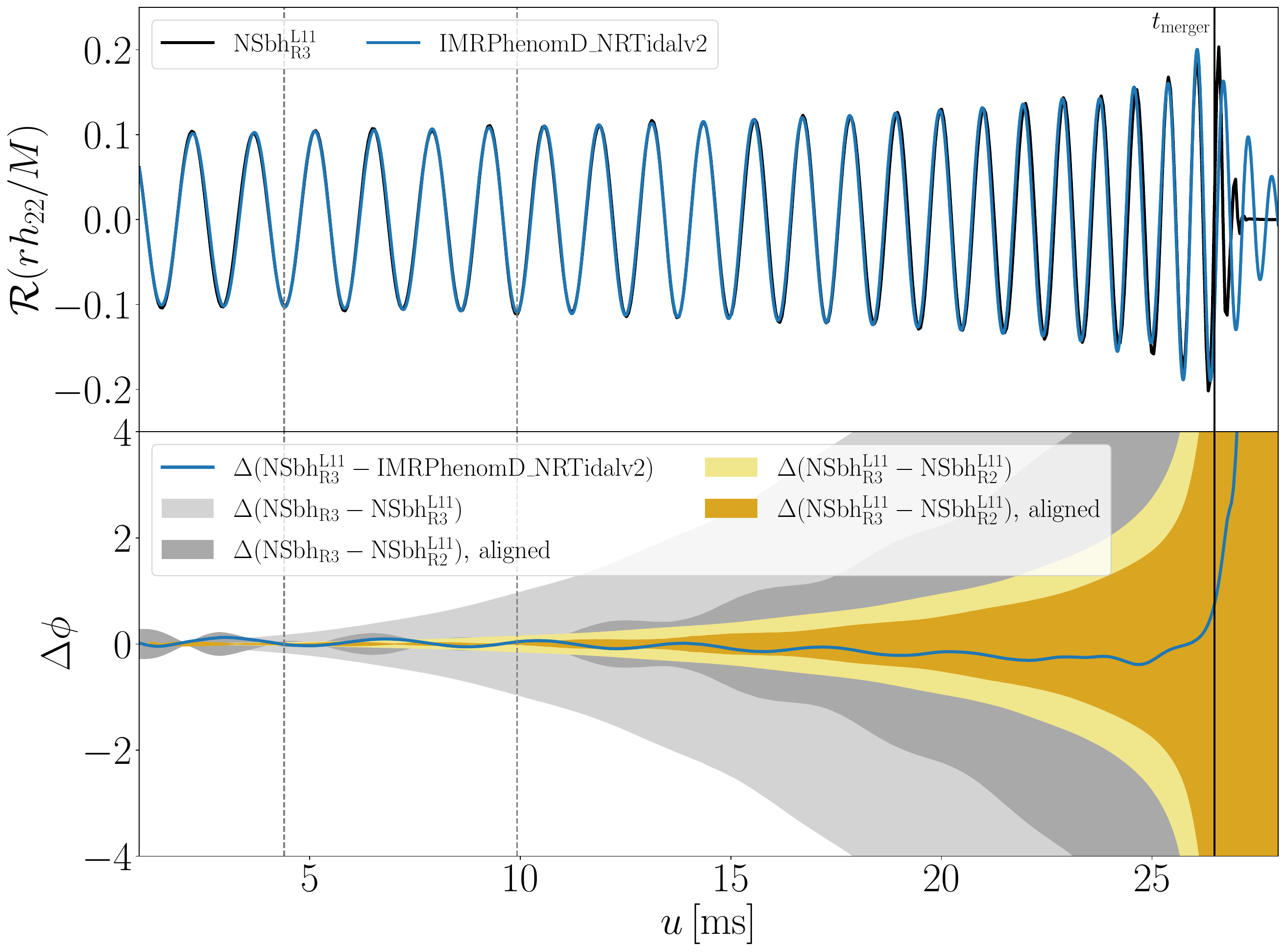}
    \caption{{\it Top}: The simulated $\mathrm{NSbh_{R3}^{L11}}$ waveform compared with the \imrphenomdnrtidal\ model. {\it Bottom}: The phase differences between the NR simulation and waveform model. The phase difference is within the resolution errors $\Delta(\mathrm{NSbh_{R3}} - \mathrm{NSbh_{R2}})$.}
    \label{fig:waveformcomparison2}
\end{figure}

We also compare the highest-resolution waveform for $L =11$ ($\rm{NSbh_{R3}^{L11}}$) with \imrphenomdnrtidal, and the results are shown in Fig.~\ref{fig:waveformcomparison2}. As in Fig.~\ref{fig:waveformcomparison}, we show different colored bands corresponding to the different errors for this system. Curiously, the NR waveform is of similar length to the model waveform, and the behavior of the phase difference is such that it falls within the tolerance given by the two highest resolutions in this level, $\Delta(\mathrm{NSbh_{R3}} - \mathrm{NSbh_{R2}})$. Further simulations will be needed to understand if this is a pure coincidence or has a more profound meaning.

\section{Input for Radiative Transfer Simulations with \possis}
\label{app:possis}

\begin{figure}[t]
\centering
\includegraphics[width=\linewidth]{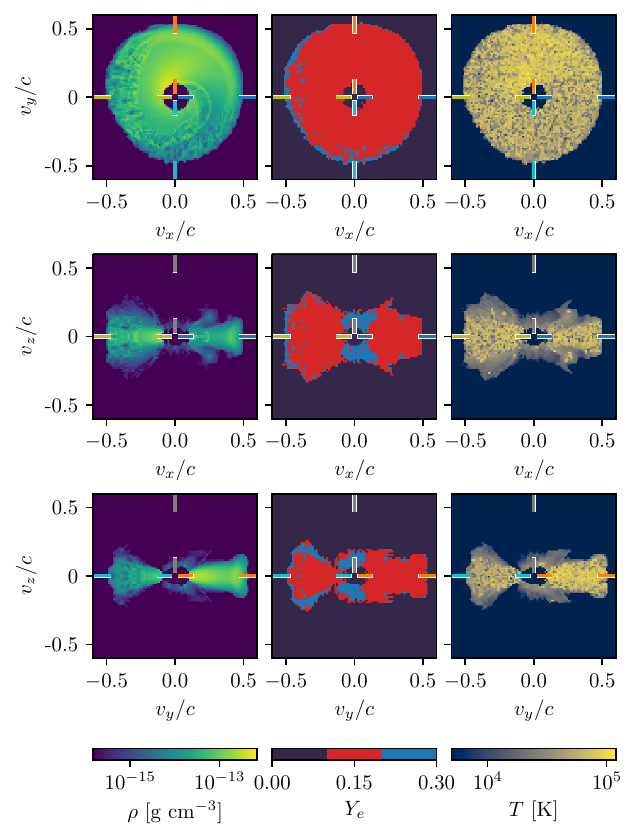}
\caption{Maps of matter density $\rho$, electron fraction $Y_e$, and temperature $T$ in the $v_x$-$v_y$ plane (top row), the $v_x$-$v_z$ plane (middle row) and in the $v_y$-$v_z$ plane (bottom row). The maps show the configuration after $1$\,day of homologous expansion and represent the input data for the radiative transfer simulations with \possis ~extracted from NSbh$_{\rm R2}$ of $l=1$ at $39.16$\,ms, i.e., about $8$\,ms after the merger. We mark chosen viewing angles we used for the light curves in Fig.~\ref{fig:lc} in the respective colors, i.e., blue for $\Theta=90^\circ$ and $\Phi = 0^\circ$, orange for $\Theta=90^\circ$ and $\Phi = 90^\circ$, green for $\Theta=90^\circ$ and $\Phi = 180^\circ$, cyan for $\Theta=90^\circ$ and $\Phi = 270^\circ$, and gray for $\Theta=0^\circ$ and $\Phi = 0^\circ$.}
\label{fig:ejecta}
\end{figure}

To produce the kilonova light curves, we use the Monte Carlo radiative transfer code \possis~\cite{Bulla:2019muo,Bulla:2022mwo}. \possis\ allows for using 3D ejecta data as input to generate a grid. The required input data represents a snapshot of the ejecta at a reference time $t_0$ and includes the density and electron fraction of the ejected material. We use here the unbound rest-mass density $D_u$. For the performed NSbh simulation, the electron fraction $Y_e$ is unavailable since it is not evolved for piecewise polytrope EoSs. The evolution of $Y_e$ using tabulated EoS was implemented in \bam\ only after our simulation had begun; cf.~\cite{Gieg:2022mut}. Hence, we determine the electron fraction $Y_e$ by considering an entropy indicator $\hat{S} = p/p\left(T=0\right)$. The entropy indicator $\hat{S}$ is generally high when the thermal component of the pressure $p_{\rm th}$ is high. Therefore, we assume a higher $\hat{S}$ for ejecta caused by shock heating than for ejecta caused by tidal disruption. Accordingly, we set a higher or lower electron fraction $Y_e$; similar to previous studies where we used \possis\ to compute light curves from \bam\ data (see, e.g.,  \cite{Neuweiler:2022eum}), we choose a threshold $\hat{S}_{\rm th} = 50$ and set the electron fraction of grid cells with $\hat{S} > \hat{S}_{\rm th}$ to $Y_e = 0.3$ and for $\hat{S} < \hat{S}_{\rm th}$ to $Y_e = 0.15$. Since tidal disruption is the main source of ejecta for the NSbh merger, the majority of the grid cells have $Y_e$ set to the lower value. 

In principle, tidal tails in a BHNS merger can have lower $Y_e$ values. However, the opacity grid used here,~\cite{Tanaka:2019iqp}, is limited to $Y_e \geq 0.1$, and opacities for $Y_e \approx 0.10$ are underestimated since the contribution from actinides is not included. Therefore, we chose $Y_e = 0.15$ as the lowest value. 

In Fig.~\ref{fig:ejecta}, maps of density, electron fraction, and temperature used as input for \possis\ are shown in the $x-y$, $x - z$, and $y - z$ plane.\par 

The grid is evolved for each time step $t_j$ following a homologous expansion: the velocity ${\bf v}_i$ of each fluid cell $i$ remains constant, while the grid coordinates evolve for each time step by ${\bf r}_{ij} = {\bf v}_i \left(t_j - t_{\rm merger}\right)/\left(t_0 - t_{\rm merger}\right)$. \par 

At each time step, \possis\ generates photon packets and propagates them throughout the ejecta material. Each packet is assigned energy, frequency, and propagation direction. The initial energy is determined by adopting heating rates libraries from \cite{Rosswog2022} and thermalization efficiencies calculated as in \cite{Barnes:2016umi,Wollaeger:2017ahm}. The total energy is then divided equally among all generated photon packets. The initial frequency of each photon packet is obtained by sampling through the thermal emissivity following Kirchhoff's law. We use bound-bound and electron-scattering opacities from \cite{Tanaka:2019iqp} and, specifically, wavelength- and time-dependent opacities $\kappa_{\rm \lambda}$($\rho_{ij},T_{ij},Y_{e,ij}$) as a function of local densities $\rho_{ij}$, temperatures $T_{ij}$, and $Y_{e,ij}$ within the ejecta. For propagating the photon packets through the ejecta, interactions such as electron-scattering and bound-bound absorption are considered, which change the properties of the respective photon packet. Finally, synthetic observables such as flux and polarization spectra are calculated ``on the fly" using an event-based technique for different observation angles \cite{Bulla:2015eza}. For more detailed information, see \cite{Bulla:2019muo,Bulla:2022mwo}. The radiative transfer simulations are performed with $N_{\rm ph} = 10^6$ photon packets. 

\section{Carbon footprint}
\label{app:carbon_footprint}
We provide a rough estimate of the amount of greenhouse gas emissions produced by this work. To calculate those, we use the Thermal Design Power (TDP) of the CPU cores, assuming that it dominates the total node energy consumption. TDP is only an order-of-magnitude indicator, while in reality, the CPU clock frequency is adjusted, and other power-saving mechanisms might be in place, cf.~\cite{osti_1838264}. 

For the NSbh evolution runs with \bam, we used Lise at HLRN with Intel Cascade Lake Platinum 9242 CPUs, which have a TDP of $\rm{7.3\ W}$ per core~\cite{IntelARK_XeonPlatinum9242}. Using the average emission factor of the German electricity grid in 2021 of $420\ \rm{gCO_2\ kWh^{-1}}$~\cite{Umweltbundesamt2021} and the total CPU time of our runs of 8.25 MCPUh (cf. Tab.~\ref{tab:grid}), we estimate the produced emission to be 25.3 $\rm{tCO_2}$. To compensate for these emissions, $\rm{CO_{2}}$ sequestering by a young forest with an area of 3000 ha (similar to the size of Grunewald near Berlin) for $\sim17$ hours would be needed. Here, we adopted a generic $\rm{CO_{2}}$ sequestering rate of 4.5 $\rm{tCO_2\ ha^{-1} yr^{-1}}$~\cite{Bernal2018}.

\bibliography{refs.bib}

\begin{thebibliography}{106}%
\makeatletter
\providecommand \@ifxundefined [1]{%
 \@ifx{#1\undefined}
}%
\providecommand \@ifnum [1]{%
 \ifnum #1\expandafter \@firstoftwo
 \else \expandafter \@secondoftwo
 \fi
}%
\providecommand \@ifx [1]{%
 \ifx #1\expandafter \@firstoftwo
 \else \expandafter \@secondoftwo
 \fi
}%
\providecommand \natexlab [1]{#1}%
\providecommand \enquote  [1]{``#1''}%
\providecommand \bibnamefont  [1]{#1}%
\providecommand \bibfnamefont [1]{#1}%
\providecommand \citenamefont [1]{#1}%
\providecommand \href@noop [0]{\@secondoftwo}%
\providecommand \href [0]{\begingroup \@sanitize@url \@href}%
\providecommand \@href[1]{\@@startlink{#1}\@@href}%
\providecommand \@@href[1]{\endgroup#1\@@endlink}%
\providecommand \@sanitize@url [0]{\catcode `\\12\catcode `\$12\catcode
  `\&12\catcode `\#12\catcode `\^12\catcode `\_12\catcode `\%12\relax}%
\providecommand \@@startlink[1]{}%
\providecommand \@@endlink[0]{}%
\providecommand \url  [0]{\begingroup\@sanitize@url \@url }%
\providecommand \@url [1]{\endgroup\@href {#1}{\urlprefix }}%
\providecommand \urlprefix  [0]{URL }%
\providecommand \Eprint [0]{\href }%
\providecommand \doibase [0]{http://dx.doi.org/}%
\providecommand \selectlanguage [0]{\@gobble}%
\providecommand \bibinfo  [0]{\@secondoftwo}%
\providecommand \bibfield  [0]{\@secondoftwo}%
\providecommand \translation [1]{[#1]}%
\providecommand \BibitemOpen [0]{}%
\providecommand \bibitemStop [0]{}%
\providecommand \bibitemNoStop [0]{.\EOS\space}%
\providecommand \EOS [0]{\spacefactor3000\relax}%
\providecommand \BibitemShut  [1]{\csname bibitem#1\endcsname}%
\let\auto@bib@innerbib\@empty
\bibitem [{\citenamefont {Abbott}\ \emph {et~al.}(2016)\citenamefont {Abbott}
  \emph {et~al.}}]{LIGOScientific:2016aoc}%
  \BibitemOpen
  \bibfield  {author} {\bibinfo {author} {\bibfnamefont {B.~P.}\ \bibnamefont
  {Abbott}} \emph {et~al.} (\bibinfo {collaboration} {LIGO Scientific,
  Virgo}),\ }\href {\doibase 10.1103/PhysRevLett.116.061102} {\bibfield
  {journal} {\bibinfo  {journal} {Phys. Rev. Lett.}\ }\textbf {\bibinfo
  {volume} {116}},\ \bibinfo {pages} {061102} (\bibinfo {year} {2016})},\
  \Eprint {http://arxiv.org/abs/1602.03837} {arXiv:1602.03837 [gr-qc]}
  \BibitemShut {NoStop}%
\bibitem [{\citenamefont {Abbott}\ \emph
  {et~al.}(2017{\natexlab{a}})\citenamefont {Abbott} \emph
  {et~al.}}]{LIGOScientific:2017vwq}%
  \BibitemOpen
  \bibfield  {author} {\bibinfo {author} {\bibfnamefont {B.~P.}\ \bibnamefont
  {Abbott}} \emph {et~al.} (\bibinfo {collaboration} {LIGO Scientific,
  Virgo}),\ }\href {\doibase 10.1103/PhysRevLett.119.161101} {\bibfield
  {journal} {\bibinfo  {journal} {Phys. Rev. Lett.}\ }\textbf {\bibinfo
  {volume} {119}},\ \bibinfo {pages} {161101} (\bibinfo {year}
  {2017}{\natexlab{a}})},\ \Eprint {http://arxiv.org/abs/1710.05832}
  {arXiv:1710.05832 [gr-qc]} \BibitemShut {NoStop}%
\bibitem [{\citenamefont {Abbott}\ \emph
  {et~al.}(2017{\natexlab{b}})\citenamefont {Abbott} \emph
  {et~al.}}]{LIGOScientific:2017ync}%
  \BibitemOpen
  \bibfield  {author} {\bibinfo {author} {\bibfnamefont {B.~P.}\ \bibnamefont
  {Abbott}} \emph {et~al.} (\bibinfo {collaboration} {LIGO Scientific, Virgo,
  Fermi GBM, INTEGRAL, IceCube, AstroSat Cadmium Zinc Telluride Imager Team,
  IPN, Insight-Hxmt, ANTARES, Swift, AGILE Team, 1M2H Team, Dark Energy Camera
  GW-EM, DES, DLT40, GRAWITA, Fermi-LAT, ATCA, ASKAP, Las Cumbres Observatory
  Group, OzGrav, DWF (Deeper Wider Faster Program), AST3, CAASTRO, VINROUGE,
  MASTER, J-GEM, GROWTH, JAGWAR, CaltechNRAO, TTU-NRAO, NuSTAR, Pan-STARRS,
  MAXI Team, TZAC Consortium, KU, Nordic Optical Telescope, ePESSTO, GROND,
  Texas Tech University, SALT Group, TOROS, BOOTES, MWA, CALET, IKI-GW
  Follow-up, H.E.S.S., LOFAR, LWA, HAWC, Pierre Auger, ALMA, Euro VLBI Team, Pi
  of Sky, Chandra Team at McGill University, DFN, ATLAS Telescopes, High Time
  Resolution Universe Survey, RIMAS, RATIR, SKA South Africa/MeerKAT}),\ }\href
  {\doibase 10.3847/2041-8213/aa91c9} {\bibfield  {journal} {\bibinfo
  {journal} {Astrophys. J. Lett.}\ }\textbf {\bibinfo {volume} {848}},\
  \bibinfo {pages} {L12} (\bibinfo {year} {2017}{\natexlab{b}})},\ \Eprint
  {http://arxiv.org/abs/1710.05833} {arXiv:1710.05833 [astro-ph.HE]}
  \BibitemShut {NoStop}%
\bibitem [{\citenamefont {Abbott}\ \emph
  {et~al.}(2021{\natexlab{a}})\citenamefont {Abbott} \emph
  {et~al.}}]{LIGOScientific:2021qlt}%
  \BibitemOpen
  \bibfield  {author} {\bibinfo {author} {\bibfnamefont {R.}~\bibnamefont
  {Abbott}} \emph {et~al.} (\bibinfo {collaboration} {LIGO Scientific, VIRGO,
  KAGRA}),\ }\href {\doibase 10.3847/2041-8213/ac082e} {\bibfield  {journal}
  {\bibinfo  {journal} {Astrophys. J. Lett.}\ }\textbf {\bibinfo {volume}
  {915}},\ \bibinfo {pages} {L5} (\bibinfo {year} {2021}{\natexlab{a}})},\
  \Eprint {http://arxiv.org/abs/2106.15163} {arXiv:2106.15163 [astro-ph.HE]}
  \BibitemShut {NoStop}%
\bibitem [{\citenamefont {Aasi}\ \emph {et~al.}(2015)\citenamefont {Aasi} \emph
  {et~al.}}]{LIGOScientific:2014pky}%
  \BibitemOpen
  \bibfield  {author} {\bibinfo {author} {\bibfnamefont {J.}~\bibnamefont
  {Aasi}} \emph {et~al.} (\bibinfo {collaboration} {LIGO Scientific}),\ }\href
  {\doibase 10.1088/0264-9381/32/7/074001} {\bibfield  {journal} {\bibinfo
  {journal} {Class. Quant. Grav.}\ }\textbf {\bibinfo {volume} {32}},\ \bibinfo
  {pages} {074001} (\bibinfo {year} {2015})},\ \Eprint
  {http://arxiv.org/abs/1411.4547} {arXiv:1411.4547 [gr-qc]} \BibitemShut
  {NoStop}%
\bibitem [{\citenamefont {Acernese}\ \emph {et~al.}(2015)\citenamefont
  {Acernese} \emph {et~al.}}]{VIRGO:2014yos}%
  \BibitemOpen
  \bibfield  {author} {\bibinfo {author} {\bibfnamefont {F.}~\bibnamefont
  {Acernese}} \emph {et~al.} (\bibinfo {collaboration} {VIRGO}),\ }\href
  {\doibase 10.1088/0264-9381/32/2/024001} {\bibfield  {journal} {\bibinfo
  {journal} {Class. Quant. Grav.}\ }\textbf {\bibinfo {volume} {32}},\ \bibinfo
  {pages} {024001} (\bibinfo {year} {2015})},\ \Eprint
  {http://arxiv.org/abs/1408.3978} {arXiv:1408.3978 [gr-qc]} \BibitemShut
  {NoStop}%
\bibitem [{\citenamefont {Punturo}\ \emph {et~al.}(2010)\citenamefont {Punturo}
  \emph {et~al.}}]{Punturo:2010zz}%
  \BibitemOpen
  \bibfield  {author} {\bibinfo {author} {\bibfnamefont {M.}~\bibnamefont
  {Punturo}} \emph {et~al.},\ }\href {\doibase 10.1088/0264-9381/27/19/194002}
  {\bibfield  {journal} {\bibinfo  {journal} {Classical Quantum Gravity}\
  }\textbf {\bibinfo {volume} {27}},\ \bibinfo {pages} {194002} (\bibinfo
  {year} {2010})}\BibitemShut {NoStop}%
\bibitem [{\citenamefont {Reitze}\ \emph {et~al.}(2019)\citenamefont {Reitze}
  \emph {et~al.}}]{Reitze:2019iox}%
  \BibitemOpen
  \bibfield  {author} {\bibinfo {author} {\bibfnamefont {D.}~\bibnamefont
  {Reitze}} \emph {et~al.},\ }\href@noop {} {\bibfield  {journal} {\bibinfo
  {journal} {Bull. Am. Astron. Soc.}\ }\textbf {\bibinfo {volume} {51}},\
  \bibinfo {pages} {035} (\bibinfo {year} {2019})},\ \Eprint
  {http://arxiv.org/abs/1907.04833} {arXiv:1907.04833 [astro-ph.IM]}
  \BibitemShut {NoStop}%
\bibitem [{\citenamefont {Adhikari}\ \emph {et~al.}(2020)\citenamefont
  {Adhikari} \emph {et~al.}}]{LIGO:2020xsf}%
  \BibitemOpen
  \bibfield  {author} {\bibinfo {author} {\bibfnamefont {R.~X.}\ \bibnamefont
  {Adhikari}} \emph {et~al.} (\bibinfo {collaboration} {LIGO}),\ }\href
  {\doibase 10.1088/1361-6382/ab9143} {\bibfield  {journal} {\bibinfo
  {journal} {Class. Quant. Grav.}\ }\textbf {\bibinfo {volume} {37}},\ \bibinfo
  {pages} {165003} (\bibinfo {year} {2020})},\ \Eprint
  {http://arxiv.org/abs/2001.11173} {arXiv:2001.11173 [astro-ph.IM]}
  \BibitemShut {NoStop}%
\bibitem [{\citenamefont {Ackley}\ \emph {et~al.}(2020)\citenamefont {Ackley}
  \emph {et~al.}}]{Ackley:2020atn}%
  \BibitemOpen
  \bibfield  {author} {\bibinfo {author} {\bibfnamefont {K.}~\bibnamefont
  {Ackley}} \emph {et~al.},\ }\href {\doibase 10.1017/pasa.2020.39} {\bibfield
  {journal} {\bibinfo  {journal} {Publ. Astron. Soc. Austral.}\ }\textbf
  {\bibinfo {volume} {37}},\ \bibinfo {pages} {e047} (\bibinfo {year}
  {2020})},\ \Eprint {http://arxiv.org/abs/2007.03128} {arXiv:2007.03128
  [astro-ph.HE]} \BibitemShut {NoStop}%
\bibitem [{\citenamefont {Ozel}\ \emph {et~al.}(2010)\citenamefont {Ozel},
  \citenamefont {Psaltis}, \citenamefont {Narayan},\ and\ \citenamefont
  {McClintock}}]{Ozel:2010su}%
  \BibitemOpen
  \bibfield  {author} {\bibinfo {author} {\bibfnamefont {F.}~\bibnamefont
  {Ozel}}, \bibinfo {author} {\bibfnamefont {D.}~\bibnamefont {Psaltis}},
  \bibinfo {author} {\bibfnamefont {R.}~\bibnamefont {Narayan}}, \ and\
  \bibinfo {author} {\bibfnamefont {J.~E.}\ \bibnamefont {McClintock}},\ }\href
  {\doibase 10.1088/0004-637X/725/2/1918} {\bibfield  {journal} {\bibinfo
  {journal} {Astrophys. J.}\ }\textbf {\bibinfo {volume} {725}},\ \bibinfo
  {pages} {1918} (\bibinfo {year} {2010})},\ \Eprint
  {http://arxiv.org/abs/1006.2834} {arXiv:1006.2834 [astro-ph.GA]} \BibitemShut
  {NoStop}%
\bibitem [{\citenamefont {Farr}\ \emph {et~al.}(2011)\citenamefont {Farr},
  \citenamefont {Sravan}, \citenamefont {Cantrell}, \citenamefont {Kreidberg},
  \citenamefont {Bailyn}, \citenamefont {Mandel},\ and\ \citenamefont
  {Kalogera}}]{Farr:2010tu}%
  \BibitemOpen
  \bibfield  {author} {\bibinfo {author} {\bibfnamefont {W.~M.}\ \bibnamefont
  {Farr}}, \bibinfo {author} {\bibfnamefont {N.}~\bibnamefont {Sravan}},
  \bibinfo {author} {\bibfnamefont {A.}~\bibnamefont {Cantrell}}, \bibinfo
  {author} {\bibfnamefont {L.}~\bibnamefont {Kreidberg}}, \bibinfo {author}
  {\bibfnamefont {C.~D.}\ \bibnamefont {Bailyn}}, \bibinfo {author}
  {\bibfnamefont {I.}~\bibnamefont {Mandel}}, \ and\ \bibinfo {author}
  {\bibfnamefont {V.}~\bibnamefont {Kalogera}},\ }\href {\doibase
  10.1088/0004-637X/741/2/103} {\bibfield  {journal} {\bibinfo  {journal}
  {Astrophys. J.}\ }\textbf {\bibinfo {volume} {741}},\ \bibinfo {pages} {103}
  (\bibinfo {year} {2011})},\ \Eprint {http://arxiv.org/abs/1011.1459}
  {arXiv:1011.1459 [astro-ph.GA]} \BibitemShut {NoStop}%
\bibitem [{\citenamefont {Jonker}\ \emph {et~al.}(2021)\citenamefont {Jonker},
  \citenamefont {Kaur}, \citenamefont {Stone},\ and\ \citenamefont
  {Torres}}]{Jonker:2021rkb}%
  \BibitemOpen
  \bibfield  {author} {\bibinfo {author} {\bibfnamefont {P.~G.}\ \bibnamefont
  {Jonker}}, \bibinfo {author} {\bibfnamefont {K.}~\bibnamefont {Kaur}},
  \bibinfo {author} {\bibfnamefont {N.}~\bibnamefont {Stone}}, \ and\ \bibinfo
  {author} {\bibfnamefont {M.~A.~P.}\ \bibnamefont {Torres}},\ }\href {\doibase
  10.3847/1538-4357/ac2839} {\bibfield  {journal} {\bibinfo  {journal}
  {Astrophys. J.}\ }\textbf {\bibinfo {volume} {921}},\ \bibinfo {pages} {131}
  (\bibinfo {year} {2021})},\ \Eprint {http://arxiv.org/abs/2104.03596}
  {arXiv:2104.03596 [astro-ph.HE]} \BibitemShut {NoStop}%
\bibitem [{\citenamefont {Abbott}\ \emph
  {et~al.}(2019{\natexlab{a}})\citenamefont {Abbott} \emph
  {et~al.}}]{LIGOScientific:2018mvr}%
  \BibitemOpen
  \bibfield  {author} {\bibinfo {author} {\bibfnamefont {B.~P.}\ \bibnamefont
  {Abbott}} \emph {et~al.} (\bibinfo {collaboration} {LIGO Scientific,
  Virgo}),\ }\href {\doibase 10.1103/PhysRevX.9.031040} {\bibfield  {journal}
  {\bibinfo  {journal} {Phys. Rev. X}\ }\textbf {\bibinfo {volume} {9}},\
  \bibinfo {pages} {031040} (\bibinfo {year} {2019}{\natexlab{a}})},\ \Eprint
  {http://arxiv.org/abs/1811.12907} {arXiv:1811.12907 [astro-ph.HE]}
  \BibitemShut {NoStop}%
\bibitem [{\citenamefont {Abbott}\ \emph
  {et~al.}(2021{\natexlab{b}})\citenamefont {Abbott} \emph
  {et~al.}}]{LIGOScientific:2020ibl}%
  \BibitemOpen
  \bibfield  {author} {\bibinfo {author} {\bibfnamefont {R.}~\bibnamefont
  {Abbott}} \emph {et~al.} (\bibinfo {collaboration} {LIGO Scientific,
  Virgo}),\ }\href {\doibase 10.1103/PhysRevX.11.021053} {\bibfield  {journal}
  {\bibinfo  {journal} {Phys. Rev. X}\ }\textbf {\bibinfo {volume} {11}},\
  \bibinfo {pages} {021053} (\bibinfo {year} {2021}{\natexlab{b}})},\ \Eprint
  {http://arxiv.org/abs/2010.14527} {arXiv:2010.14527 [gr-qc]} \BibitemShut
  {NoStop}%
\bibitem [{\citenamefont {Abbott}\ \emph
  {et~al.}(2021{\natexlab{c}})\citenamefont {Abbott} \emph
  {et~al.}}]{LIGOScientific:2021djp}%
  \BibitemOpen
  \bibfield  {author} {\bibinfo {author} {\bibfnamefont {R.}~\bibnamefont
  {Abbott}} \emph {et~al.} (\bibinfo {collaboration} {LIGO Scientific, VIRGO,
  KAGRA}),\ }\href@noop {} {\  (\bibinfo {year} {2021}{\natexlab{c}})},\
  \Eprint {http://arxiv.org/abs/2111.03606} {arXiv:2111.03606 [gr-qc]}
  \BibitemShut {NoStop}%
\bibitem [{\citenamefont {Abbott}\ \emph {et~al.}(2020)\citenamefont {Abbott}
  \emph {et~al.}}]{LIGOScientific:2020zkf}%
  \BibitemOpen
  \bibfield  {author} {\bibinfo {author} {\bibfnamefont {R.}~\bibnamefont
  {Abbott}} \emph {et~al.} (\bibinfo {collaboration} {LIGO Scientific,
  Virgo}),\ }\href {\doibase 10.3847/2041-8213/ab960f} {\bibfield  {journal}
  {\bibinfo  {journal} {Astrophys. J. Lett.}\ }\textbf {\bibinfo {volume}
  {896}},\ \bibinfo {pages} {L44} (\bibinfo {year} {2020})},\ \Eprint
  {http://arxiv.org/abs/2006.12611} {arXiv:2006.12611 [astro-ph.HE]}
  \BibitemShut {NoStop}%
\bibitem [{\citenamefont {Garcia-Bellido}\ \emph {et~al.}(1996)\citenamefont
  {Garcia-Bellido}, \citenamefont {Linde},\ and\ \citenamefont
  {Wands}}]{Garcia-Bellido:1996mdl}%
  \BibitemOpen
  \bibfield  {author} {\bibinfo {author} {\bibfnamefont {J.}~\bibnamefont
  {Garcia-Bellido}}, \bibinfo {author} {\bibfnamefont {A.~D.}\ \bibnamefont
  {Linde}}, \ and\ \bibinfo {author} {\bibfnamefont {D.}~\bibnamefont
  {Wands}},\ }\href {\doibase 10.1103/PhysRevD.54.6040} {\bibfield  {journal}
  {\bibinfo  {journal} {Phys. Rev. D}\ }\textbf {\bibinfo {volume} {54}},\
  \bibinfo {pages} {6040} (\bibinfo {year} {1996})},\ \Eprint
  {http://arxiv.org/abs/astro-ph/9605094} {arXiv:astro-ph/9605094} \BibitemShut
  {NoStop}%
\bibitem [{\citenamefont {Dolgov}\ \emph {et~al.}(2000)\citenamefont {Dolgov},
  \citenamefont {Naselsky},\ and\ \citenamefont {Novikov}}]{Dolgov2000}%
  \BibitemOpen
  \bibfield  {author} {\bibinfo {author} {\bibfnamefont {A.~D.}\ \bibnamefont
  {Dolgov}}, \bibinfo {author} {\bibfnamefont {P.~D.}\ \bibnamefont
  {Naselsky}}, \ and\ \bibinfo {author} {\bibfnamefont {I.~D.}\ \bibnamefont
  {Novikov}},\ }\href {\doibase 10.48550/ARXIV.ASTRO-PH/0009407} {\enquote
  {\bibinfo {title} {Gravitational waves, baryogenesis, and dark matter from
  primordial black holes},}\ } (\bibinfo {year} {2000})\BibitemShut {NoStop}%
\bibitem [{\citenamefont {Khlopov}(2010)}]{Khlopov:2008qy}%
  \BibitemOpen
  \bibfield  {author} {\bibinfo {author} {\bibfnamefont {M.~Y.}\ \bibnamefont
  {Khlopov}},\ }\href {\doibase 10.1088/1674-4527/10/6/001} {\bibfield
  {journal} {\bibinfo  {journal} {Res. Astron. Astrophys.}\ }\textbf {\bibinfo
  {volume} {10}},\ \bibinfo {pages} {495} (\bibinfo {year} {2010})},\ \Eprint
  {http://arxiv.org/abs/0801.0116} {arXiv:0801.0116 [astro-ph]} \BibitemShut
  {NoStop}%
\bibitem [{\citenamefont {Green}\ and\ \citenamefont
  {Kavanagh}(2021)}]{Green:2020jor}%
  \BibitemOpen
  \bibfield  {author} {\bibinfo {author} {\bibfnamefont {A.~M.}\ \bibnamefont
  {Green}}\ and\ \bibinfo {author} {\bibfnamefont {B.~J.}\ \bibnamefont
  {Kavanagh}},\ }\href {\doibase 10.1088/1361-6471/abc534} {\bibfield
  {journal} {\bibinfo  {journal} {J. Phys. G}\ }\textbf {\bibinfo {volume}
  {48}},\ \bibinfo {pages} {043001} (\bibinfo {year} {2021})},\ \Eprint
  {http://arxiv.org/abs/2007.10722} {arXiv:2007.10722 [astro-ph.CO]}
  \BibitemShut {NoStop}%
\bibitem [{\citenamefont {Jedamzik}(1997)}]{Jedamzik:1996mr}%
  \BibitemOpen
  \bibfield  {author} {\bibinfo {author} {\bibfnamefont {K.}~\bibnamefont
  {Jedamzik}},\ }\href {\doibase 10.1103/PhysRevD.55.R5871} {\bibfield
  {journal} {\bibinfo  {journal} {Phys. Rev. D}\ }\textbf {\bibinfo {volume}
  {55}},\ \bibinfo {pages} {5871} (\bibinfo {year} {1997})},\ \Eprint
  {http://arxiv.org/abs/astro-ph/9605152} {arXiv:astro-ph/9605152} \BibitemShut
  {NoStop}%
\bibitem [{\citenamefont {Byrnes}\ \emph {et~al.}(2018)\citenamefont {Byrnes},
  \citenamefont {Hindmarsh}, \citenamefont {Young},\ and\ \citenamefont
  {Hawkins}}]{Byrnes:2018clq}%
  \BibitemOpen
  \bibfield  {author} {\bibinfo {author} {\bibfnamefont {C.~T.}\ \bibnamefont
  {Byrnes}}, \bibinfo {author} {\bibfnamefont {M.}~\bibnamefont {Hindmarsh}},
  \bibinfo {author} {\bibfnamefont {S.}~\bibnamefont {Young}}, \ and\ \bibinfo
  {author} {\bibfnamefont {M.~R.~S.}\ \bibnamefont {Hawkins}},\ }\href
  {\doibase 10.1088/1475-7516/2018/08/041} {\bibfield  {journal} {\bibinfo
  {journal} {JCAP}\ }\textbf {\bibinfo {volume} {08}},\ \bibinfo {pages} {041}
  (\bibinfo {year} {2018})},\ \Eprint {http://arxiv.org/abs/1801.06138}
  {arXiv:1801.06138 [astro-ph.CO]} \BibitemShut {NoStop}%
\bibitem [{\citenamefont {Carr}\ \emph {et~al.}(2021)\citenamefont {Carr},
  \citenamefont {Clesse}, \citenamefont {Garc\'\i{}a-Bellido},\ and\
  \citenamefont {K\"uhnel}}]{Carr:2019kxo}%
  \BibitemOpen
  \bibfield  {author} {\bibinfo {author} {\bibfnamefont {B.}~\bibnamefont
  {Carr}}, \bibinfo {author} {\bibfnamefont {S.}~\bibnamefont {Clesse}},
  \bibinfo {author} {\bibfnamefont {J.}~\bibnamefont {Garc\'\i{}a-Bellido}}, \
  and\ \bibinfo {author} {\bibfnamefont {F.}~\bibnamefont {K\"uhnel}},\ }\href
  {\doibase 10.1016/j.dark.2020.100755} {\bibfield  {journal} {\bibinfo
  {journal} {Phys. Dark Univ.}\ }\textbf {\bibinfo {volume} {31}},\ \bibinfo
  {pages} {100755} (\bibinfo {year} {2021})},\ \Eprint
  {http://arxiv.org/abs/1906.08217} {arXiv:1906.08217 [astro-ph.CO]}
  \BibitemShut {NoStop}%
\bibitem [{\citenamefont {Carr}\ and\ \citenamefont
  {Kuhnel}(2020)}]{Carr:2020xqk}%
  \BibitemOpen
  \bibfield  {author} {\bibinfo {author} {\bibfnamefont {B.}~\bibnamefont
  {Carr}}\ and\ \bibinfo {author} {\bibfnamefont {F.}~\bibnamefont {Kuhnel}},\
  }\href {\doibase 10.1146/annurev-nucl-050520-125911} {\bibfield  {journal}
  {\bibinfo  {journal} {Ann. Rev. Nucl. Part. Sci.}\ }\textbf {\bibinfo
  {volume} {70}},\ \bibinfo {pages} {355} (\bibinfo {year} {2020})},\ \Eprint
  {http://arxiv.org/abs/2006.02838} {arXiv:2006.02838 [astro-ph.CO]}
  \BibitemShut {NoStop}%
\bibitem [{\citenamefont {Clesse}\ and\ \citenamefont
  {Garcia-Bellido}(2022)}]{Clesse:2020ghq}%
  \BibitemOpen
  \bibfield  {author} {\bibinfo {author} {\bibfnamefont {S.}~\bibnamefont
  {Clesse}}\ and\ \bibinfo {author} {\bibfnamefont {J.}~\bibnamefont
  {Garcia-Bellido}},\ }\href {\doibase 10.1016/j.dark.2022.101111} {\bibfield
  {journal} {\bibinfo  {journal} {Phys. Dark Univ.}\ }\textbf {\bibinfo
  {volume} {38}},\ \bibinfo {pages} {101111} (\bibinfo {year} {2022})},\
  \Eprint {http://arxiv.org/abs/2007.06481} {arXiv:2007.06481 [astro-ph.CO]}
  \BibitemShut {NoStop}%
\bibitem [{\citenamefont {Capela}\ \emph {et~al.}(2013)\citenamefont {Capela},
  \citenamefont {Pshirkov},\ and\ \citenamefont {Tinyakov}}]{Capela:2013yf}%
  \BibitemOpen
  \bibfield  {author} {\bibinfo {author} {\bibfnamefont {F.}~\bibnamefont
  {Capela}}, \bibinfo {author} {\bibfnamefont {M.}~\bibnamefont {Pshirkov}}, \
  and\ \bibinfo {author} {\bibfnamefont {P.}~\bibnamefont {Tinyakov}},\ }\href
  {\doibase 10.1103/PhysRevD.87.123524} {\bibfield  {journal} {\bibinfo
  {journal} {Phys. Rev. D}\ }\textbf {\bibinfo {volume} {87}},\ \bibinfo
  {pages} {123524} (\bibinfo {year} {2013})},\ \Eprint
  {http://arxiv.org/abs/1301.4984} {arXiv:1301.4984 [astro-ph.CO]} \BibitemShut
  {NoStop}%
\bibitem [{\citenamefont {Sasaki}\ \emph {et~al.}(2018)\citenamefont {Sasaki},
  \citenamefont {Suyama}, \citenamefont {Tanaka},\ and\ \citenamefont
  {Yokoyama}}]{Sasaki:2018dmp}%
  \BibitemOpen
  \bibfield  {author} {\bibinfo {author} {\bibfnamefont {M.}~\bibnamefont
  {Sasaki}}, \bibinfo {author} {\bibfnamefont {T.}~\bibnamefont {Suyama}},
  \bibinfo {author} {\bibfnamefont {T.}~\bibnamefont {Tanaka}}, \ and\ \bibinfo
  {author} {\bibfnamefont {S.}~\bibnamefont {Yokoyama}},\ }\href {\doibase
  10.1088/1361-6382/aaa7b4} {\bibfield  {journal} {\bibinfo  {journal} {Class.
  Quant. Grav.}\ }\textbf {\bibinfo {volume} {35}},\ \bibinfo {pages} {063001}
  (\bibinfo {year} {2018})},\ \Eprint {http://arxiv.org/abs/1801.05235}
  {arXiv:1801.05235 [astro-ph.CO]} \BibitemShut {NoStop}%
\bibitem [{\citenamefont {Bramante}\ \emph {et~al.}(2018)\citenamefont
  {Bramante}, \citenamefont {Linden},\ and\ \citenamefont
  {Tsai}}]{Bramante:2017ulk}%
  \BibitemOpen
  \bibfield  {author} {\bibinfo {author} {\bibfnamefont {J.}~\bibnamefont
  {Bramante}}, \bibinfo {author} {\bibfnamefont {T.}~\bibnamefont {Linden}}, \
  and\ \bibinfo {author} {\bibfnamefont {Y.-D.}\ \bibnamefont {Tsai}},\ }\href
  {\doibase 10.1103/PhysRevD.97.055016} {\bibfield  {journal} {\bibinfo
  {journal} {Phys. Rev. D}\ }\textbf {\bibinfo {volume} {97}},\ \bibinfo
  {pages} {055016} (\bibinfo {year} {2018})},\ \Eprint
  {http://arxiv.org/abs/1706.00001} {arXiv:1706.00001 [hep-ph]} \BibitemShut
  {NoStop}%
\bibitem [{\citenamefont {Takhistov}(2019)}]{Takhistov:2017nmt}%
  \BibitemOpen
  \bibfield  {author} {\bibinfo {author} {\bibfnamefont {V.}~\bibnamefont
  {Takhistov}},\ }\href {\doibase 10.1016/j.physletb.2018.12.043} {\bibfield
  {journal} {\bibinfo  {journal} {Phys. Lett. B}\ }\textbf {\bibinfo {volume}
  {789}},\ \bibinfo {pages} {538} (\bibinfo {year} {2019})},\ \Eprint
  {http://arxiv.org/abs/1710.09458} {arXiv:1710.09458 [astro-ph.HE]}
  \BibitemShut {NoStop}%
\bibitem [{\citenamefont {Takhistov}(2018)}]{Takhistov:2017bpt}%
  \BibitemOpen
  \bibfield  {author} {\bibinfo {author} {\bibfnamefont {V.}~\bibnamefont
  {Takhistov}},\ }\href {\doibase 10.1016/j.physletb.2018.05.026} {\bibfield
  {journal} {\bibinfo  {journal} {Phys. Lett. B}\ }\textbf {\bibinfo {volume}
  {782}},\ \bibinfo {pages} {77} (\bibinfo {year} {2018})},\ \Eprint
  {http://arxiv.org/abs/1707.05849} {arXiv:1707.05849 [astro-ph.CO]}
  \BibitemShut {NoStop}%
\bibitem [{\citenamefont {Sasaki}\ \emph {et~al.}(2022)\citenamefont {Sasaki},
  \citenamefont {Takhistov}, \citenamefont {Vardanyan},\ and\ \citenamefont
  {Zhang}}]{Sasaki:2021iuc}%
  \BibitemOpen
  \bibfield  {author} {\bibinfo {author} {\bibfnamefont {M.}~\bibnamefont
  {Sasaki}}, \bibinfo {author} {\bibfnamefont {V.}~\bibnamefont {Takhistov}},
  \bibinfo {author} {\bibfnamefont {V.}~\bibnamefont {Vardanyan}}, \ and\
  \bibinfo {author} {\bibfnamefont {Y.-l.}\ \bibnamefont {Zhang}},\ }\href
  {\doibase 10.3847/1538-4357/ac66da} {\bibfield  {journal} {\bibinfo
  {journal} {Astrophys. J.}\ }\textbf {\bibinfo {volume} {931}},\ \bibinfo
  {pages} {2} (\bibinfo {year} {2022})},\ \Eprint
  {http://arxiv.org/abs/2110.09509} {arXiv:2110.09509 [astro-ph.CO]}
  \BibitemShut {NoStop}%
\bibitem [{\citenamefont {Shandera}\ \emph {et~al.}(2018)\citenamefont
  {Shandera}, \citenamefont {Jeong},\ and\ \citenamefont
  {Gebhardt}}]{Shandera:2018xkn}%
  \BibitemOpen
  \bibfield  {author} {\bibinfo {author} {\bibfnamefont {S.}~\bibnamefont
  {Shandera}}, \bibinfo {author} {\bibfnamefont {D.}~\bibnamefont {Jeong}}, \
  and\ \bibinfo {author} {\bibfnamefont {H.~S.~G.}\ \bibnamefont {Gebhardt}},\
  }\href {\doibase 10.1103/PhysRevLett.120.241102} {\bibfield  {journal}
  {\bibinfo  {journal} {Phys. Rev. Lett.}\ }\textbf {\bibinfo {volume} {120}},\
  \bibinfo {pages} {241102} (\bibinfo {year} {2018})},\ \Eprint
  {http://arxiv.org/abs/1802.08206} {arXiv:1802.08206 [astro-ph.CO]}
  \BibitemShut {NoStop}%
\bibitem [{\citenamefont {Abbott}\ \emph {et~al.}(2018)\citenamefont {Abbott}
  \emph {et~al.}}]{LIGOScientific:2018glc}%
  \BibitemOpen
  \bibfield  {author} {\bibinfo {author} {\bibfnamefont {B.~P.}\ \bibnamefont
  {Abbott}} \emph {et~al.} (\bibinfo {collaboration} {LIGO Scientific,
  Virgo}),\ }\href {\doibase 10.1103/PhysRevLett.121.231103} {\bibfield
  {journal} {\bibinfo  {journal} {Phys. Rev. Lett.}\ }\textbf {\bibinfo
  {volume} {121}},\ \bibinfo {pages} {231103} (\bibinfo {year} {2018})},\
  \Eprint {http://arxiv.org/abs/1808.04771} {arXiv:1808.04771 [astro-ph.CO]}
  \BibitemShut {NoStop}%
\bibitem [{\citenamefont {Abbott}\ \emph
  {et~al.}(2019{\natexlab{b}})\citenamefont {Abbott} \emph
  {et~al.}}]{LIGOScientific:2019kan}%
  \BibitemOpen
  \bibfield  {author} {\bibinfo {author} {\bibfnamefont {B.~P.}\ \bibnamefont
  {Abbott}} \emph {et~al.} (\bibinfo {collaboration} {LIGO Scientific,
  Virgo}),\ }\href {\doibase 10.1103/PhysRevLett.123.161102} {\bibfield
  {journal} {\bibinfo  {journal} {Phys. Rev. Lett.}\ }\textbf {\bibinfo
  {volume} {123}},\ \bibinfo {pages} {161102} (\bibinfo {year}
  {2019}{\natexlab{b}})},\ \Eprint {http://arxiv.org/abs/1904.08976}
  {arXiv:1904.08976 [astro-ph.CO]} \BibitemShut {NoStop}%
\bibitem [{\citenamefont {Nitz}\ and\ \citenamefont
  {Wang}(2021)}]{Nitz:2021vqh}%
  \BibitemOpen
  \bibfield  {author} {\bibinfo {author} {\bibfnamefont {A.~H.}\ \bibnamefont
  {Nitz}}\ and\ \bibinfo {author} {\bibfnamefont {Y.-F.}\ \bibnamefont
  {Wang}},\ }\href {\doibase 10.1103/PhysRevLett.127.151101} {\bibfield
  {journal} {\bibinfo  {journal} {Phys. Rev. Lett.}\ }\textbf {\bibinfo
  {volume} {127}},\ \bibinfo {pages} {151101} (\bibinfo {year} {2021})},\
  \Eprint {http://arxiv.org/abs/2106.08979} {arXiv:2106.08979 [astro-ph.HE]}
  \BibitemShut {NoStop}%
\bibitem [{\citenamefont {Abbott}\ \emph {et~al.}(2022)\citenamefont {Abbott}
  \emph {et~al.}}]{LIGOScientific:2021job}%
  \BibitemOpen
  \bibfield  {author} {\bibinfo {author} {\bibfnamefont {R.}~\bibnamefont
  {Abbott}} \emph {et~al.} (\bibinfo {collaboration} {LIGO Scientific, VIRGO,
  KAGRA}),\ }\href {\doibase 10.1103/PhysRevLett.129.061104} {\bibfield
  {journal} {\bibinfo  {journal} {Phys. Rev. Lett.}\ }\textbf {\bibinfo
  {volume} {129}},\ \bibinfo {pages} {061104} (\bibinfo {year} {2022})},\
  \Eprint {http://arxiv.org/abs/2109.12197} {arXiv:2109.12197 [astro-ph.CO]}
  \BibitemShut {NoStop}%
\bibitem [{\citenamefont {Nitz}\ and\ \citenamefont
  {Wang}(2022)}]{Nitz:2022ltl}%
  \BibitemOpen
  \bibfield  {author} {\bibinfo {author} {\bibfnamefont {A.~H.}\ \bibnamefont
  {Nitz}}\ and\ \bibinfo {author} {\bibfnamefont {Y.-F.}\ \bibnamefont
  {Wang}},\ }\href {\doibase 10.1103/PhysRevD.106.023024} {\bibfield  {journal}
  {\bibinfo  {journal} {Phys. Rev. D}\ }\textbf {\bibinfo {volume} {106}},\
  \bibinfo {pages} {023024} (\bibinfo {year} {2022})},\ \Eprint
  {http://arxiv.org/abs/2202.11024} {arXiv:2202.11024 [astro-ph.HE]}
  \BibitemShut {NoStop}%
\bibitem [{\citenamefont {Hinderer}\ \emph {et~al.}(2016)\citenamefont
  {Hinderer} \emph {et~al.}}]{Hinderer:2016eia}%
  \BibitemOpen
  \bibfield  {author} {\bibinfo {author} {\bibfnamefont {T.}~\bibnamefont
  {Hinderer}} \emph {et~al.},\ }\href {\doibase 10.1103/PhysRevLett.116.181101}
  {\bibfield  {journal} {\bibinfo  {journal} {Phys. Rev. Lett.}\ }\textbf
  {\bibinfo {volume} {116}},\ \bibinfo {pages} {181101} (\bibinfo {year}
  {2016})},\ \Eprint {http://arxiv.org/abs/1602.00599} {arXiv:1602.00599
  [gr-qc]} \BibitemShut {NoStop}%
\bibitem [{\citenamefont {Nagar}\ \emph {et~al.}(2018)\citenamefont {Nagar}
  \emph {et~al.}}]{Nagar:2018zoe}%
  \BibitemOpen
  \bibfield  {author} {\bibinfo {author} {\bibfnamefont {A.}~\bibnamefont
  {Nagar}} \emph {et~al.},\ }\href {\doibase 10.1103/PhysRevD.98.104052}
  {\bibfield  {journal} {\bibinfo  {journal} {Phys. Rev. D}\ }\textbf {\bibinfo
  {volume} {98}},\ \bibinfo {pages} {104052} (\bibinfo {year} {2018})},\
  \Eprint {http://arxiv.org/abs/1806.01772} {arXiv:1806.01772 [gr-qc]}
  \BibitemShut {NoStop}%
\bibitem [{\citenamefont {Thompson}\ \emph {et~al.}(2020)\citenamefont
  {Thompson}, \citenamefont {Fauchon-Jones}, \citenamefont {Khan},
  \citenamefont {Nitoglia}, \citenamefont {Pannarale}, \citenamefont
  {Dietrich},\ and\ \citenamefont {Hannam}}]{Thompson:2020nei}%
  \BibitemOpen
  \bibfield  {author} {\bibinfo {author} {\bibfnamefont {J.~E.}\ \bibnamefont
  {Thompson}}, \bibinfo {author} {\bibfnamefont {E.}~\bibnamefont
  {Fauchon-Jones}}, \bibinfo {author} {\bibfnamefont {S.}~\bibnamefont {Khan}},
  \bibinfo {author} {\bibfnamefont {E.}~\bibnamefont {Nitoglia}}, \bibinfo
  {author} {\bibfnamefont {F.}~\bibnamefont {Pannarale}}, \bibinfo {author}
  {\bibfnamefont {T.}~\bibnamefont {Dietrich}}, \ and\ \bibinfo {author}
  {\bibfnamefont {M.}~\bibnamefont {Hannam}},\ }\href {\doibase
  10.1103/PhysRevD.101.124059} {\bibfield  {journal} {\bibinfo  {journal}
  {Phys. Rev. D}\ }\textbf {\bibinfo {volume} {101}},\ \bibinfo {pages}
  {124059} (\bibinfo {year} {2020})},\ \Eprint
  {http://arxiv.org/abs/2002.08383} {arXiv:2002.08383 [gr-qc]} \BibitemShut
  {NoStop}%
\bibitem [{\citenamefont {Matas}\ \emph {et~al.}(2020)\citenamefont {Matas}
  \emph {et~al.}}]{Matas:2020wab}%
  \BibitemOpen
  \bibfield  {author} {\bibinfo {author} {\bibfnamefont {A.}~\bibnamefont
  {Matas}} \emph {et~al.},\ }\href {\doibase 10.1103/PhysRevD.102.043023}
  {\bibfield  {journal} {\bibinfo  {journal} {Phys. Rev. D}\ }\textbf {\bibinfo
  {volume} {102}},\ \bibinfo {pages} {043023} (\bibinfo {year} {2020})},\
  \Eprint {http://arxiv.org/abs/2004.10001} {arXiv:2004.10001 [gr-qc]}
  \BibitemShut {NoStop}%
\bibitem [{\citenamefont {Chaurasia}\ \emph {et~al.}(2021)\citenamefont
  {Chaurasia}, \citenamefont {Dietrich},\ and\ \citenamefont
  {Rosswog}}]{Chaurasia:2021zgt}%
  \BibitemOpen
  \bibfield  {author} {\bibinfo {author} {\bibfnamefont {S.~V.}\ \bibnamefont
  {Chaurasia}}, \bibinfo {author} {\bibfnamefont {T.}~\bibnamefont {Dietrich}},
  \ and\ \bibinfo {author} {\bibfnamefont {S.}~\bibnamefont {Rosswog}},\ }\href
  {\doibase 10.1103/PhysRevD.104.084010} {\bibfield  {journal} {\bibinfo
  {journal} {Phys. Rev. D}\ }\textbf {\bibinfo {volume} {104}},\ \bibinfo
  {pages} {084010} (\bibinfo {year} {2021})},\ \Eprint
  {http://arxiv.org/abs/2107.08752} {arXiv:2107.08752 [gr-qc]} \BibitemShut
  {NoStop}%
\bibitem [{\citenamefont {Rashti}\ \emph {et~al.}(2022)\citenamefont {Rashti},
  \citenamefont {Fabbri}, \citenamefont {Br\"ugmann}, \citenamefont
  {Chaurasia}, \citenamefont {Dietrich}, \citenamefont {Ujevic},\ and\
  \citenamefont {Tichy}}]{Rashti:2021ihv}%
  \BibitemOpen
  \bibfield  {author} {\bibinfo {author} {\bibfnamefont {A.}~\bibnamefont
  {Rashti}}, \bibinfo {author} {\bibfnamefont {F.~M.}\ \bibnamefont {Fabbri}},
  \bibinfo {author} {\bibfnamefont {B.}~\bibnamefont {Br\"ugmann}}, \bibinfo
  {author} {\bibfnamefont {S.~V.}\ \bibnamefont {Chaurasia}}, \bibinfo {author}
  {\bibfnamefont {T.}~\bibnamefont {Dietrich}}, \bibinfo {author}
  {\bibfnamefont {M.}~\bibnamefont {Ujevic}}, \ and\ \bibinfo {author}
  {\bibfnamefont {W.}~\bibnamefont {Tichy}},\ }\href {\doibase
  10.1103/PhysRevD.105.104027} {\bibfield  {journal} {\bibinfo  {journal}
  {Phys. Rev. D}\ }\textbf {\bibinfo {volume} {105}},\ \bibinfo {pages}
  {104027} (\bibinfo {year} {2022})},\ \Eprint
  {http://arxiv.org/abs/2109.14511} {arXiv:2109.14511 [gr-qc]} \BibitemShut
  {NoStop}%
\bibitem [{\citenamefont {Papenfort}\ \emph {et~al.}(2021)\citenamefont
  {Papenfort}, \citenamefont {Tootle}, \citenamefont {Grandcl\'ement},
  \citenamefont {Most},\ and\ \citenamefont {Rezzolla}}]{Papenfort:2021hod}%
  \BibitemOpen
  \bibfield  {author} {\bibinfo {author} {\bibfnamefont {L.~J.}\ \bibnamefont
  {Papenfort}}, \bibinfo {author} {\bibfnamefont {S.~D.}\ \bibnamefont
  {Tootle}}, \bibinfo {author} {\bibfnamefont {P.}~\bibnamefont
  {Grandcl\'ement}}, \bibinfo {author} {\bibfnamefont {E.~R.}\ \bibnamefont
  {Most}}, \ and\ \bibinfo {author} {\bibfnamefont {L.}~\bibnamefont
  {Rezzolla}},\ }\href {\doibase 10.1103/PhysRevD.104.024057} {\bibfield
  {journal} {\bibinfo  {journal} {Phys. Rev. D}\ }\textbf {\bibinfo {volume}
  {104}},\ \bibinfo {pages} {024057} (\bibinfo {year} {2021})},\ \Eprint
  {http://arxiv.org/abs/2103.09911} {arXiv:2103.09911 [gr-qc]} \BibitemShut
  {NoStop}%
\bibitem [{\citenamefont {Wilson}\ and\ \citenamefont
  {Mathews}(1995)}]{Wilson:1995uh}%
  \BibitemOpen
  \bibfield  {author} {\bibinfo {author} {\bibfnamefont {J.}~\bibnamefont
  {Wilson}}\ and\ \bibinfo {author} {\bibfnamefont {G.}~\bibnamefont
  {Mathews}},\ }\href {\doibase 10.1103/PhysRevLett.75.4161} {\bibfield
  {journal} {\bibinfo  {journal} {Phys. Rev. Lett.}\ }\textbf {\bibinfo
  {volume} {75}},\ \bibinfo {pages} {4161} (\bibinfo {year}
  {1995})}\BibitemShut {NoStop}%
\bibitem [{\citenamefont {Wilson}\ \emph {et~al.}(1996)\citenamefont {Wilson},
  \citenamefont {Mathews},\ and\ \citenamefont {Marronetti}}]{Wilson:1996ty}%
  \BibitemOpen
  \bibfield  {author} {\bibinfo {author} {\bibfnamefont {J.}~\bibnamefont
  {Wilson}}, \bibinfo {author} {\bibfnamefont {G.}~\bibnamefont {Mathews}}, \
  and\ \bibinfo {author} {\bibfnamefont {P.}~\bibnamefont {Marronetti}},\
  }\href {\doibase 10.1103/PhysRevD.54.1317} {\bibfield  {journal} {\bibinfo
  {journal} {Phys. Rev. D}\ }\textbf {\bibinfo {volume} {54}},\ \bibinfo
  {pages} {1317} (\bibinfo {year} {1996})},\ \Eprint
  {http://arxiv.org/abs/gr-qc/9601017} {arXiv:gr-qc/9601017 [gr-qc]}
  \BibitemShut {NoStop}%
\bibitem [{\citenamefont {York~Jr.}(1999)}]{York:1998hy}%
  \BibitemOpen
  \bibfield  {author} {\bibinfo {author} {\bibfnamefont {J.~W.}\ \bibnamefont
  {York~Jr.}},\ }\href {\doibase 10.1103/PhysRevLett.82.1350} {\bibfield
  {journal} {\bibinfo  {journal} {Phys. Rev. Lett.}\ }\textbf {\bibinfo
  {volume} {82}},\ \bibinfo {pages} {1350} (\bibinfo {year} {1999})},\ \Eprint
  {http://arxiv.org/abs/gr-qc/9810051} {arXiv:gr-qc/9810051 [gr-qc]}
  \BibitemShut {NoStop}%
\bibitem [{\citenamefont {Gourgoulhon}\ \emph {et~al.}()\citenamefont
  {Gourgoulhon}, \citenamefont {Grandcl\'{e}ment}, \citenamefont {Marck},
  \citenamefont {Novak},\ and\ \citenamefont {Taniguchi}}]{LoreneCode}%
  \BibitemOpen
  \bibfield  {author} {\bibinfo {author} {\bibfnamefont {E.}~\bibnamefont
  {Gourgoulhon}}, \bibinfo {author} {\bibfnamefont {P.}~\bibnamefont
  {Grandcl\'{e}ment}}, \bibinfo {author} {\bibfnamefont {J.-A.}\ \bibnamefont
  {Marck}}, \bibinfo {author} {\bibfnamefont {J.}~\bibnamefont {Novak}}, \ and\
  \bibinfo {author} {\bibfnamefont {K.}~\bibnamefont {Taniguchi}},\ }\href@noop
  {} {\enquote {\bibinfo {title} {Lorene code},}\ }\bibinfo {howpublished}
  {\url{http://www.lorene.obspm.fr/}}\BibitemShut {NoStop}%
\bibitem [{\citenamefont {Demircik}\ \emph {et~al.}(2022)\citenamefont
  {Demircik}, \citenamefont {Ecker}, \citenamefont {J\"arvinen}, \citenamefont
  {Rezzolla}, \citenamefont {Tootle},\ and\ \citenamefont
  {Topolski}}]{Demircik:2022uol}%
  \BibitemOpen
  \bibfield  {author} {\bibinfo {author} {\bibfnamefont {T.}~\bibnamefont
  {Demircik}}, \bibinfo {author} {\bibfnamefont {C.}~\bibnamefont {Ecker}},
  \bibinfo {author} {\bibfnamefont {M.}~\bibnamefont {J\"arvinen}}, \bibinfo
  {author} {\bibfnamefont {L.}~\bibnamefont {Rezzolla}}, \bibinfo {author}
  {\bibfnamefont {S.}~\bibnamefont {Tootle}}, \ and\ \bibinfo {author}
  {\bibfnamefont {K.}~\bibnamefont {Topolski}},\ }\href {\doibase
  10.1051/epjconf/202227407006} {\bibfield  {journal} {\bibinfo  {journal} {EPJ
  Web Conf.}\ }\textbf {\bibinfo {volume} {274}},\ \bibinfo {pages} {07006}
  (\bibinfo {year} {2022})},\ \Eprint {http://arxiv.org/abs/2211.10118}
  {arXiv:2211.10118 [astro-ph.HE]} \BibitemShut {NoStop}%
\bibitem [{\citenamefont {Papenfort}\ \emph {et~al.}(2022)\citenamefont
  {Papenfort}, \citenamefont {Most}, \citenamefont {Tootle},\ and\
  \citenamefont {Rezzolla}}]{Papenfort:2022ywx}%
  \BibitemOpen
  \bibfield  {author} {\bibinfo {author} {\bibfnamefont {L.~J.}\ \bibnamefont
  {Papenfort}}, \bibinfo {author} {\bibfnamefont {E.~R.}\ \bibnamefont {Most}},
  \bibinfo {author} {\bibfnamefont {S.}~\bibnamefont {Tootle}}, \ and\ \bibinfo
  {author} {\bibfnamefont {L.}~\bibnamefont {Rezzolla}},\ }\href {\doibase
  10.1093/mnras/stac964} {\bibfield  {journal} {\bibinfo  {journal} {Mon. Not.
  Roy. Astron. Soc.}\ }\textbf {\bibinfo {volume} {513}},\ \bibinfo {pages}
  {3646} (\bibinfo {year} {2022})},\ \Eprint {http://arxiv.org/abs/2201.03632}
  {arXiv:2201.03632 [astro-ph.HE]} \BibitemShut {NoStop}%
\bibitem [{\citenamefont {Tootle}\ \emph {et~al.}(2021)\citenamefont {Tootle},
  \citenamefont {Papenfort}, \citenamefont {Most},\ and\ \citenamefont
  {Rezzolla}}]{Tootle:2021umi}%
  \BibitemOpen
  \bibfield  {author} {\bibinfo {author} {\bibfnamefont {S.~D.}\ \bibnamefont
  {Tootle}}, \bibinfo {author} {\bibfnamefont {L.~J.}\ \bibnamefont
  {Papenfort}}, \bibinfo {author} {\bibfnamefont {E.~R.}\ \bibnamefont {Most}},
  \ and\ \bibinfo {author} {\bibfnamefont {L.}~\bibnamefont {Rezzolla}},\
  }\href {\doibase 10.3847/2041-8213/ac350d} {\bibfield  {journal} {\bibinfo
  {journal} {Astrophys. J. Lett.}\ }\textbf {\bibinfo {volume} {922}},\
  \bibinfo {pages} {L19} (\bibinfo {year} {2021})},\ \Eprint
  {http://arxiv.org/abs/2109.00940} {arXiv:2109.00940 [gr-qc]} \BibitemShut
  {NoStop}%
\bibitem [{\citenamefont {Campanelli}\ \emph {et~al.}(2006)\citenamefont
  {Campanelli}, \citenamefont {Lousto}, \citenamefont {Marronetti},\ and\
  \citenamefont {Zlochower}}]{Campanelli:2005dd}%
  \BibitemOpen
  \bibfield  {author} {\bibinfo {author} {\bibfnamefont {M.}~\bibnamefont
  {Campanelli}}, \bibinfo {author} {\bibfnamefont {C.~O.}\ \bibnamefont
  {Lousto}}, \bibinfo {author} {\bibfnamefont {P.}~\bibnamefont {Marronetti}},
  \ and\ \bibinfo {author} {\bibfnamefont {Y.}~\bibnamefont {Zlochower}},\
  }\href {\doibase 10.1103/PhysRevLett.96.111101} {\bibfield  {journal}
  {\bibinfo  {journal} {Phys. Rev. Lett.}\ }\textbf {\bibinfo {volume} {96}},\
  \bibinfo {pages} {111101} (\bibinfo {year} {2006})},\ \Eprint
  {http://arxiv.org/abs/gr-qc/0511048} {arXiv:gr-qc/0511048} \BibitemShut
  {NoStop}%
\bibitem [{\citenamefont {Read}\ \emph {et~al.}(2009)\citenamefont {Read},
  \citenamefont {Lackey}, \citenamefont {Owen},\ and\ \citenamefont
  {Friedman}}]{Read:2008iy}%
  \BibitemOpen
  \bibfield  {author} {\bibinfo {author} {\bibfnamefont {J.~S.}\ \bibnamefont
  {Read}}, \bibinfo {author} {\bibfnamefont {B.~D.}\ \bibnamefont {Lackey}},
  \bibinfo {author} {\bibfnamefont {B.~J.}\ \bibnamefont {Owen}}, \ and\
  \bibinfo {author} {\bibfnamefont {J.~L.}\ \bibnamefont {Friedman}},\ }\href
  {\doibase 10.1103/PhysRevD.79.124032} {\bibfield  {journal} {\bibinfo
  {journal} {Phys. Rev. D}\ }\textbf {\bibinfo {volume} {79}},\ \bibinfo
  {pages} {124032} (\bibinfo {year} {2009})},\ \Eprint
  {http://arxiv.org/abs/0812.2163} {arXiv:0812.2163 [astro-ph]} \BibitemShut
  {NoStop}%
\bibitem [{\citenamefont {Dietrich}\ \emph {et~al.}(2015)\citenamefont
  {Dietrich}, \citenamefont {Moldenhauer}, \citenamefont {Johnson-McDaniel},
  \citenamefont {Bernuzzi}, \citenamefont {Markakis}, \citenamefont
  {Br{\"u}gmann},\ and\ \citenamefont {Tichy}}]{Dietrich:2015pxa}%
  \BibitemOpen
  \bibfield  {author} {\bibinfo {author} {\bibfnamefont {T.}~\bibnamefont
  {Dietrich}}, \bibinfo {author} {\bibfnamefont {N.}~\bibnamefont
  {Moldenhauer}}, \bibinfo {author} {\bibfnamefont {N.~K.}\ \bibnamefont
  {Johnson-McDaniel}}, \bibinfo {author} {\bibfnamefont {S.}~\bibnamefont
  {Bernuzzi}}, \bibinfo {author} {\bibfnamefont {C.~M.}\ \bibnamefont
  {Markakis}}, \bibinfo {author} {\bibfnamefont {B.}~\bibnamefont
  {Br{\"u}gmann}}, \ and\ \bibinfo {author} {\bibfnamefont {W.}~\bibnamefont
  {Tichy}},\ }\href {\doibase 10.1103/PhysRevD.92.124007} {\bibfield  {journal}
  {\bibinfo  {journal} {Phys. Rev. D}\ }\textbf {\bibinfo {volume} {92}},\
  \bibinfo {pages} {124007} (\bibinfo {year} {2015})},\ \Eprint
  {http://arxiv.org/abs/1507.07100} {arXiv:1507.07100 [gr-qc]} \BibitemShut
  {NoStop}%
\bibitem [{\citenamefont {Bruegmann}\ \emph {et~al.}(2008)\citenamefont
  {Bruegmann}, \citenamefont {Gonzalez}, \citenamefont {Hannam}, \citenamefont
  {Husa}, \citenamefont {Sperhake},\ and\ \citenamefont
  {Tichy}}]{Bruegmann:2006ulg}%
  \BibitemOpen
  \bibfield  {author} {\bibinfo {author} {\bibfnamefont {B.}~\bibnamefont
  {Bruegmann}}, \bibinfo {author} {\bibfnamefont {J.~A.}\ \bibnamefont
  {Gonzalez}}, \bibinfo {author} {\bibfnamefont {M.}~\bibnamefont {Hannam}},
  \bibinfo {author} {\bibfnamefont {S.}~\bibnamefont {Husa}}, \bibinfo {author}
  {\bibfnamefont {U.}~\bibnamefont {Sperhake}}, \ and\ \bibinfo {author}
  {\bibfnamefont {W.}~\bibnamefont {Tichy}},\ }\href {\doibase
  10.1103/PhysRevD.77.024027} {\bibfield  {journal} {\bibinfo  {journal} {Phys.
  Rev. D}\ }\textbf {\bibinfo {volume} {77}},\ \bibinfo {pages} {024027}
  (\bibinfo {year} {2008})},\ \Eprint {http://arxiv.org/abs/gr-qc/0610128}
  {arXiv:gr-qc/0610128} \BibitemShut {NoStop}%
\bibitem [{\citenamefont {Hilditch}\ \emph {et~al.}(2013)\citenamefont
  {Hilditch}, \citenamefont {Bernuzzi}, \citenamefont {Thierfelder},
  \citenamefont {Cao}, \citenamefont {Tichy} \emph {et~al.}}]{Hilditch:2012fp}%
  \BibitemOpen
  \bibfield  {author} {\bibinfo {author} {\bibfnamefont {D.}~\bibnamefont
  {Hilditch}}, \bibinfo {author} {\bibfnamefont {S.}~\bibnamefont {Bernuzzi}},
  \bibinfo {author} {\bibfnamefont {M.}~\bibnamefont {Thierfelder}}, \bibinfo
  {author} {\bibfnamefont {Z.}~\bibnamefont {Cao}}, \bibinfo {author}
  {\bibfnamefont {W.}~\bibnamefont {Tichy}},  \emph {et~al.},\ }\href {\doibase
  10.1103/PhysRevD.88.084057} {\bibfield  {journal} {\bibinfo  {journal} {Phys.
  Rev. D}\ }\textbf {\bibinfo {volume} {88}},\ \bibinfo {pages} {084057}
  (\bibinfo {year} {2013})},\ \Eprint {http://arxiv.org/abs/1212.2901}
  {arXiv:1212.2901 [gr-qc]} \BibitemShut {NoStop}%
\bibitem [{\citenamefont {Bernuzzi}\ and\ \citenamefont
  {Hilditch}(2010)}]{Bernuzzi:2009ex}%
  \BibitemOpen
  \bibfield  {author} {\bibinfo {author} {\bibfnamefont {S.}~\bibnamefont
  {Bernuzzi}}\ and\ \bibinfo {author} {\bibfnamefont {D.}~\bibnamefont
  {Hilditch}},\ }\href {\doibase 10.1103/PhysRevD.81.084003} {\bibfield
  {journal} {\bibinfo  {journal} {Phys. Rev. D}\ }\textbf {\bibinfo {volume}
  {81}},\ \bibinfo {pages} {084003} (\bibinfo {year} {2010})},\ \Eprint
  {http://arxiv.org/abs/0912.2920} {arXiv:0912.2920 [gr-qc]} \BibitemShut
  {NoStop}%
\bibitem [{\citenamefont {Markin}\ and\ \citenamefont
  {Chaurasia}(2023)}]{NSbhDensityVideo}%
  \BibitemOpen
  \bibfield  {author} {\bibinfo {author} {\bibfnamefont {I.}~\bibnamefont
  {Markin}}\ and\ \bibinfo {author} {\bibfnamefont {S.~V.}\ \bibnamefont
  {Chaurasia}},\ }\href {\doibase 10.5281/zenodo.7821033} {\enquote {\bibinfo
  {title} {{General-Relativistic Hydrodynamics Simulation of a Neutron Star —
  Sub-Solar-Mass Black Hole Merger - 3D Density Visualization}},}\ } (\bibinfo
  {year} {2023}),\ \bibinfo {note} {\relax{YouTube link:}
  \url{https://youtu.be/C8UL_fktipQ}}\BibitemShut {NoStop}%
\bibitem [{\citenamefont {Thierfelder}\ \emph {et~al.}(2011)\citenamefont
  {Thierfelder}, \citenamefont {Bernuzzi}, \citenamefont {Hilditch},
  \citenamefont {Br{\"u}gmann},\ and\ \citenamefont
  {Rezzolla}}]{Thierfelder:2010dv}%
  \BibitemOpen
  \bibfield  {author} {\bibinfo {author} {\bibfnamefont {M.}~\bibnamefont
  {Thierfelder}}, \bibinfo {author} {\bibfnamefont {S.}~\bibnamefont
  {Bernuzzi}}, \bibinfo {author} {\bibfnamefont {D.}~\bibnamefont {Hilditch}},
  \bibinfo {author} {\bibfnamefont {B.}~\bibnamefont {Br{\"u}gmann}}, \ and\
  \bibinfo {author} {\bibfnamefont {L.}~\bibnamefont {Rezzolla}},\ }\href
  {\doibase 10.1103/PhysRevD.83.064022} {\bibfield  {journal} {\bibinfo
  {journal} {Phys. Rev. D}\ }\textbf {\bibinfo {volume} {83}},\ \bibinfo
  {pages} {064022} (\bibinfo {year} {2011})},\ \Eprint
  {http://arxiv.org/abs/1012.3703} {arXiv:1012.3703 [gr-qc]} \BibitemShut
  {NoStop}%
\bibitem [{\citenamefont {Pannarale}(2014)}]{Pannarale:2013jua}%
  \BibitemOpen
  \bibfield  {author} {\bibinfo {author} {\bibfnamefont {F.}~\bibnamefont
  {Pannarale}},\ }\href {\doibase 10.1103/PhysRevD.89.044045} {\bibfield
  {journal} {\bibinfo  {journal} {Phys. Rev. D}\ }\textbf {\bibinfo {volume}
  {89}},\ \bibinfo {pages} {044045} (\bibinfo {year} {2014})},\ \Eprint
  {http://arxiv.org/abs/1311.5931} {arXiv:1311.5931 [gr-qc]} \BibitemShut
  {NoStop}%
\bibitem [{\citenamefont {Taylor}\ and\ \citenamefont
  {Varma}(2020)}]{Taylor:2020bmj}%
  \BibitemOpen
  \bibfield  {author} {\bibinfo {author} {\bibfnamefont {A.}~\bibnamefont
  {Taylor}}\ and\ \bibinfo {author} {\bibfnamefont {V.}~\bibnamefont {Varma}},\
  }\href {\doibase 10.1103/PhysRevD.102.104047} {\bibfield  {journal} {\bibinfo
   {journal} {Phys. Rev. D}\ }\textbf {\bibinfo {volume} {102}},\ \bibinfo
  {pages} {104047} (\bibinfo {year} {2020})},\ \Eprint
  {http://arxiv.org/abs/2010.00120} {arXiv:2010.00120 [gr-qc]} \BibitemShut
  {NoStop}%
\bibitem [{\citenamefont {Varma}\ \emph {et~al.}(2019)\citenamefont {Varma},
  \citenamefont {Field}, \citenamefont {Scheel}, \citenamefont {Blackman},
  \citenamefont {Gerosa}, \citenamefont {Stein}, \citenamefont {Kidder},\ and\
  \citenamefont {Pfeiffer}}]{Varma:2019csw}%
  \BibitemOpen
  \bibfield  {author} {\bibinfo {author} {\bibfnamefont {V.}~\bibnamefont
  {Varma}}, \bibinfo {author} {\bibfnamefont {S.~E.}\ \bibnamefont {Field}},
  \bibinfo {author} {\bibfnamefont {M.~A.}\ \bibnamefont {Scheel}}, \bibinfo
  {author} {\bibfnamefont {J.}~\bibnamefont {Blackman}}, \bibinfo {author}
  {\bibfnamefont {D.}~\bibnamefont {Gerosa}}, \bibinfo {author} {\bibfnamefont
  {L.~C.}\ \bibnamefont {Stein}}, \bibinfo {author} {\bibfnamefont {L.~E.}\
  \bibnamefont {Kidder}}, \ and\ \bibinfo {author} {\bibfnamefont {H.~P.}\
  \bibnamefont {Pfeiffer}},\ }\href {\doibase 10.1103/PhysRevResearch.1.033015}
  {\bibfield  {journal} {\bibinfo  {journal} {Phys. Rev. Research.}\ }\textbf
  {\bibinfo {volume} {1}},\ \bibinfo {pages} {033015} (\bibinfo {year}
  {2019})},\ \Eprint {http://arxiv.org/abs/1905.09300} {arXiv:1905.09300
  [gr-qc]} \BibitemShut {NoStop}%
\bibitem [{\citenamefont {Foucart}\ \emph {et~al.}(2014)\citenamefont
  {Foucart}, \citenamefont {Deaton}, \citenamefont {Duez}, \citenamefont
  {O'Connor}, \citenamefont {Ott}, \citenamefont {Haas}, \citenamefont
  {Kidder}, \citenamefont {Pfeiffer}, \citenamefont {Scheel},\ and\
  \citenamefont {Szilagyi}}]{Foucart:2014nda}%
  \BibitemOpen
  \bibfield  {author} {\bibinfo {author} {\bibfnamefont {F.}~\bibnamefont
  {Foucart}}, \bibinfo {author} {\bibfnamefont {M.~B.}\ \bibnamefont {Deaton}},
  \bibinfo {author} {\bibfnamefont {M.~D.}\ \bibnamefont {Duez}}, \bibinfo
  {author} {\bibfnamefont {E.}~\bibnamefont {O'Connor}}, \bibinfo {author}
  {\bibfnamefont {C.~D.}\ \bibnamefont {Ott}}, \bibinfo {author} {\bibfnamefont
  {R.}~\bibnamefont {Haas}}, \bibinfo {author} {\bibfnamefont {L.~E.}\
  \bibnamefont {Kidder}}, \bibinfo {author} {\bibfnamefont {H.~P.}\
  \bibnamefont {Pfeiffer}}, \bibinfo {author} {\bibfnamefont {M.~A.}\
  \bibnamefont {Scheel}}, \ and\ \bibinfo {author} {\bibfnamefont
  {B.}~\bibnamefont {Szilagyi}},\ }\href {\doibase 10.1103/PhysRevD.90.024026}
  {\bibfield  {journal} {\bibinfo  {journal} {Phys. Rev. D}\ }\textbf {\bibinfo
  {volume} {90}},\ \bibinfo {pages} {024026} (\bibinfo {year} {2014})},\
  \Eprint {http://arxiv.org/abs/1405.1121} {arXiv:1405.1121 [astro-ph.HE]}
  \BibitemShut {NoStop}%
\bibitem [{\citenamefont {Kyutoku}\ \emph {et~al.}(2015)\citenamefont
  {Kyutoku}, \citenamefont {Ioka}, \citenamefont {Okawa}, \citenamefont
  {Shibata},\ and\ \citenamefont {Taniguchi}}]{Kyutoku:2015gda}%
  \BibitemOpen
  \bibfield  {author} {\bibinfo {author} {\bibfnamefont {K.}~\bibnamefont
  {Kyutoku}}, \bibinfo {author} {\bibfnamefont {K.}~\bibnamefont {Ioka}},
  \bibinfo {author} {\bibfnamefont {H.}~\bibnamefont {Okawa}}, \bibinfo
  {author} {\bibfnamefont {M.}~\bibnamefont {Shibata}}, \ and\ \bibinfo
  {author} {\bibfnamefont {K.}~\bibnamefont {Taniguchi}},\ }\href {\doibase
  10.1103/PhysRevD.92.044028} {\bibfield  {journal} {\bibinfo  {journal} {Phys.
  Rev. D}\ }\textbf {\bibinfo {volume} {92}},\ \bibinfo {pages} {044028}
  (\bibinfo {year} {2015})},\ \Eprint {http://arxiv.org/abs/1502.05402}
  {arXiv:1502.05402 [astro-ph.HE]} \BibitemShut {NoStop}%
\bibitem [{\citenamefont {Most}\ \emph {et~al.}(2021)\citenamefont {Most},
  \citenamefont {Papenfort}, \citenamefont {Tootle},\ and\ \citenamefont
  {Rezzolla}}]{Most:2020exl}%
  \BibitemOpen
  \bibfield  {author} {\bibinfo {author} {\bibfnamefont {E.~R.}\ \bibnamefont
  {Most}}, \bibinfo {author} {\bibfnamefont {L.~J.}\ \bibnamefont {Papenfort}},
  \bibinfo {author} {\bibfnamefont {S.}~\bibnamefont {Tootle}}, \ and\ \bibinfo
  {author} {\bibfnamefont {L.}~\bibnamefont {Rezzolla}},\ }\href {\doibase
  10.3847/1538-4357/abf0a5} {\bibfield  {journal} {\bibinfo  {journal}
  {Astrophys. J.}\ }\textbf {\bibinfo {volume} {912}},\ \bibinfo {pages} {80}
  (\bibinfo {year} {2021})},\ \Eprint {http://arxiv.org/abs/2012.03896}
  {arXiv:2012.03896 [astro-ph.HE]} \BibitemShut {NoStop}%
\bibitem [{\citenamefont {Hayashi}\ \emph {et~al.}(2023)\citenamefont
  {Hayashi}, \citenamefont {Kiuchi}, \citenamefont {Kyutoku}, \citenamefont
  {Sekiguchi},\ and\ \citenamefont {Shibata}}]{Hayashi:2022cdq}%
  \BibitemOpen
  \bibfield  {author} {\bibinfo {author} {\bibfnamefont {K.}~\bibnamefont
  {Hayashi}}, \bibinfo {author} {\bibfnamefont {K.}~\bibnamefont {Kiuchi}},
  \bibinfo {author} {\bibfnamefont {K.}~\bibnamefont {Kyutoku}}, \bibinfo
  {author} {\bibfnamefont {Y.}~\bibnamefont {Sekiguchi}}, \ and\ \bibinfo
  {author} {\bibfnamefont {M.}~\bibnamefont {Shibata}},\ }\href {\doibase
  10.1103/PhysRevD.107.123001} {\bibfield  {journal} {\bibinfo  {journal}
  {Phys. Rev. D}\ }\textbf {\bibinfo {volume} {107}},\ \bibinfo {pages}
  {123001} (\bibinfo {year} {2023})},\ \Eprint
  {http://arxiv.org/abs/2211.07158} {arXiv:2211.07158 [astro-ph.HE]}
  \BibitemShut {NoStop}%
\bibitem [{\citenamefont {Kawaguchi}\ \emph {et~al.}(2016)\citenamefont
  {Kawaguchi}, \citenamefont {Kyutoku}, \citenamefont {Shibata},\ and\
  \citenamefont {Tanaka}}]{Kawaguchi:2016ana}%
  \BibitemOpen
  \bibfield  {author} {\bibinfo {author} {\bibfnamefont {K.}~\bibnamefont
  {Kawaguchi}}, \bibinfo {author} {\bibfnamefont {K.}~\bibnamefont {Kyutoku}},
  \bibinfo {author} {\bibfnamefont {M.}~\bibnamefont {Shibata}}, \ and\
  \bibinfo {author} {\bibfnamefont {M.}~\bibnamefont {Tanaka}},\ }\href
  {\doibase 10.3847/0004-637X/825/1/52} {\bibfield  {journal} {\bibinfo
  {journal} {Astrophys. J.}\ }\textbf {\bibinfo {volume} {825}},\ \bibinfo
  {pages} {52} (\bibinfo {year} {2016})},\ \Eprint
  {http://arxiv.org/abs/1601.07711} {arXiv:1601.07711 [astro-ph.HE]}
  \BibitemShut {NoStop}%
\bibitem [{\citenamefont {Foucart}\ \emph {et~al.}(2018)\citenamefont
  {Foucart}, \citenamefont {Hinderer},\ and\ \citenamefont
  {Nissanke}}]{Foucart:2018rjc}%
  \BibitemOpen
  \bibfield  {author} {\bibinfo {author} {\bibfnamefont {F.}~\bibnamefont
  {Foucart}}, \bibinfo {author} {\bibfnamefont {T.}~\bibnamefont {Hinderer}}, \
  and\ \bibinfo {author} {\bibfnamefont {S.}~\bibnamefont {Nissanke}},\ }\href
  {\doibase 10.1103/PhysRevD.98.081501} {\bibfield  {journal} {\bibinfo
  {journal} {Phys. Rev. D}\ }\textbf {\bibinfo {volume} {98}},\ \bibinfo
  {pages} {081501} (\bibinfo {year} {2018})},\ \Eprint
  {http://arxiv.org/abs/1807.00011} {arXiv:1807.00011 [astro-ph.HE]}
  \BibitemShut {NoStop}%
\bibitem [{\citenamefont {Kr\"uger}\ and\ \citenamefont
  {Foucart}(2020)}]{Kruger:2020gig}%
  \BibitemOpen
  \bibfield  {author} {\bibinfo {author} {\bibfnamefont {C.~J.}\ \bibnamefont
  {Kr\"uger}}\ and\ \bibinfo {author} {\bibfnamefont {F.}~\bibnamefont
  {Foucart}},\ }\href {\doibase 10.1103/PhysRevD.101.103002} {\bibfield
  {journal} {\bibinfo  {journal} {Phys. Rev. D}\ }\textbf {\bibinfo {volume}
  {101}},\ \bibinfo {pages} {103002} (\bibinfo {year} {2020})},\ \Eprint
  {http://arxiv.org/abs/2002.07728} {arXiv:2002.07728 [astro-ph.HE]}
  \BibitemShut {NoStop}%
\bibitem [{\citenamefont {Bernuzzi}\ \emph {et~al.}(2012)\citenamefont
  {Bernuzzi}, \citenamefont {Nagar}, \citenamefont {Thierfelder},\ and\
  \citenamefont {Br{\"u}gmann}}]{Bernuzzi:2012ci}%
  \BibitemOpen
  \bibfield  {author} {\bibinfo {author} {\bibfnamefont {S.}~\bibnamefont
  {Bernuzzi}}, \bibinfo {author} {\bibfnamefont {A.}~\bibnamefont {Nagar}},
  \bibinfo {author} {\bibfnamefont {M.}~\bibnamefont {Thierfelder}}, \ and\
  \bibinfo {author} {\bibfnamefont {B.}~\bibnamefont {Br{\"u}gmann}},\ }\href
  {\doibase 10.1103/PhysRevD.86.044030} {\bibfield  {journal} {\bibinfo
  {journal} {Phys. Rev. D}\ }\textbf {\bibinfo {volume} {86}},\ \bibinfo
  {pages} {044030} (\bibinfo {year} {2012})},\ \Eprint
  {http://arxiv.org/abs/1205.3403} {arXiv:1205.3403 [gr-qc]} \BibitemShut
  {NoStop}%
\bibitem [{\citenamefont {Bernuzzi}\ and\ \citenamefont
  {Dietrich}(2016)}]{Bernuzzi:2016pie}%
  \BibitemOpen
  \bibfield  {author} {\bibinfo {author} {\bibfnamefont {S.}~\bibnamefont
  {Bernuzzi}}\ and\ \bibinfo {author} {\bibfnamefont {T.}~\bibnamefont
  {Dietrich}},\ }\href {\doibase 10.1103/PhysRevD.94.064062} {\bibfield
  {journal} {\bibinfo  {journal} {Phys. Rev. D}\ }\textbf {\bibinfo {volume}
  {94}},\ \bibinfo {pages} {064062} (\bibinfo {year} {2016})},\ \Eprint
  {http://arxiv.org/abs/1604.07999} {arXiv:1604.07999 [gr-qc]} \BibitemShut
  {NoStop}%
\bibitem [{\citenamefont {Dietrich}\ \emph
  {et~al.}(2019{\natexlab{a}})\citenamefont {Dietrich} \emph
  {et~al.}}]{Dietrich:2018uni}%
  \BibitemOpen
  \bibfield  {author} {\bibinfo {author} {\bibfnamefont {T.}~\bibnamefont
  {Dietrich}} \emph {et~al.},\ }\href {\doibase 10.1103/PhysRevD.99.024029}
  {\bibfield  {journal} {\bibinfo  {journal} {Phys. Rev. D}\ }\textbf {\bibinfo
  {volume} {99}},\ \bibinfo {pages} {024029} (\bibinfo {year}
  {2019}{\natexlab{a}})},\ \Eprint {http://arxiv.org/abs/1804.02235}
  {arXiv:1804.02235 [gr-qc]} \BibitemShut {NoStop}%
\bibitem [{\citenamefont {Dietrich}\ \emph {et~al.}(2017)\citenamefont
  {Dietrich}, \citenamefont {Bernuzzi},\ and\ \citenamefont
  {Tichy}}]{Dietrich:2017aum}%
  \BibitemOpen
  \bibfield  {author} {\bibinfo {author} {\bibfnamefont {T.}~\bibnamefont
  {Dietrich}}, \bibinfo {author} {\bibfnamefont {S.}~\bibnamefont {Bernuzzi}},
  \ and\ \bibinfo {author} {\bibfnamefont {W.}~\bibnamefont {Tichy}},\ }\href
  {\doibase 10.1103/PhysRevD.96.121501} {\bibfield  {journal} {\bibinfo
  {journal} {Phys. Rev. D}\ }\textbf {\bibinfo {volume} {96}},\ \bibinfo
  {pages} {121501} (\bibinfo {year} {2017})},\ \Eprint
  {http://arxiv.org/abs/1706.02969} {arXiv:1706.02969 [gr-qc]} \BibitemShut
  {NoStop}%
\bibitem [{\citenamefont {Dietrich}\ \emph
  {et~al.}(2019{\natexlab{b}})\citenamefont {Dietrich}, \citenamefont
  {Samajdar}, \citenamefont {Khan}, \citenamefont {Johnson-McDaniel},
  \citenamefont {Dudi},\ and\ \citenamefont {Tichy}}]{Dietrich:2019kaq}%
  \BibitemOpen
  \bibfield  {author} {\bibinfo {author} {\bibfnamefont {T.}~\bibnamefont
  {Dietrich}}, \bibinfo {author} {\bibfnamefont {A.}~\bibnamefont {Samajdar}},
  \bibinfo {author} {\bibfnamefont {S.}~\bibnamefont {Khan}}, \bibinfo {author}
  {\bibfnamefont {N.~K.}\ \bibnamefont {Johnson-McDaniel}}, \bibinfo {author}
  {\bibfnamefont {R.}~\bibnamefont {Dudi}}, \ and\ \bibinfo {author}
  {\bibfnamefont {W.}~\bibnamefont {Tichy}},\ }\href {\doibase
  10.1103/PhysRevD.100.044003} {\bibfield  {journal} {\bibinfo  {journal}
  {Phys. Rev. D}\ }\textbf {\bibinfo {volume} {100}},\ \bibinfo {pages}
  {044003} (\bibinfo {year} {2019}{\natexlab{b}})},\ \Eprint
  {http://arxiv.org/abs/1905.06011} {arXiv:1905.06011 [gr-qc]} \BibitemShut
  {NoStop}%
\bibitem [{\citenamefont {{LIGO Scientific Collaboration}}(2018)}]{lalsuite}%
  \BibitemOpen
  \bibfield  {author} {\bibinfo {author} {\bibnamefont {{LIGO Scientific
  Collaboration}}},\ }\href {\doibase 10.7935/GT1W-FZ16} {\enquote {\bibinfo
  {title} {{LIGO} {A}lgorithm {L}ibrary - {LALS}uite},}\ }\bibinfo
  {howpublished} {free software (GPL)} (\bibinfo {year} {2018})\BibitemShut
  {NoStop}%
\bibitem [{\citenamefont {Khan}\ \emph {et~al.}(2016)\citenamefont {Khan},
  \citenamefont {Husa}, \citenamefont {Hannam}, \citenamefont {Ohme},
  \citenamefont {P{\"u}rrer}, \citenamefont {Jim{\'e}nez~Forteza},\ and\
  \citenamefont {Boh{\'e}}}]{Khan:2015jqa}%
  \BibitemOpen
  \bibfield  {author} {\bibinfo {author} {\bibfnamefont {S.}~\bibnamefont
  {Khan}}, \bibinfo {author} {\bibfnamefont {S.}~\bibnamefont {Husa}}, \bibinfo
  {author} {\bibfnamefont {M.}~\bibnamefont {Hannam}}, \bibinfo {author}
  {\bibfnamefont {F.}~\bibnamefont {Ohme}}, \bibinfo {author} {\bibfnamefont
  {M.}~\bibnamefont {P{\"u}rrer}}, \bibinfo {author} {\bibfnamefont
  {X.}~\bibnamefont {Jim{\'e}nez~Forteza}}, \ and\ \bibinfo {author}
  {\bibfnamefont {A.}~\bibnamefont {Boh{\'e}}},\ }\href {\doibase
  10.1103/PhysRevD.93.044007} {\bibfield  {journal} {\bibinfo  {journal} {Phys.
  Rev. D}\ }\textbf {\bibinfo {volume} {93}},\ \bibinfo {pages} {044007}
  (\bibinfo {year} {2016})},\ \Eprint {http://arxiv.org/abs/1508.07253}
  {arXiv:1508.07253 [gr-qc]} \BibitemShut {NoStop}%
\bibitem [{\citenamefont {Hotokezaka}\ \emph {et~al.}(2016)\citenamefont
  {Hotokezaka}, \citenamefont {Kyutoku}, \citenamefont {Sekiguchi},\ and\
  \citenamefont {Shibata}}]{Hotokezaka:2016bzh}%
  \BibitemOpen
  \bibfield  {author} {\bibinfo {author} {\bibfnamefont {K.}~\bibnamefont
  {Hotokezaka}}, \bibinfo {author} {\bibfnamefont {K.}~\bibnamefont {Kyutoku}},
  \bibinfo {author} {\bibfnamefont {Y.-i.}\ \bibnamefont {Sekiguchi}}, \ and\
  \bibinfo {author} {\bibfnamefont {M.}~\bibnamefont {Shibata}},\ }\href
  {\doibase 10.1103/PhysRevD.93.064082} {\bibfield  {journal} {\bibinfo
  {journal} {Phys. Rev. D}\ }\textbf {\bibinfo {volume} {93}},\ \bibinfo
  {pages} {064082} (\bibinfo {year} {2016})},\ \Eprint
  {http://arxiv.org/abs/1603.01286} {arXiv:1603.01286 [gr-qc]} \BibitemShut
  {NoStop}%
\bibitem [{\citenamefont {Lackey}\ \emph {et~al.}(2019)\citenamefont {Lackey},
  \citenamefont {P\"urrer}, \citenamefont {Taracchini},\ and\ \citenamefont
  {Marsat}}]{Lackey:2018zvw}%
  \BibitemOpen
  \bibfield  {author} {\bibinfo {author} {\bibfnamefont {B.~D.}\ \bibnamefont
  {Lackey}}, \bibinfo {author} {\bibfnamefont {M.}~\bibnamefont {P\"urrer}},
  \bibinfo {author} {\bibfnamefont {A.}~\bibnamefont {Taracchini}}, \ and\
  \bibinfo {author} {\bibfnamefont {S.}~\bibnamefont {Marsat}},\ }\href
  {\doibase 10.1103/PhysRevD.100.024002} {\bibfield  {journal} {\bibinfo
  {journal} {Phys. Rev. D}\ }\textbf {\bibinfo {volume} {100}},\ \bibinfo
  {pages} {024002} (\bibinfo {year} {2019})},\ \Eprint
  {http://arxiv.org/abs/1812.08643} {arXiv:1812.08643 [gr-qc]} \BibitemShut
  {NoStop}%
\bibitem [{\citenamefont {Riemenschneider}\ \emph {et~al.}(2021)\citenamefont
  {Riemenschneider}, \citenamefont {Rettegno}, \citenamefont {Breschi},
  \citenamefont {Albertini}, \citenamefont {Gamba}, \citenamefont {Bernuzzi},\
  and\ \citenamefont {Nagar}}]{Riemenschneider:2021ppj}%
  \BibitemOpen
  \bibfield  {author} {\bibinfo {author} {\bibfnamefont {G.}~\bibnamefont
  {Riemenschneider}}, \bibinfo {author} {\bibfnamefont {P.}~\bibnamefont
  {Rettegno}}, \bibinfo {author} {\bibfnamefont {M.}~\bibnamefont {Breschi}},
  \bibinfo {author} {\bibfnamefont {A.}~\bibnamefont {Albertini}}, \bibinfo
  {author} {\bibfnamefont {R.}~\bibnamefont {Gamba}}, \bibinfo {author}
  {\bibfnamefont {S.}~\bibnamefont {Bernuzzi}}, \ and\ \bibinfo {author}
  {\bibfnamefont {A.}~\bibnamefont {Nagar}},\ }\href {\doibase
  10.1103/PhysRevD.104.104045} {\bibfield  {journal} {\bibinfo  {journal}
  {Phys. Rev. D}\ }\textbf {\bibinfo {volume} {104}},\ \bibinfo {pages}
  {104045} (\bibinfo {year} {2021})},\ \Eprint
  {http://arxiv.org/abs/2104.07533} {arXiv:2104.07533 [gr-qc]} \BibitemShut
  {NoStop}%
\bibitem [{\citenamefont {Bulla}(2019)}]{Bulla:2019muo}%
  \BibitemOpen
  \bibfield  {author} {\bibinfo {author} {\bibfnamefont {M.}~\bibnamefont
  {Bulla}},\ }\href {\doibase 10.1093/mnras/stz2495} {\bibfield  {journal}
  {\bibinfo  {journal} {Mon. Not. Roy. Astron. Soc.}\ }\textbf {\bibinfo
  {volume} {489}},\ \bibinfo {pages} {5037} (\bibinfo {year} {2019})},\ \Eprint
  {http://arxiv.org/abs/1906.04205} {arXiv:1906.04205 [astro-ph.HE]}
  \BibitemShut {NoStop}%
\bibitem [{\citenamefont {Bulla}(2023)}]{Bulla:2022mwo}%
  \BibitemOpen
  \bibfield  {author} {\bibinfo {author} {\bibfnamefont {M.}~\bibnamefont
  {Bulla}},\ }\href {\doibase 10.1093/mnras/stad232} {\bibfield  {journal}
  {\bibinfo  {journal} {Mon. Not. Roy. Astron. Soc.}\ }\textbf {\bibinfo
  {volume} {520}},\ \bibinfo {pages} {2558} (\bibinfo {year} {2023})},\ \Eprint
  {http://arxiv.org/abs/2211.14348} {arXiv:2211.14348 [astro-ph.HE]}
  \BibitemShut {NoStop}%
\bibitem [{\citenamefont {{Rosswog}}\ and\ \citenamefont
  {{Korobkin}}(2022)}]{Rosswog2022}%
  \BibitemOpen
  \bibfield  {author} {\bibinfo {author} {\bibfnamefont {S.}~\bibnamefont
  {{Rosswog}}}\ and\ \bibinfo {author} {\bibfnamefont {O.}~\bibnamefont
  {{Korobkin}}},\ }\href@noop {} {\bibfield  {journal} {\bibinfo  {journal}
  {arXiv e-prints}\ ,\ \bibinfo {eid} {arXiv:2208.14026}} (\bibinfo {year}
  {2022})},\ \Eprint {http://arxiv.org/abs/2208.14026} {arXiv:2208.14026
  [astro-ph.HE]} \BibitemShut {NoStop}%
\bibitem [{\citenamefont {Barnes}\ \emph {et~al.}(2016)\citenamefont {Barnes},
  \citenamefont {Kasen}, \citenamefont {Wu},\ and\ \citenamefont
  {Martínez-Pinedo}}]{Barnes:2016umi}%
  \BibitemOpen
  \bibfield  {author} {\bibinfo {author} {\bibfnamefont {J.}~\bibnamefont
  {Barnes}}, \bibinfo {author} {\bibfnamefont {D.}~\bibnamefont {Kasen}},
  \bibinfo {author} {\bibfnamefont {M.-R.}\ \bibnamefont {Wu}}, \ and\ \bibinfo
  {author} {\bibfnamefont {G.}~\bibnamefont {Martínez-Pinedo}},\ }\href
  {\doibase 10.3847/0004-637X/829/2/110} {\bibfield  {journal} {\bibinfo
  {journal} {Astrophys. J.}\ }\textbf {\bibinfo {volume} {829}},\ \bibinfo
  {pages} {110} (\bibinfo {year} {2016})},\ \Eprint
  {http://arxiv.org/abs/1605.07218} {arXiv:1605.07218 [astro-ph.HE]}
  \BibitemShut {NoStop}%
\bibitem [{\citenamefont {Tanaka}\ \emph {et~al.}(2020)\citenamefont {Tanaka},
  \citenamefont {Kato}, \citenamefont {Gaigalas},\ and\ \citenamefont
  {Kawaguchi}}]{Tanaka:2019iqp}%
  \BibitemOpen
  \bibfield  {author} {\bibinfo {author} {\bibfnamefont {M.}~\bibnamefont
  {Tanaka}}, \bibinfo {author} {\bibfnamefont {D.}~\bibnamefont {Kato}},
  \bibinfo {author} {\bibfnamefont {G.}~\bibnamefont {Gaigalas}}, \ and\
  \bibinfo {author} {\bibfnamefont {K.}~\bibnamefont {Kawaguchi}},\ }\href
  {\doibase 10.1093/mnras/staa1576} {\bibfield  {journal} {\bibinfo  {journal}
  {Mon. Not. Roy. Astron. Soc.}\ }\textbf {\bibinfo {volume} {496}},\ \bibinfo
  {pages} {1369} (\bibinfo {year} {2020})},\ \Eprint
  {http://arxiv.org/abs/1906.08914} {arXiv:1906.08914 [astro-ph.HE]}
  \BibitemShut {NoStop}%
\bibitem [{\citenamefont {Kawaguchi}\ \emph {et~al.}(2021)\citenamefont
  {Kawaguchi}, \citenamefont {Fujibayashi}, \citenamefont {Shibata},
  \citenamefont {Tanaka},\ and\ \citenamefont {Wanajo}}]{Kawaguchi:2020vbf}%
  \BibitemOpen
  \bibfield  {author} {\bibinfo {author} {\bibfnamefont {K.}~\bibnamefont
  {Kawaguchi}}, \bibinfo {author} {\bibfnamefont {S.}~\bibnamefont
  {Fujibayashi}}, \bibinfo {author} {\bibfnamefont {M.}~\bibnamefont
  {Shibata}}, \bibinfo {author} {\bibfnamefont {M.}~\bibnamefont {Tanaka}}, \
  and\ \bibinfo {author} {\bibfnamefont {S.}~\bibnamefont {Wanajo}},\ }\href
  {\doibase 10.3847/1538-4357/abf3bc} {\bibfield  {journal} {\bibinfo
  {journal} {Astrophys. J.}\ }\textbf {\bibinfo {volume} {913}},\ \bibinfo
  {pages} {100} (\bibinfo {year} {2021})},\ \Eprint
  {http://arxiv.org/abs/2012.14711} {arXiv:2012.14711 [astro-ph.HE]}
  \BibitemShut {NoStop}%
\bibitem [{\citenamefont {Kawaguchi}\ \emph {et~al.}(2022)\citenamefont
  {Kawaguchi}, \citenamefont {Fujibayashi}, \citenamefont {Hotokezaka},
  \citenamefont {Shibata},\ and\ \citenamefont {Wanajo}}]{Kawaguchi:2022bub}%
  \BibitemOpen
  \bibfield  {author} {\bibinfo {author} {\bibfnamefont {K.}~\bibnamefont
  {Kawaguchi}}, \bibinfo {author} {\bibfnamefont {S.}~\bibnamefont
  {Fujibayashi}}, \bibinfo {author} {\bibfnamefont {K.}~\bibnamefont
  {Hotokezaka}}, \bibinfo {author} {\bibfnamefont {M.}~\bibnamefont {Shibata}},
  \ and\ \bibinfo {author} {\bibfnamefont {S.}~\bibnamefont {Wanajo}},\ }\href
  {\doibase 10.3847/1538-4357/ac6ef7} {\bibfield  {journal} {\bibinfo
  {journal} {Astrophys. J.}\ }\textbf {\bibinfo {volume} {933}},\ \bibinfo
  {pages} {22} (\bibinfo {year} {2022})},\ \Eprint
  {http://arxiv.org/abs/2202.13149} {arXiv:2202.13149 [astro-ph.HE]}
  \BibitemShut {NoStop}%
\bibitem [{\citenamefont {Neuweiler}\ \emph {et~al.}(2023)\citenamefont
  {Neuweiler}, \citenamefont {Dietrich}, \citenamefont {Bulla}, \citenamefont
  {Chaurasia}, \citenamefont {Rosswog},\ and\ \citenamefont
  {Ujevic}}]{Neuweiler:2022eum}%
  \BibitemOpen
  \bibfield  {author} {\bibinfo {author} {\bibfnamefont {A.}~\bibnamefont
  {Neuweiler}}, \bibinfo {author} {\bibfnamefont {T.}~\bibnamefont {Dietrich}},
  \bibinfo {author} {\bibfnamefont {M.}~\bibnamefont {Bulla}}, \bibinfo
  {author} {\bibfnamefont {S.~V.}\ \bibnamefont {Chaurasia}}, \bibinfo {author}
  {\bibfnamefont {S.}~\bibnamefont {Rosswog}}, \ and\ \bibinfo {author}
  {\bibfnamefont {M.}~\bibnamefont {Ujevic}},\ }\href {\doibase
  10.1103/PhysRevD.107.023016} {\bibfield  {journal} {\bibinfo  {journal}
  {Phys. Rev. D}\ }\textbf {\bibinfo {volume} {107}},\ \bibinfo {pages}
  {023016} (\bibinfo {year} {2023})},\ \Eprint
  {http://arxiv.org/abs/2208.13460} {arXiv:2208.13460 [astro-ph.HE]}
  \BibitemShut {NoStop}%
\bibitem [{\citenamefont {Darbha}\ \emph {et~al.}(2021)\citenamefont {Darbha},
  \citenamefont {Kasen}, \citenamefont {Foucart},\ and\ \citenamefont
  {Price}}]{Darbha:2021rqj}%
  \BibitemOpen
  \bibfield  {author} {\bibinfo {author} {\bibfnamefont {S.}~\bibnamefont
  {Darbha}}, \bibinfo {author} {\bibfnamefont {D.}~\bibnamefont {Kasen}},
  \bibinfo {author} {\bibfnamefont {F.}~\bibnamefont {Foucart}}, \ and\
  \bibinfo {author} {\bibfnamefont {D.~J.}\ \bibnamefont {Price}},\ }\href
  {\doibase 10.3847/1538-4357/abff5d} {\  (\bibinfo {year} {2021}),\
  10.3847/1538-4357/abff5d},\ \Eprint {http://arxiv.org/abs/2103.03378}
  {arXiv:2103.03378 [astro-ph.HE]} \BibitemShut {NoStop}%
\bibitem [{\citenamefont {Collins}\ \emph {et~al.}(2023)\citenamefont
  {Collins}, \citenamefont {Bauswein}, \citenamefont {Sim}, \citenamefont
  {Vijayan}, \citenamefont {Mart\'\i{}nez-Pinedo}, \citenamefont {Just},
  \citenamefont {Shingles},\ and\ \citenamefont {Kromer}}]{Collins:2022ocl}%
  \BibitemOpen
  \bibfield  {author} {\bibinfo {author} {\bibfnamefont {C.~E.}\ \bibnamefont
  {Collins}}, \bibinfo {author} {\bibfnamefont {A.}~\bibnamefont {Bauswein}},
  \bibinfo {author} {\bibfnamefont {S.~A.}\ \bibnamefont {Sim}}, \bibinfo
  {author} {\bibfnamefont {V.}~\bibnamefont {Vijayan}}, \bibinfo {author}
  {\bibfnamefont {G.}~\bibnamefont {Mart\'\i{}nez-Pinedo}}, \bibinfo {author}
  {\bibfnamefont {O.}~\bibnamefont {Just}}, \bibinfo {author} {\bibfnamefont
  {L.~J.}\ \bibnamefont {Shingles}}, \ and\ \bibinfo {author} {\bibfnamefont
  {M.}~\bibnamefont {Kromer}},\ }\href {\doibase 10.1093/mnras/stad606}
  {\bibfield  {journal} {\bibinfo  {journal} {Mon. Not. Roy. Astron. Soc.}\
  }\textbf {\bibinfo {volume} {521}},\ \bibinfo {pages} {1858} (\bibinfo {year}
  {2023})},\ \Eprint {http://arxiv.org/abs/2209.05246} {arXiv:2209.05246
  [astro-ph.HE]} \BibitemShut {NoStop}%
\bibitem [{\citenamefont {Kasen}\ \emph {et~al.}(2015)\citenamefont {Kasen},
  \citenamefont {Fernandez},\ and\ \citenamefont {Metzger}}]{Kasen:2014toa}%
  \BibitemOpen
  \bibfield  {author} {\bibinfo {author} {\bibfnamefont {D.}~\bibnamefont
  {Kasen}}, \bibinfo {author} {\bibfnamefont {R.}~\bibnamefont {Fernandez}}, \
  and\ \bibinfo {author} {\bibfnamefont {B.}~\bibnamefont {Metzger}},\ }\href
  {\doibase 10.1093/mnras/stv721} {\bibfield  {journal} {\bibinfo  {journal}
  {Mon. Not. Roy. Astron. Soc.}\ }\textbf {\bibinfo {volume} {450}},\ \bibinfo
  {pages} {1777} (\bibinfo {year} {2015})},\ \Eprint
  {http://arxiv.org/abs/1411.3726} {arXiv:1411.3726 [astro-ph.HE]} \BibitemShut
  {NoStop}%
\bibitem [{\citenamefont {Curtis}\ \emph {et~al.}(2022)\citenamefont {Curtis},
  \citenamefont {M\"osta}, \citenamefont {Wu}, \citenamefont {Radice},
  \citenamefont {Roberts}, \citenamefont {Ricigliano},\ and\ \citenamefont
  {Perego}}]{Curtis:2021guz}%
  \BibitemOpen
  \bibfield  {author} {\bibinfo {author} {\bibfnamefont {S.}~\bibnamefont
  {Curtis}}, \bibinfo {author} {\bibfnamefont {P.}~\bibnamefont {M\"osta}},
  \bibinfo {author} {\bibfnamefont {Z.}~\bibnamefont {Wu}}, \bibinfo {author}
  {\bibfnamefont {D.}~\bibnamefont {Radice}}, \bibinfo {author} {\bibfnamefont
  {L.}~\bibnamefont {Roberts}}, \bibinfo {author} {\bibfnamefont
  {G.}~\bibnamefont {Ricigliano}}, \ and\ \bibinfo {author} {\bibfnamefont
  {A.}~\bibnamefont {Perego}},\ }\href {\doibase 10.1093/mnras/stac3128}
  {\bibfield  {journal} {\bibinfo  {journal} {Mon. Not. Roy. Astron. Soc.}\
  }\textbf {\bibinfo {volume} {518}},\ \bibinfo {pages} {5313} (\bibinfo {year}
  {2022})},\ \Eprint {http://arxiv.org/abs/2112.00772} {arXiv:2112.00772
  [astro-ph.HE]} \BibitemShut {NoStop}%
\bibitem [{\citenamefont {Just}\ \emph {et~al.}(2022)\citenamefont {Just},
  \citenamefont {Kullmann}, \citenamefont {Goriely}, \citenamefont {Bauswein},
  \citenamefont {Janka},\ and\ \citenamefont {Collins}}]{Just:2021vzy}%
  \BibitemOpen
  \bibfield  {author} {\bibinfo {author} {\bibfnamefont {O.}~\bibnamefont
  {Just}}, \bibinfo {author} {\bibfnamefont {I.}~\bibnamefont {Kullmann}},
  \bibinfo {author} {\bibfnamefont {S.}~\bibnamefont {Goriely}}, \bibinfo
  {author} {\bibfnamefont {A.}~\bibnamefont {Bauswein}}, \bibinfo {author}
  {\bibfnamefont {H.-T.}\ \bibnamefont {Janka}}, \ and\ \bibinfo {author}
  {\bibfnamefont {C.~E.}\ \bibnamefont {Collins}},\ }\href {\doibase
  10.1093/mnras/stab3327} {\bibfield  {journal} {\bibinfo  {journal} {Mon. Not.
  Roy. Astron. Soc.}\ }\textbf {\bibinfo {volume} {510}},\ \bibinfo {pages}
  {2820} (\bibinfo {year} {2022})},\ \Eprint {http://arxiv.org/abs/2109.14617}
  {arXiv:2109.14617 [astro-ph.HE]} \BibitemShut {NoStop}%
\bibitem [{\citenamefont {Klion}\ \emph {et~al.}(2022)\citenamefont {Klion},
  \citenamefont {Tchekhovskoy}, \citenamefont {Kasen}, \citenamefont
  {Kathirgamaraju}, \citenamefont {Quataert},\ and\ \citenamefont
  {Fern\'andez}}]{Klion:2021jzr}%
  \BibitemOpen
  \bibfield  {author} {\bibinfo {author} {\bibfnamefont {H.}~\bibnamefont
  {Klion}}, \bibinfo {author} {\bibfnamefont {A.}~\bibnamefont {Tchekhovskoy}},
  \bibinfo {author} {\bibfnamefont {D.}~\bibnamefont {Kasen}}, \bibinfo
  {author} {\bibfnamefont {A.}~\bibnamefont {Kathirgamaraju}}, \bibinfo
  {author} {\bibfnamefont {E.}~\bibnamefont {Quataert}}, \ and\ \bibinfo
  {author} {\bibfnamefont {R.}~\bibnamefont {Fern\'andez}},\ }\href {\doibase
  10.1093/mnras/stab3583} {\bibfield  {journal} {\bibinfo  {journal} {Mon. Not.
  Roy. Astron. Soc.}\ }\textbf {\bibinfo {volume} {510}},\ \bibinfo {pages}
  {2968} (\bibinfo {year} {2022})},\ \Eprint {http://arxiv.org/abs/2108.04251}
  {arXiv:2108.04251 [astro-ph.HE]} \BibitemShut {NoStop}%
\bibitem [{\citenamefont {Wu}\ \emph {et~al.}(2022)\citenamefont {Wu},
  \citenamefont {Ricigliano}, \citenamefont {Kashyap}, \citenamefont {Perego},\
  and\ \citenamefont {Radice}}]{Wu:2021ibi}%
  \BibitemOpen
  \bibfield  {author} {\bibinfo {author} {\bibfnamefont {Z.}~\bibnamefont
  {Wu}}, \bibinfo {author} {\bibfnamefont {G.}~\bibnamefont {Ricigliano}},
  \bibinfo {author} {\bibfnamefont {R.}~\bibnamefont {Kashyap}}, \bibinfo
  {author} {\bibfnamefont {A.}~\bibnamefont {Perego}}, \ and\ \bibinfo {author}
  {\bibfnamefont {D.}~\bibnamefont {Radice}},\ }\href {\doibase
  10.1093/mnras/stac399} {\bibfield  {journal} {\bibinfo  {journal} {Mon. Not.
  Roy. Astron. Soc.}\ }\textbf {\bibinfo {volume} {512}},\ \bibinfo {pages}
  {328} (\bibinfo {year} {2022})},\ \Eprint {http://arxiv.org/abs/2111.06870}
  {arXiv:2111.06870 [astro-ph.HE]} \BibitemShut {NoStop}%
\bibitem [{\citenamefont {Kedia}\ \emph {et~al.}(2023)\citenamefont {Kedia},
  \citenamefont {Ristic}, \citenamefont {O'Shaughnessy}, \citenamefont
  {Yelikar}, \citenamefont {Wollaeger}, \citenamefont {Korobkin}, \citenamefont
  {Chase}, \citenamefont {Fryer},\ and\ \citenamefont
  {Fontes}}]{Kedia:2022onl}%
  \BibitemOpen
  \bibfield  {author} {\bibinfo {author} {\bibfnamefont {A.}~\bibnamefont
  {Kedia}}, \bibinfo {author} {\bibfnamefont {M.}~\bibnamefont {Ristic}},
  \bibinfo {author} {\bibfnamefont {R.}~\bibnamefont {O'Shaughnessy}}, \bibinfo
  {author} {\bibfnamefont {A.~B.}\ \bibnamefont {Yelikar}}, \bibinfo {author}
  {\bibfnamefont {R.~T.}\ \bibnamefont {Wollaeger}}, \bibinfo {author}
  {\bibfnamefont {O.}~\bibnamefont {Korobkin}}, \bibinfo {author}
  {\bibfnamefont {E.~A.}\ \bibnamefont {Chase}}, \bibinfo {author}
  {\bibfnamefont {C.~L.}\ \bibnamefont {Fryer}}, \ and\ \bibinfo {author}
  {\bibfnamefont {C.~J.}\ \bibnamefont {Fontes}},\ }\href {\doibase
  10.1103/PhysRevResearch.5.013168} {\bibfield  {journal} {\bibinfo  {journal}
  {Phys. Rev. Res.}\ }\textbf {\bibinfo {volume} {5}},\ \bibinfo {pages}
  {013168} (\bibinfo {year} {2023})},\ \Eprint
  {http://arxiv.org/abs/2211.04363} {arXiv:2211.04363 [astro-ph.HE]}
  \BibitemShut {NoStop}%
\bibitem [{\citenamefont {Markin}\ \emph {et~al.}(2023)\citenamefont {Markin},
  \citenamefont {Bulla}, \citenamefont {Neuweiler},\ and\ \citenamefont
  {Chaurasia}}]{NSbhLuminosityVideo}%
  \BibitemOpen
  \bibfield  {author} {\bibinfo {author} {\bibfnamefont {I.}~\bibnamefont
  {Markin}}, \bibinfo {author} {\bibfnamefont {M.}~\bibnamefont {Bulla}},
  \bibinfo {author} {\bibfnamefont {A.}~\bibnamefont {Neuweiler}}, \ and\
  \bibinfo {author} {\bibfnamefont {S.~V.}\ \bibnamefont {Chaurasia}},\ }\href
  {\doibase 10.5281/zenodo.7826447} {\enquote {\bibinfo {title}
  {{General-Relativistic Hydrodynamics Simulation of a Neutron Star —
  Sub-Solar-Mass Black Hole Merger - Kilonova Luminosity Evolution
  Visualization}},}\ } (\bibinfo {year} {2023})\BibitemShut {NoStop}%
\bibitem [{\citenamefont {Chaurasia}\ \emph {et~al.}(2023)\citenamefont
  {Chaurasia}, \citenamefont {Ujevic}, \citenamefont {Abac},\ and\
  \citenamefont {Markin}}]{NSbh_waveform}%
  \BibitemOpen
  \bibfield  {author} {\bibinfo {author} {\bibfnamefont {S.~V.}\ \bibnamefont
  {Chaurasia}}, \bibinfo {author} {\bibfnamefont {M.}~\bibnamefont {Ujevic}},
  \bibinfo {author} {\bibfnamefont {A.}~\bibnamefont {Abac}}, \ and\ \bibinfo
  {author} {\bibfnamefont {I.}~\bibnamefont {Markin}},\ }\href {\doibase
  10.5281/zenodo.7916719} {\enquote {\bibinfo {title} {{General-Relativistic
  Hydrodynamics Simulation of a Neutron Star — Sub-Solar-Mass Black Hole
  Merger - Gravitational Waveform}},}\ } (\bibinfo {year} {2023})\BibitemShut
  {NoStop}%
\bibitem [{\citenamefont {Chaurasia}\ and\ \citenamefont
  {Markin}(2023)}]{NSbh_3D_data}%
  \BibitemOpen
  \bibfield  {author} {\bibinfo {author} {\bibfnamefont {S.~V.}\ \bibnamefont
  {Chaurasia}}\ and\ \bibinfo {author} {\bibfnamefont {I.}~\bibnamefont
  {Markin}},\ }\href {\doibase 10.5281/zenodo.7915747} {\enquote {\bibinfo
  {title} {{General-Relativistic Hydrodynamics Simulation of a Neutron Star —
  Sub-Solar-Mass Black Hole Merger - 3D Ejecta Data}},}\ } (\bibinfo {year}
  {2023})\BibitemShut {NoStop}%
\bibitem [{\citenamefont {Gieg}\ \emph {et~al.}(2022)\citenamefont {Gieg},
  \citenamefont {Schianchi}, \citenamefont {Dietrich},\ and\ \citenamefont
  {Ujevic}}]{Gieg:2022mut}%
  \BibitemOpen
  \bibfield  {author} {\bibinfo {author} {\bibfnamefont {H.}~\bibnamefont
  {Gieg}}, \bibinfo {author} {\bibfnamefont {F.}~\bibnamefont {Schianchi}},
  \bibinfo {author} {\bibfnamefont {T.}~\bibnamefont {Dietrich}}, \ and\
  \bibinfo {author} {\bibfnamefont {M.}~\bibnamefont {Ujevic}},\ }\href
  {\doibase 10.3390/universe8070370} {\bibfield  {journal} {\bibinfo  {journal}
  {Universe}\ }\textbf {\bibinfo {volume} {8}},\ \bibinfo {pages} {370}
  (\bibinfo {year} {2022})},\ \Eprint {http://arxiv.org/abs/2206.01337}
  {arXiv:2206.01337 [gr-qc]} \BibitemShut {NoStop}%
\bibitem [{\citenamefont {Wollaeger}\ \emph {et~al.}(2018)\citenamefont
  {Wollaeger}, \citenamefont {Korobkin}, \citenamefont {Fontes}, \citenamefont
  {Rosswog}, \citenamefont {Even}, \citenamefont {Fryer}, \citenamefont
  {Sollerman}, \citenamefont {Hungerford}, \citenamefont {van Rossum},\ and\
  \citenamefont {Wollaber}}]{Wollaeger:2017ahm}%
  \BibitemOpen
  \bibfield  {author} {\bibinfo {author} {\bibfnamefont {R.~T.}\ \bibnamefont
  {Wollaeger}}, \bibinfo {author} {\bibfnamefont {O.}~\bibnamefont {Korobkin}},
  \bibinfo {author} {\bibfnamefont {C.~J.}\ \bibnamefont {Fontes}}, \bibinfo
  {author} {\bibfnamefont {S.~K.}\ \bibnamefont {Rosswog}}, \bibinfo {author}
  {\bibfnamefont {W.~P.}\ \bibnamefont {Even}}, \bibinfo {author}
  {\bibfnamefont {C.~L.}\ \bibnamefont {Fryer}}, \bibinfo {author}
  {\bibfnamefont {J.}~\bibnamefont {Sollerman}}, \bibinfo {author}
  {\bibfnamefont {A.~L.}\ \bibnamefont {Hungerford}}, \bibinfo {author}
  {\bibfnamefont {D.~R.}\ \bibnamefont {van Rossum}}, \ and\ \bibinfo {author}
  {\bibfnamefont {A.~B.}\ \bibnamefont {Wollaber}},\ }\href {\doibase
  10.1093/mnras/sty1018} {\bibfield  {journal} {\bibinfo  {journal} {Mon. Not.
  Roy. Astron. Soc.}\ }\textbf {\bibinfo {volume} {478}},\ \bibinfo {pages}
  {3298} (\bibinfo {year} {2018})},\ \Eprint {http://arxiv.org/abs/1705.07084}
  {arXiv:1705.07084 [astro-ph.HE]} \BibitemShut {NoStop}%
\bibitem [{\citenamefont {Bulla}\ \emph {et~al.}(2015)\citenamefont {Bulla},
  \citenamefont {Sim},\ and\ \citenamefont {Kromer}}]{Bulla:2015eza}%
  \BibitemOpen
  \bibfield  {author} {\bibinfo {author} {\bibfnamefont {M.}~\bibnamefont
  {Bulla}}, \bibinfo {author} {\bibfnamefont {S.~A.}\ \bibnamefont {Sim}}, \
  and\ \bibinfo {author} {\bibfnamefont {M.}~\bibnamefont {Kromer}},\ }\href
  {\doibase 10.1093/mnras/stv657} {\bibfield  {journal} {\bibinfo  {journal}
  {Mon. Not. Roy. Astron. Soc.}\ }\textbf {\bibinfo {volume} {450}},\ \bibinfo
  {pages} {967} (\bibinfo {year} {2015})},\ \Eprint
  {http://arxiv.org/abs/1503.07002} {arXiv:1503.07002 [astro-ph.HE]}
  \BibitemShut {NoStop}%
\bibitem [{\citenamefont {Ryujin}\ \emph {et~al.}(2022)\citenamefont {Ryujin},
  \citenamefont {Vargas}, \citenamefont {Karlin}, \citenamefont {Dawson},
  \citenamefont {Weiss}, \citenamefont {Bertsch}, \citenamefont {McKinley},
  \citenamefont {Collette}, \citenamefont {Hammond}, \citenamefont {Pedretti},\
  and\ \citenamefont {Rieben}}]{osti_1838264}%
  \BibitemOpen
  \bibfield  {author} {\bibinfo {author} {\bibfnamefont {B.~S.}\ \bibnamefont
  {Ryujin}}, \bibinfo {author} {\bibfnamefont {A.}~\bibnamefont {Vargas}},
  \bibinfo {author} {\bibfnamefont {I.}~\bibnamefont {Karlin}}, \bibinfo
  {author} {\bibfnamefont {S.~A.}\ \bibnamefont {Dawson}}, \bibinfo {author}
  {\bibfnamefont {K.}~\bibnamefont {Weiss}}, \bibinfo {author} {\bibfnamefont
  {A.}~\bibnamefont {Bertsch}}, \bibinfo {author} {\bibfnamefont {M.~S.}\
  \bibnamefont {McKinley}}, \bibinfo {author} {\bibfnamefont {M.~R.}\
  \bibnamefont {Collette}}, \bibinfo {author} {\bibfnamefont {S.~D.}\
  \bibnamefont {Hammond}}, \bibinfo {author} {\bibfnamefont {K.}~\bibnamefont
  {Pedretti}}, \ and\ \bibinfo {author} {\bibfnamefont {R.~N.}\ \bibnamefont
  {Rieben}},\ }\href {\doibase 10.2172/1838264} {\  (\bibinfo {year} {2022}),\
  10.2172/1838264}\BibitemShut {NoStop}%
\bibitem [{\citenamefont {\relax
  Intel~Corporation}()}]{IntelARK_XeonPlatinum9242}%
  \BibitemOpen
  \bibfield  {author} {\bibinfo {author} {\bibnamefont {\relax
  Intel~Corporation}},\ }\href
  {https://ark.intel.com/content/www/us/en/ark/products/194145/intel-xeon-platinum-9242-processor-71-5m-cache-2-30-ghz.html}
  {\enquote {\bibinfo {title} {{Intel Xeon Platinum 9242 Processor
  Specifications}},}\ }\bibinfo {note} {Accessed 29 March 2023}\BibitemShut
  {NoStop}%
\bibitem [{\citenamefont {Icha}(2022)}]{Umweltbundesamt2021}%
  \BibitemOpen
  \bibfield  {author} {\bibinfo {author} {\bibfnamefont {P.}~\bibnamefont
  {Icha}},\ }\href
  {https://www.umweltbundesamt.de/sites/default/files/medien/1410/publikationen/2022-04-13_cc_15-2022_strommix_2022_fin_bf.pdf}
  {\enquote {\bibinfo {title} {{Entwicklung der spezifischen
  Treibhausgas-Emissionen des deutschen Strommix in den Jahren 1990 - 2021}},}\
  } (\bibinfo {year} {2022})\BibitemShut {NoStop}%
\bibitem [{\citenamefont {Bernal}\ \emph {et~al.}(2018)\citenamefont {Bernal},
  \citenamefont {Murray},\ and\ \citenamefont {Pearson}}]{Bernal2018}%
  \BibitemOpen
  \bibfield  {author} {\bibinfo {author} {\bibfnamefont {B.}~\bibnamefont
  {Bernal}}, \bibinfo {author} {\bibfnamefont {L.~T.}\ \bibnamefont {Murray}},
  \ and\ \bibinfo {author} {\bibfnamefont {T.~R.~H.}\ \bibnamefont {Pearson}},\
  }\href {\doibase 10.1186/s13021-018-0110-8} {\bibfield  {journal} {\bibinfo
  {journal} {Carbon Balance and Management}\ }\textbf {\bibinfo {volume} {13}}
  (\bibinfo {year} {2018}),\ 10.1186/s13021-018-0110-8}\BibitemShut {NoStop}%
\end{thebibliography}%

\end{document}